\documentclass[graybox, nosecnum]{svmult}

\usepackage{mathptmx}       
\usepackage{helvet}         
\usepackage{courier}        
\usepackage{type1cm}        
\usepackage{amssymb}        
\DeclareMathAlphabet{\mathcal}{OMS}{cmsy}{m}{n}
\SetMathAlphabet{\mathcal}{bold}{OMS}{cmsy}{b}{n}
\usepackage{makeidx}         
\usepackage{graphicx}        
\usepackage{multicol}        
\usepackage{array}
\usepackage[bottom]{footmisc}
\usepackage{hyperref}        
\usepackage{soul}            
\hypersetup{colorlinks=true,urlcolor=blue}
\usepackage[square,numbers]{natbib}
\usepackage{comment}

\usepackage{xcolor}
\definecolor{purple2}{rgb}{0.5, 0.0, 0.5}

\definecolor{sapphire}{rgb}{0.03, 0.15, 0.4}

\definecolor{teal}{rgb}{0.0, 0.5, 0.5}
\newcommand{\DMH}[1]{\textcolor{teal}{\textbf{[DH: #1]}}}
\definecolor{robineggblue}{rgb}{0.0, 0.8, 0.8}

\newcommand{\mrm}[1]{\mathrm{#1}}
\newcommand{\nuc}[2]{$\mrm{^{#2}#1}$}

\makeindex             

\begin{document}
\title*{Telescope Concepts in Gamma-Ray Astronomy}
\author{Thomas Siegert, Deirdre Horan, Gottfried Kanbach}
\institute{Thomas Siegert \at Institut f\"ur Theoretische Physik und Astrophysik, Universit\"at W\"urzburg, Campus Hubland Nord, Emil-Fischer-Str. 31, 97074 W\"urzburg, Germany \email{thomas.siegert@uni-wuerzburg.de}
\and
Deirdre Horan \at Laboratoire Leprince-Ringuet, CNRS/IN2P3, Institut Polytechnique de Paris, F-91128 Palaiseau, France \email{deirdre@llr.in2p3.fr} 
\and 
Gottfried Kanbach \at Max Planck Institute for Extraterrestrial Physics, Giessenbachstrasse 1, 85748 Garching, Germany, \email{gok@mpe.mpg.de}} 
%
%
\maketitle
\abstract{
This chapter outlines the general principles for the detection and characterisation of high-energy $\gamma$-ray photons in the energy range from MeV to GeV.
Applications of these fundamental photon--matter interaction processes to the construction of instruments for $\gamma$-ray astronomy are described, including a short review of past and present realisations of telescopes.
The constraints encountered in operating telescopes on high-altitude balloon and satellite platforms are described in the context of the strong instrumental background from cosmic rays as well as astrophysical sources.
The basic telescope concepts start from the general collimator aperture in the MeV range over its improvements through coded-mask and Compton telescopes, to pair production telescopes in the GeV range.
Other apertures as well as understanding the measurement principles of $\gamma$-ray astrophysics from simulations to calibrations are also provided.
}


\pagebreak

\section{Introduction}\label{sec:intro}
%
Gamma rays ($\gamma$-rays) are traditionally defined as penetrating electromagnetic radiation that arise from the radioactive decay of an atomic nucleus and, indeed, for $\gamma$-rays produced naturally on Earth, this is the case.
Gamma rays constitute the electromagnetic radiation having energy of $\gtrsim 100$\,keV and, therefore, have energies that traverse more than ten decades of the electromagnetic spectrum.
In addition to those $\gamma$-rays coming from radioactive decays, the extra-terrestrial $\gamma$-rays incident on Earth are produced in a variety of different astrophysical scenarios.
These include when extremely energetic charged particles accelerate in magnetic fields or upscatter ambient radiation to $\gamma$-ray energies, hadronic processes such as cosmic-ray (CR) interactions in the Galaxy and, indeed, possibly in more exotic interactions such as, for example, the self-annihilation of dark matter particles (Sec.\,\ref{sec:astrophysical_sources}). 

In this Chapter an outline of the various techniques and instruments for the detection and characterisation of $\gamma$-rays will be presented.
The limitations and advantages of each particular detection technique, the backgrounds that must be overcome, and the environmental circumstances that must be considered will be reviewed.
Only the direct detection of $\gamma$-rays will be discussed here thus effectively limiting the upper energy range to approximately 100\,GeV.
For the detection techniques at so-called very-high energy (VHE; $E_{\gamma} \gtrsim 100$\,GeV) the reader is referred to other chapters of this book.

In order to detect $\gamma$-rays, we rely on one of their three interactions with matter: the photoelectric effect, Compton scattering and pair production, whose respective cross sections depend on the energy of the $\gamma$-ray and on the material in which it interacts (Sec.\,\ref{sec:cross_sections}).
Since $\gamma$-rays cannot penetrate the Earth's atmosphere, it having an equivalent thickness of approximately 0.9\,m of lead, a detector needs to be placed above the atmosphere in order to detect $\gamma$-rays directly (Sec.\,\ref{sec:measurement_conditions}). 
Like no other energy range the $\gamma$-ray regime is dominated by an irreducible instrumental background (MeV energies) and limited by collection area (GeV energies) and telemetry (MeV \& GeV energies; Sec.\,\ref{sec:instrumental_background}).
Due to these factors and to the energy-dependant cross section of a $\gamma$-ray's interaction with matter, the discussions in this Chapter will be split according to the detection technique being employed, which is itself a function of the energy range of the $\gamma$-rays being studied.

Because of CR bombardment, MeV telescopes suffer from a high level of secondary $\gamma$ radiation.
This includes electron bremsstrahlung, spallation, nuclear excitation, delayed decay, and annihilation, all of which contribute to the instrumental background in the MeV range.
This orders of magnitude enhanced rate of unwanted events leads to a worse instrument sensitivity -- the `MeV Sensitivity Gap' -- compared to neighboring photon energy bands (Sec.\,\ref{sec:MeV_gap}).
The correct identification of background photons from celestial emission leads to an artificial split in the science cases (Sec.\,\ref{sec:science_cases}) because $\gamma$-ray instruments are built to observe either in the MeV or in the GeV.
Even though astrophysical high-energy sources can span several decades in the electromagnetic spectrum, the MeV regime is often omitted because the sensitivities of current instruments rarely add much spectral information.
In this Chapter we therefore handle MeV- and GeV-type instruments separately.
We note, however, that leaving out information, even though it appears weak in the first place, leads to a biased view of the astrophysics to be understood.

In the pair-production regime at tens of MeV to GeV energies, the spectra of $\gamma$-ray sources phenomenologically follow a power law such that the flux changes rapidly as a function of energy.
With the exception of grazing incident mirrors used for hard X-rays in the energy range up to $\sim 100$\,keV (Sec.\,\,\ref{sec:other_apertures}), $\gamma$-rays cannot be focused and, therefore, in order to be detected, need to enter the detector and interact with it.
So, unlike optical telescopes where a large effective area can be achieved by using a huge mirror to focus the optical photons onto a small detector, a large collection area can only be achieved at $\gamma$-ray energies by having a large detector volume.
This requirement coupled with the constraints of launching a large mass high enough in the atmosphere that it can detect sufficient $\gamma$-rays (Sec.\,\ref{sec:atmospheric_effects}) limit the upper bound energy at which $\gamma$-rays can effectively be detected directly: for the direct detection of $\gamma$-rays the physical volume of the detector is always larger than its effecting detecting volume and, therefore, its effective area.
The requirement to observe from space has the advantage of fewer constraints with respect to observing schedules (e.g., no day and night cycles), however limits the sensitive collecting areas of the telescopes because of the mass that can be transferred into an orbit.

In this Chapter, we will introduce the only 60-year long history of $\gamma$-ray observations (Sec.\,\ref{sec:historical}), describe the least explored range of the electromagnetic spectrum, the MeV range (Sec.\,\ref{sec:MeV_gap}), and present the basic interactions of high-energy photons with matter that are used in all $\gamma$-ray telescopes (Sec.\,\ref{sec:cross_sections}).
Details about the instruments' current capabilities and requirements to study high-energy sources are given in Sec.\,\ref{sec:instrument_requirements}, followed by an extensive discussion of instrumental background origins in Sec.\,\ref{sec:instrumental_background}, and state-of-the-art suppression mechanisms (Sec.\,\ref{sec:background_suppression}).
Based on the different science cases in the MeV and GeV range (Sec.\,\ref{sec:science_cases}), we detail principal instrument designs in Sec.\,\ref{sec:instrument_designs}.
This includes the basic collimator (Sec.\,\ref{sec:general_collimator}), coded mask telescopes (Sec.\,\ref{sec:temporal_spatial_apertures}), Compton telescopes (Sec.\,\ref{sec:compton_telescopes}), pair creation telescopes (Sec.\,\ref{sec:pair_tracking_telescopes}), $\gamma$-ray polarimeters (Sec.\,\ref{sec:polarimeters}), as well as other, alternative, but not necessarily yet realised instruments (Sec.\,\ref{sec:other_apertures}).
Gamma-ray detectors are briefly explained in Sec.\,\ref{sec:detectors}, followed by how $\gamma$-ray measurements are to be understood, evaluated (Sec.\,\ref{sec:measurements}), and compared to simulations (Sec.\,\ref{sec:simulations}) and calibrations (Sec.\,\ref{sec:calibration}).
We close this Chapter with an outlook about future, and possible more advanced concepts in Sec.\,\ref{sec:outlook}.

\subsection{Historical Perspective}\label{sec:historical}
\subsubsection{First Observations}\label{sec:first_observations}
Gamma-ray astronomy, the highest-energy range of multi-wavelength astronomy, was already recognised in the 1950's as having the potential to provide direct insight into astrophysical processes with particles, fields, and dynamics of extreme conditions in the Universe.
The principal source processes for high-energy $\gamma$ radiation in space (beyond the energy range of radioactivity) were studied first.
These include synchrotron radiation \citep{Iwanenko1944_betatron}, Compton scattering \citep{Feenberg1948_ComptonSun}, meson production and the decay of $\pi ^{0}\rightarrow 2 \gamma$ \citep{Hayakawa1952_Pions}, and bremsstrahlung \citep{Hutchinson1952_CRbremsstrahlung}.

The status of particle physics, CR research, and radio astronomy in the 1950's raised widely-debated questions such as:
`\textit{Where do CRs come from and how are they produced?}', `\textit{What is the photon fraction in the CR beam?}', `\textit{What powers the strong Galactic radio emission?}', `\textit{Are there discrete $\gamma$-ray sources in the sky and what could be the nature of such sources?}', `\textit{Do the observed particle emissions from solar flares lead to $\gamma$-ray emissions (nuclear lines and continua)?}', and `\textit{Is there antimatter around?}'.

\begin{figure}[b]
    \centering
    \includegraphics[width=0.7\textwidth,trim=0.0cm 9.0cm 0.0cm 0.0cm,clip=True]{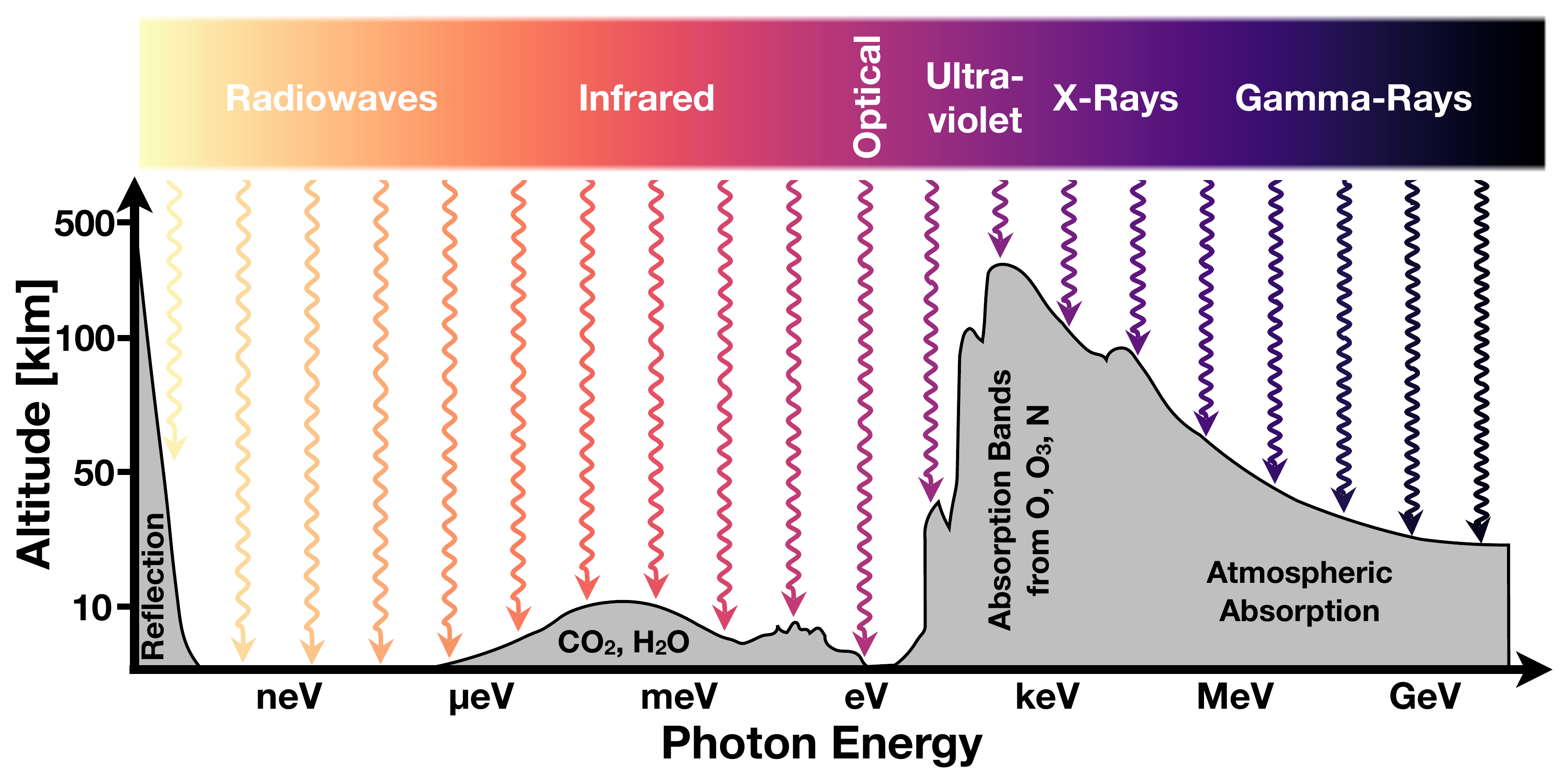}
    \caption{Absorption of cosmic electromagnetic radiation in Earth's atmosphere.}
    \label{fig:atmosphere}
\end{figure}

Estimates for the strength of cosmic $\gamma$-ray sources were mainly based on the contemporary knowledge of CRs, the distribution and density of the Galactic interstellar medium (HI radio emission), and the observations of radio emission from individual objects like the Crab nebula or radio galaxies.
Morrison (1958, \citep{morrison1958}) estimated that the active Sun would emit $0.1$--$1\,\mathrm{ph\,cm^{-2}\,s^{-1}}$ between 10 and 100\,MeV, and $1$-$100\,\mathrm{ph\,cm^{-2}\,s^{-1}}$ in the neutron-proton capture line at 2.23\,MeV.
The Crab nebula (the pulsar was unknown at the time) and typical radio galaxies should have intensities of $10^{-2}\,\mathrm{ph\,cm^{-2}\,s^{-1}}$.
A thorough study of $\gamma$-ray production by CRs interacting with the interstellar medium in the Galaxy by Pollack \& Fazio (1963, \citep{Pollack1963_pion511}) predicted a flux from the Galactic Centre of $\sim 10^{-4}\,\mathrm{ph\,cm^{-2}\,s^{-1}\,sr^{-1}}$ and half that intensity from the Anticentre.
All of these flux estimates turned out to be much too high, but nevertheless many experiments were started to detect celestial $\gamma$-rays.
Short exposures on balloons and a very strong environmental background prevented significant detection of $\gamma$-rays from the Milky Way or from discrete sources.
The beginning of the space age in 1958 finally provided the facilities to operate $\gamma$-ray experiments above the atmosphere.
The clear, unabsorbed view of the sky, the longer exposures, the absence of the atmospheric background, and advanced instruments succeeded in establishing $\gamma$-ray astronomy as a new and promising branch of astrophysics.

Gamma-ray astronomy is a discipline that depends on the technical resources of the space age.
The ground level of Earth-bound observatories is shielded from cosmic $\gamma$ radiation by the atmosphere (Fig.\,\ref{fig:atmosphere}), with roughly 20 (resp. 60) mean free path lengths of attenuation at 1\,GeV (resp. 1\,MeV)
Furthermore, the charged fraction of cosmic radiation, which dominates primary $\gamma$-rays by roughly a factor of $10^4$, generates a high level of secondary $\gamma$-ray background in the atmosphere and in detector equipment.
It is therefore essential to expose a $\gamma$-ray telescope in a low level of external background, i.e. above the atmosphere but below the Earth’s radiation belts, for long periods of observation to obtain the necessary detection statistics.
For satellites this is best achieved in a low Earth orbit (LEO) above the equator with an altitude of 400--500\,km.
Equally important is the design of the telescope so as to suppress the recording of unwanted charged particles (veto systems) and local $\gamma$ radiation (material selections).

Both requirements directly impact the sensitivity of a $\gamma$-ray telescope with detector exposure area $A$, detection efficiency $\epsilon$, and angular resolution elements of size $\theta$.
Discrete cosmic sources with fluxes of $F_{\gamma}$ embedded in a `quasi’ continuous background intensity $I_B$ observed for an exposure time $t_{\rm obs}$ are then detected with a statistical sensitivity of $S$: 
\begin{equation}
\label{eq:sensitivity}
S = { {F_{\gamma} A \epsilon t_{obs}} \over { {(I_B  A \epsilon t_{\rm obs} \pi \theta ^2)}^{1/2}}} = {{F_{\gamma} \over \theta} \left( {{A \epsilon t_{\rm obs}} \over {I_B \pi}} \right)^{1/2}} 
\end{equation}
\noindent It is evident from Eq.\,(\ref{eq:sensitivity}) that high sensitivity is the result of a large effective area, $A_{\rm eff} = A\epsilon$, and long observation times, combined with small angular resolution and low background intensities.
Of course this formulation is extremely simplified compared to more appropriate analysis tools using proper instrument response functions for effective detector area, angular and energy resolution, and detailed models for the background radiation, all as a function of primary energy and incidence direction.
%

\begin{figure}[ht]
    \centering
    \includegraphics[width=0.7\textwidth,trim=0.0cm 0.33cm 0.0cm 0.0cm,clip=True]{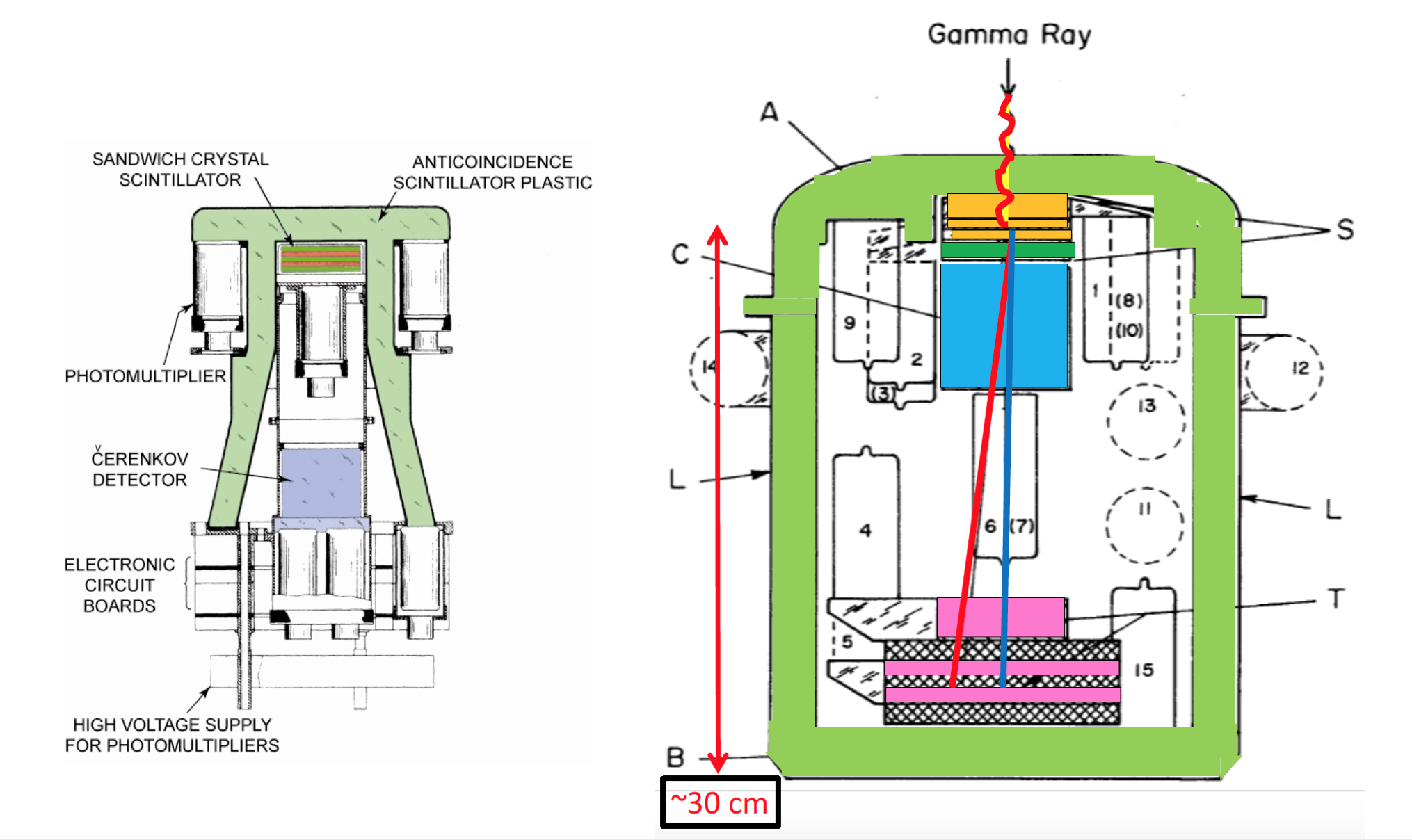}
    \caption{
        \textbf{\textit{Left}}: Explorer XI (Apr--Nov, 1961): The small effective area ($\sim 5\,\mathrm{cm^2}$), and a useful observation time of $\sim 6$\,d resulted in only 22 detected high-energy $\gamma$-rays, mostly from Earth's atmosphere.
        \textbf{\textit{Right}}: OSO-3 (1967--69): The successor of Explorer XI. A typical scintillator counter telescope with a multi-layer conversion detector (CsI), a fast trigger system and a calorimeter with layers of NaI and tungsten. The instrument is surrounded by an anti coincidence veto counter made of plastic scintillator. A Galactic map of 621 events ($E>50$\,MeV) was derived from $\sim 16$ months of observations \citep{Kraushaar1972_OSO3gammarays}, but the coarse angular resolution ($15^{\circ}$) prevented the detection of point-sources.
    }
    \label{fig:Explorer-OSO}
\end{figure}

\begin{figure}[b]
    \centering
    \includegraphics[width=0.6\textwidth]{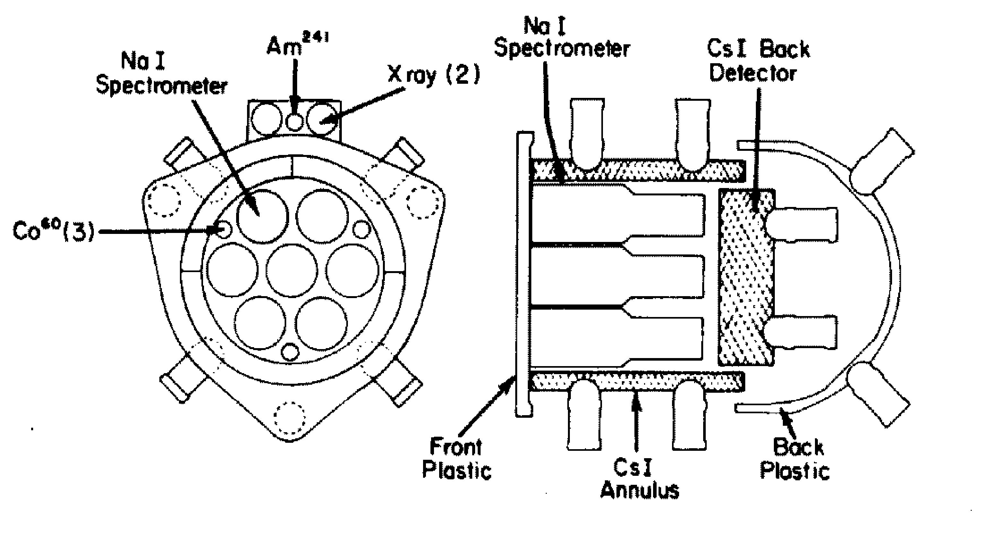}
    \caption{
         SMM/GRS (1981--90): an actively shielded multi-crystal scintillation spectrometer, sensitive to photons in the range $0.3$--$100$\,MeV \citep{Forrest1980_SMM}. SMM was continuously pointed at the Sun. The open acceptance angle of about ($135^{\circ}$) in the forward direction prevented the identification of individual sources, but allowed the instrument to monitor the temporal signatures of solar flares.
    }
    \label{fig:SMM}
\end{figure}

\subsubsection{Missions 1960-1990}

The first successful satellite detectors for high-energy $\gamma$-radiation were small scintillation \v{C}erenkov counter assemblies with anticoincidence shields.
As depicted in Fig.\,\ref{fig:Explorer-OSO} they had to fit on the satellites of the 1960's and could only transmit data with limited rates.
The emission of $>100$\,MeV photons from the inner Galaxy was, however, clearly established by the OSO-3 measurements \citep{Kraushaar1972_OSO3gammarays} and confirmed by a spark-chamber imaging balloon experiment \citep{Fichtel1969_gammarayballoon}.
Here, it is interesting to note a performance comparison between the scintillator telescope and the pointed balloon instrument: both could achieve similar results on the Galactic emission with an effective area of $2$--$8\,\mathrm{cm^2}$ even though the former required $\sim 16$ months of observation time whereas the balloon flight only required several hours.

Gamma-ray instruments for the low-energy range $<10$\,MeV can be based on Compton interactions (e.g., {\it{CGRO}}-COMPTEL) or be designed as spectrometers for more narrowly defined targets and scientific objectives.
In order to restrict the acceptance angle of a scintillator or solid-state spectrometer, with its omni-directional response,  a massive `well-type' collimator is placed around a central detector.
Collimators can be either active radiation detectors, for example made of BGO or CsI scintillators, or passive structures made of high-Z metals.
Two examples of successful instruments are the $\gamma$-ray spectrometer (GRS) on the Solar Maximum Mission (SMM, 1981--90; Fig.\,\ref{fig:SMM}), and the Oriented Scintillation-Spectrometer Experiment (OSSE) on the {\it{Compton Gamma-Ray Observatory}} ({\it{CGRO}}, 1991--2000).

The advantage of an imaging telescope for astronomical observations was clearly established and led to the next generation of high-energy detectors, SAS-2 and COS-B.
Both high-energy satellite telescopes were based on digital readout spark-chambers that allowed for the reconstruction of pair-creation events by tracking the electron-positron pairs.
Around the central tracker a charged particle anti-coincidence shield made of plastic scintillator and, for COS-B, a calorimeter to measure the deposited pair energy, were used.
SAS-2 was developed on the basis of previous balloon detectors at NASA/GSFC and the COS-B instrument was built by a European consortium of research institutes.

\subsection{\textit{The `MeV Sensitivity' Gap}}\label{sec:MeV_gap}
\begin{figure}[b]
    \centering
    \includegraphics[width=0.75\textwidth]{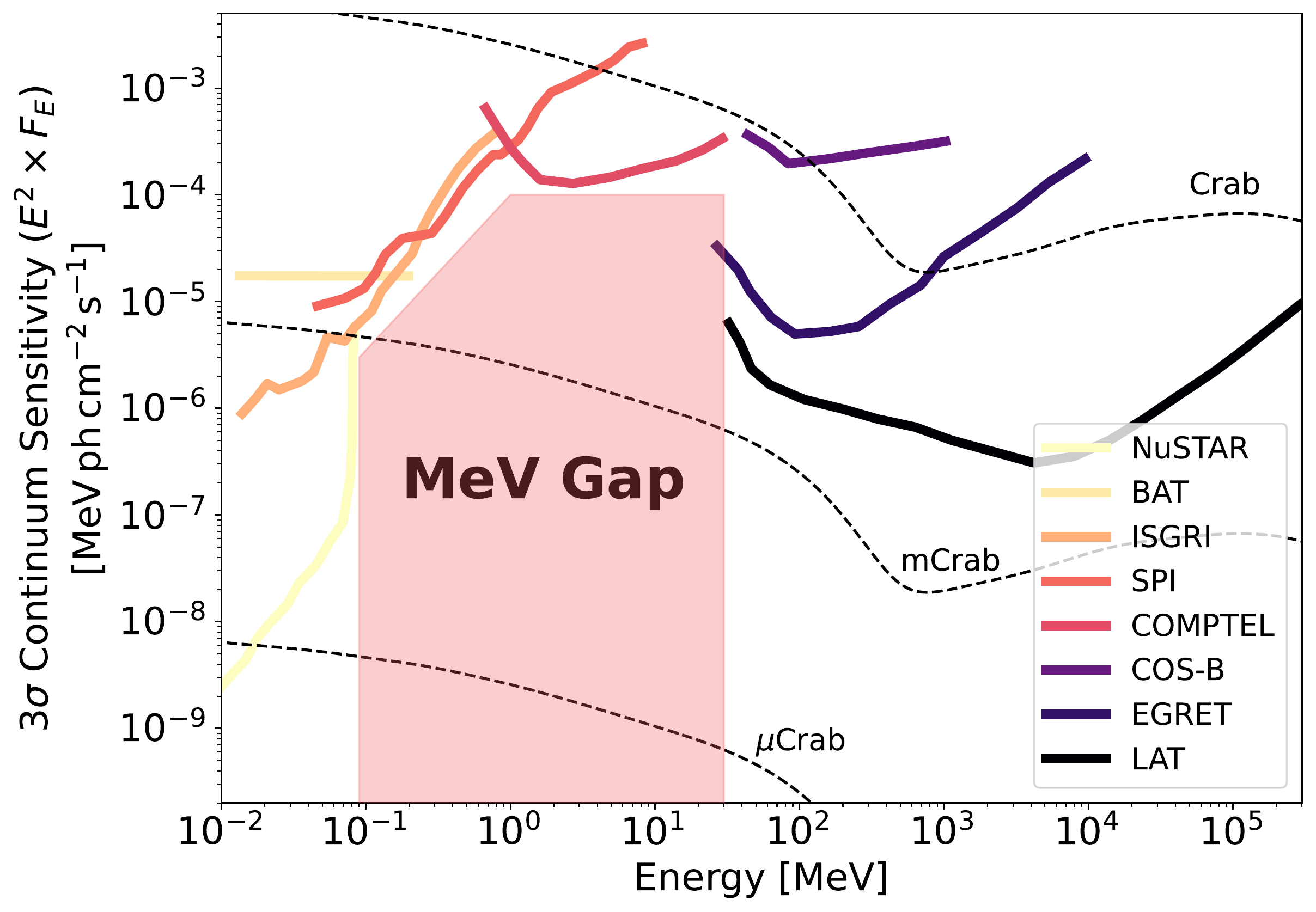}
    \caption{Continuum sensitivities of hard X-ray to high-energy $\gamma$-ray instruments. Shown is the $3\sigma$ sensitivity for an observation time of $1\,\mathrm{Ms}$. The Crab's spectral energy distribution from \citep{Meyer2010_Crab} is shown with respect to the sensitivities as it is the `standard candle' of high-energy sources. MilliCrab (mCrab) flux levels can only be seen with deep exposures in the 0.3--100\,GeV range or below 100\,keV. This defines the `MeV Gap' of instrument sensitivities (red shaded area) -- the least explored region in the electromagnetic spectrum.}
    \label{fig:sensitivities}
\end{figure}

Fig.\,\ref{fig:sensitivities} shows the sensitivities for past and current $\gamma$-ray instruments in the range between $10^{-2}$ and $10^5$\,MeV.
While sensitivity should be defined case by case, i.e. depending on the source spectrum, its spatial distribution, and position in the instrument field of view, an order of magnitude estimate of the instrument performance can be given assuming a generic spectral shape at each photon energy.
When provided with background estimates, the effective area, and a typical exposure time (here 1\,Ms), the sensitivity can be calculated from Eq.\,(\ref{eq:sensitivity}) to the desired level (here $3\sigma$).
In general, the lower the sensitivity the better the instrument performs.
It is evident that the currently flying telescopes {\it{NuSTAR}} ($<80$\,keV), INTEGRAL's ISGRI and SPI (0.02--8\,MeV), and {\textit{Fermi}}-LAT ($>0.03$\,GeV) shape a region in sensitivity space that peaks in the MeV.
This several orders of magnitude worse sensitivity is called the `MeV Sensitivity Gap', and is the direct result of a small collection area (Sec.\,\ref{sec:instrument_designs}) combined with a high instrumental background (Sec.\,\ref{sec:instrumental_background}).
Reducing this MeV gap is a currently an active field of technological, methodological, and conceptual development, and attempts to alleviate the problems in the MeV range are described in Secs.\,\ref{sec:other_apertures} and \ref{sec:outlook}.

In addition to this MeV gap for continuum emission, there is also a similar problem for nuclear $\gamma$-ray lines.
While COMPTEL on {\it{CGRO}} could, for example, identify narrow line emission at 1.8\,MeV, its spectral resolution was only 10\,\% (FWHM) so that many lines blended together to form one broad feature.
High spectral resolution in the MeV range can be achieved by the use of germanium detectors (Sec.\,\ref{sec:detectors}), such as in {\it{RHESSI}} or SPI.
While increased spectral resolution helps to identify background features more easily, the small collecting area still prohibits the investigation of many potential astrophysical sources.
As of now, only a dozen nuclear lines of astrophysical origin have been observed with {\it{HEAO-3}}, COMPTEL, {\it{RHESSI}}, SPI (and {\it{NuSTAR}}).
These include the positron annihilation line from the centre of the Galaxy at 511\,keV \citep[e.g.,][]{1994ApJS...92..387M,Purcell1997_511,Jean2006_511,Churazov2011_511,Siegert2016_511,Siegert2021_BDHanalysis}, short- and long-lived ejecta from massive stars and their supernovae such as \nuc{Ti}{44} \citep[e.g.,][at 68, 78, 1157\,keV]{Iyudin1997_CasA,Renaud2006_CasA,Grefenstette2014_CasA,Boggs2015_SN1987A,Siegert2015_CasA,Weinberger2020_CasA}, \nuc{Al}{26} \citep[e.g.,][at 1809\,keV]{Mahoney1984_26Al,Diehl2006_26Al,Kretschmer2013_26Al,Siegert2016_Orion26Al,Pleintinger2019_26Al}, and \nuc{Fe}{60} \citep[e.g.,][at 1173 and 1332\,keV]{Harris2005_Fe60,Wang2007_60Fe,Wang2020_Fe60}, short-lived isotopes powering the early light-curves of type Ia supernovae \citep[e.g.,][with \nuc{Ni}{56} and \nuc{Co}{56} at 158, 812, and 847, 1238\,keV, respectively]{Diehl2014_SN2014J_Ni,Diehl2015_SN2014J_Co,Churazov2014_SN2014J,Isern2016_SN2014J}, as well as nuclear excitation lines from solar flares \citep[e.g.,][with  511\,keV from electron-positron annihilation, \nuc{H}{2} at 2223, \nuc{C}{12} at 4438, and \nuc{O}{16} at 6129\,keV, among others]{Gros2004_solarflare,Kiener2006_solarflare}.
With a factor of ten improvement in the line sensitivity, the number of detected lines, and therefore the science enabled by this, could increase by the same order of magnitude, eventually finding CR excitation of interstellar medium material, ejecta from classical novae, and multiple supernova lines \citep[e.g.,][]{Timmes2019_RA2020}.
The advantage of nuclear line studies is the possibility of finding an absolute measure of ejecta masses, CR fluxes, and kinematics, which may be biased by using observations at other wavelengths.
\subsection{Interactions of Light with Matter}\label{sec:cross_sections}
While for longer-wavelength light, most interactions with matter are either of refractive, reflective or diffractive nature owing to the wave character of light, higher energy photons experience processes prone to particles instead of waves.
These are used to determine the energy of the incoming light by measuring their partial or total deposits in the detecting material.
While more processes can occur, the most relevant reactions for X- and $\gamma$-ray photons are photoelectric absorption (photo-effect), Compton scattering, and pair production.

The photo effect \citep{Einstein1905_PE} describes a photon undergoing an interaction with an atom in which the photon deposits its total energy and is removed completely.
To conserve momentum and energy, a photoelectron is emitted by the absorbing atom.
Since the interaction is with the atom as a whole, having bound electrons in its shells, the photo effect cannot occur on free electrons.
The most probable electron to be ejected in photoelectric absorption is the one most tightly bound in the K-shell.
The photoelectron has an energy of $E_e = E_{\gamma} - E_b$ where $E_b$ is the binding energy of the electron in the atom.
The interaction probability for a $\gamma$-ray photon to undergo the photo effect is described by the cross section, typically as a function of energy,
\begin{equation}
    \sigma_{\rm PE} = \frac{16}{3} \sqrt{2} \pi r_e^2 \alpha^4 \frac{Z^5}{k^{3.5}}\mathrm{,}
    \label{eq:photo_effect_cross_section}
\end{equation}
where $r_e$ is the classical electron radius, $\alpha$ is the finestructure constant, $Z$ is the atomic charge number, and $k = E_{\gamma} / (m_ec^2)$ is the photon energy in units of electron rest masses \citep{Fornalski2018_cross_sections}.
Equation\,(\ref{eq:photo_effect_cross_section}) is a valid approximation for $k \lesssim 0.9$; for higher energies and, for more precision over large photon energy ranges, the cross section from \citep{Davisson1952_cross_sections} should be used.

At photon energies of approximately between $k=0.1$--$1.0$, depending on the material, the Compton effect \citep{Compton1923_CE} becomes the dominant interaction process of light with matter.
Compton scattering describes the process of a $\gamma$-ray undergoing scattering with an electron, assumed to be at rest.
The photon changes its path as a result of this process and transfers some of its energy to the electron which then recoils.
The deflection angle, also called the Compton scattering angle $\varphi$, is the fundamental property that determines the origin of the $\gamma$-rays in Compton telescopes (Sec.\,\ref{sec:compton_telescopes}).
Because, in principle, the range of scattering angles covers a full circle, the process of Compton scattering, i.e. the photon loses energy to enhance the kinetic energy of the electron, can also be inverted to Inverse Compton scattering, i.e. the photon gains energy by scattering with fast electrons.
In the range $k=0.2$--$20$ \citep{Fornalski2018_cross_sections}, the total cross section for Compton scattering is approximated by
\begin{equation}
    \sigma_{\rm CE} = Z 2 \pi r_e^2 \left\{ \frac{1+k}{k^2} \left[ \frac{2(1+k)}{1+2k} - \frac{\ln(1+2k)}{k} \right] + \frac{\ln(1+2k)}{2k} - \frac{1+3k}{(1+2k)^2} \right\}\mathrm{.}
    \label{eq:Compton_scattering_cross_section}
\end{equation}
Higher order corrections can again be found in \citep{Davisson1952_cross_sections}.

For $k > 2$, pair production \citep{Blackett1933_PP}, i.e. the conversion of a $\gamma$-ray into an electron-positron-pair, becomes possible.
While formally, the production of pairs starts at twice the rest mass energy of an electron of 1.022\,MeV, the interaction probability stays at a low level until the cross sections dominate, typically above $k=20$.
Pair production can occur in any electromagnetic field; for the detection of $\gamma$-rays, the Coulomb fields of nuclei are to be considered.
The $\gamma$-ray photon loses all of its energy in the process, is removed from the scheme, and replaced by a pair that carries the total energy of the photon.
The kinetic energy of the electron and positron, respectively, is symmetric about half the energy of the incident photon, minus the rest mass energy of the electron.
The interaction cross section for pair production \citep{Fornalski2018_cross_sections,Davisson1952_cross_sections} is
\begin{equation}
    \sigma_{\rm PP} = Z^2 \alpha r_e^2 \left( \frac{28}{9}\ln2k - \frac{218}{27} + \mathcal{O}(\ln k / k^2) \right)\mathrm{,}
    \label{eq:pair_production_cross_section}
\end{equation}
where higher order terms span several lines of terms.
The important feature to note here is that the cross section for pair production in the field of a nucleus increases with the charge number of the nucleus squared.

\begin{figure}
    \centering
    \includegraphics[width=0.49\textwidth]{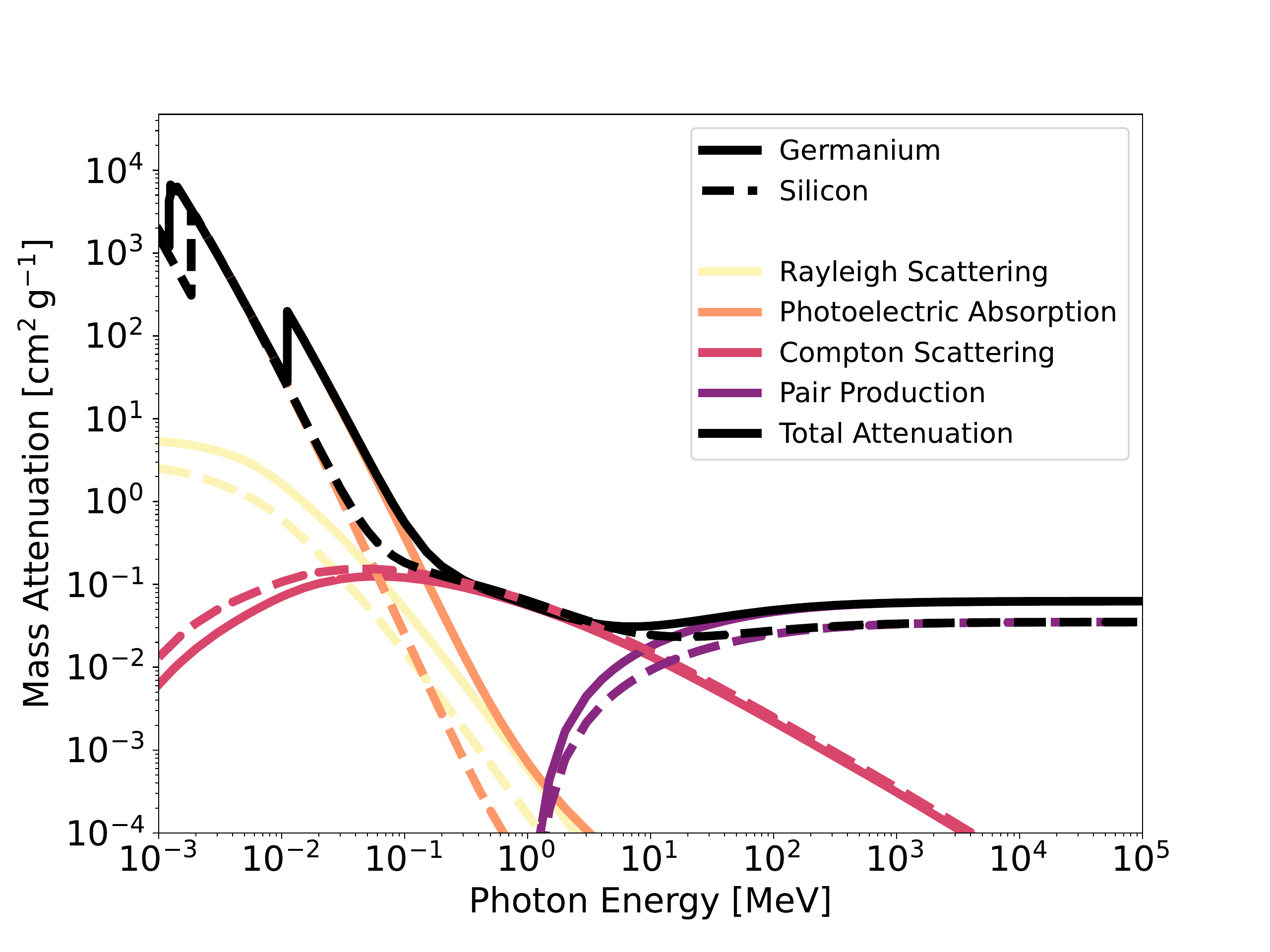}
    \includegraphics[width=0.49\textwidth]{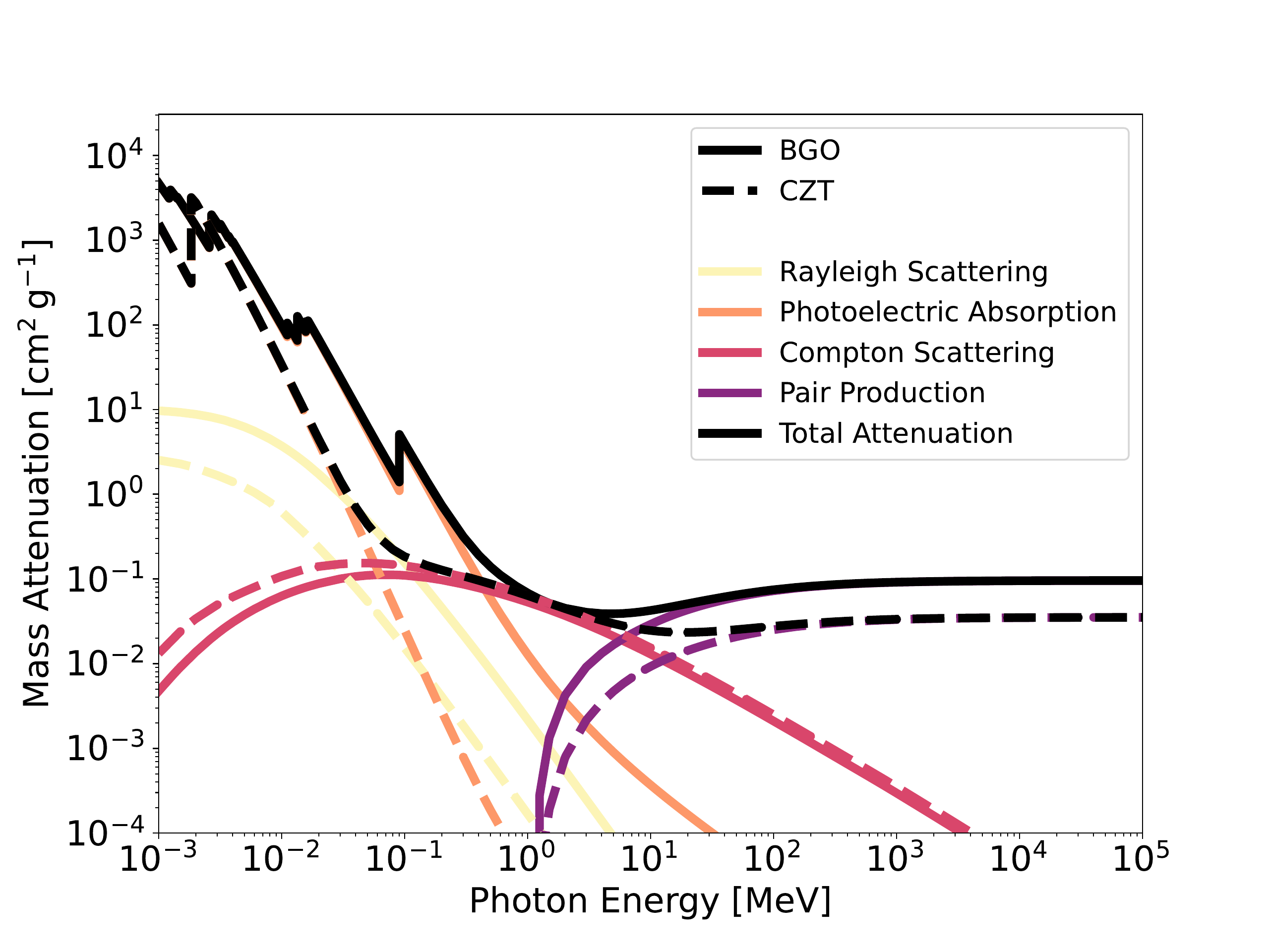}
    \caption{Mass attenuation coefficients for commonly used detector materials as a function of photon energy. The individual interaction processes are shown as coloured lines. \textit{\textbf{Left}}: Germanium (solid) and Silicon (dashed). Compton scattering dominates in the energy range $\sim 200$ keV to $\sim 10$ MeV which makes these semiconductors efficient scattering detectors. \textit{\textbf{Right}}: BGO (solid) and CZT (dashed). In these high-Z materials absorption through photo-effect or pair-creation is more pronounced.}
    \label{fig:cross_sections}
\end{figure}

In detail, the cross sections vary for different materials, compositions, and matter structures.
In Fig.\,\ref{fig:cross_sections}, the mass attenuation coefficients, $n\sigma/\rho$, with $\rho$ being the density and $n$ being the number density of the material, for $\gamma$-ray detector media that are typically used are shown.
The shapes of photo effect, Compton scattering, and pair production are similar for the elements and compounds shown, however the minuscule details change the behaviours and areas of use of the detectors.
For example, plastic shows a much broader Compton scattering regime compared to other scintillating materials (e.g., BGO), making it the scattering material of choice of classic Compton telescopes (Sec.\,\ref{sec:compton_telescopes}).

\section{Instrument Capabilities and Requirements}\label{sec:instrument_requirements}
In order to do $\gamma$-ray astronomy, the direction from which the $\gamma$-ray originated, its time of arrival, its energy and its polarisation would, ideally, be determined accurately.
Depending upon the energy of the incident $\gamma$-ray and upon the nature of the source of interest, different types of $\gamma$-ray detectors are required for this task.
As will be discussed in Sec.\,\ref{sec:science_cases}, some scientific objectives require highly-accurate energy resolution, usually achieved at the expense of positional accuracy i.e., angular resolution.
Conversely, when high angular resolution is required, the spectral accuracy of the measurement usually has to be compromised.
Gamma-ray polarimetry is an upcoming field and individual Chapters in this book are dedicated to this topic; a brief overview of measuring the polarisation of $\gamma$-rays is provided later in this current Chapter. 

\begin{figure}[t]
    \centering
    \includegraphics[width=0.49\textwidth,trim=0.2in 0.2in 0.9in 0.6in,clip=True]{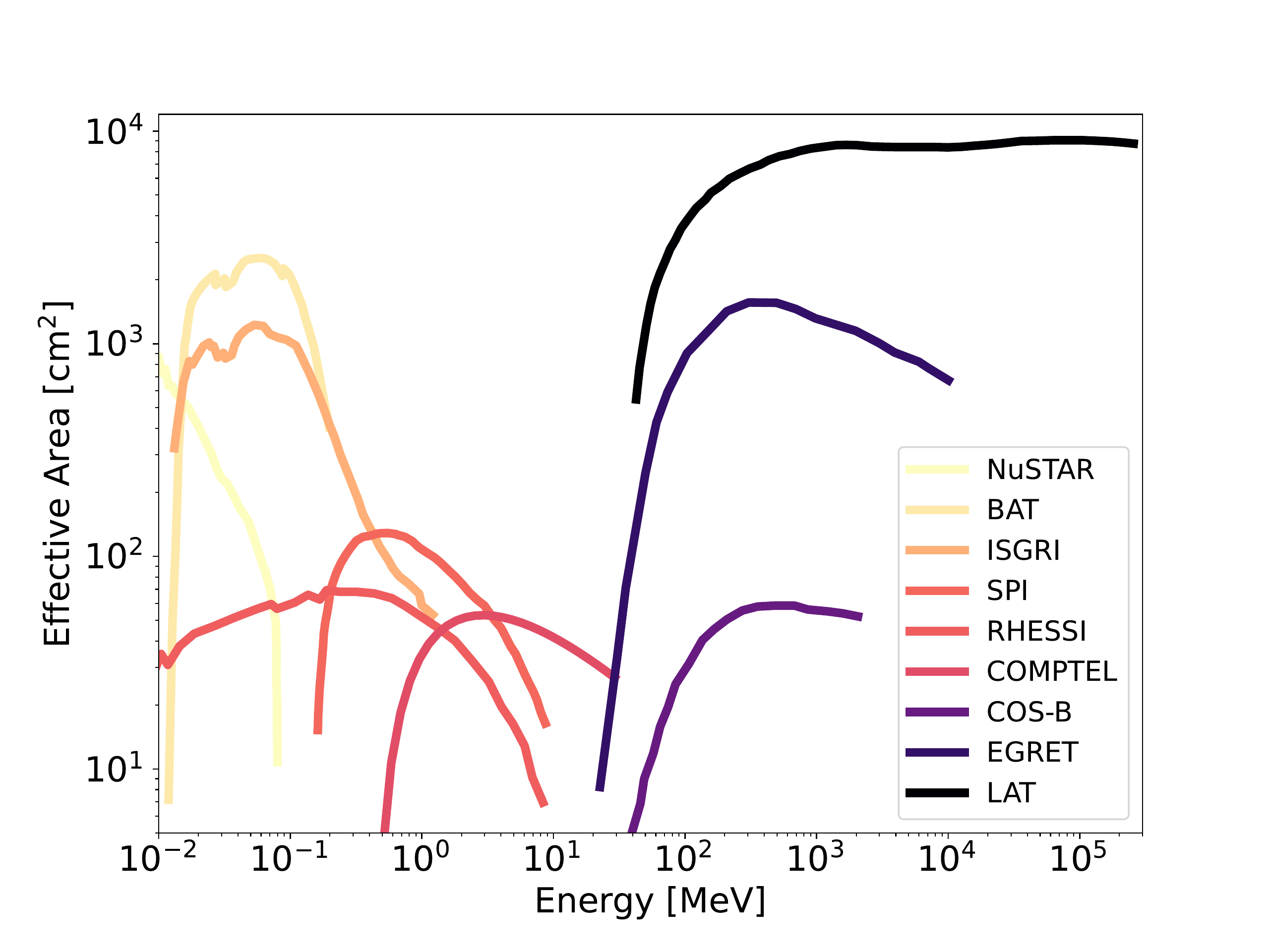}
    \includegraphics[width=0.49\textwidth,trim=0.2in 0.2in 0.9in 0.6in,clip=True]{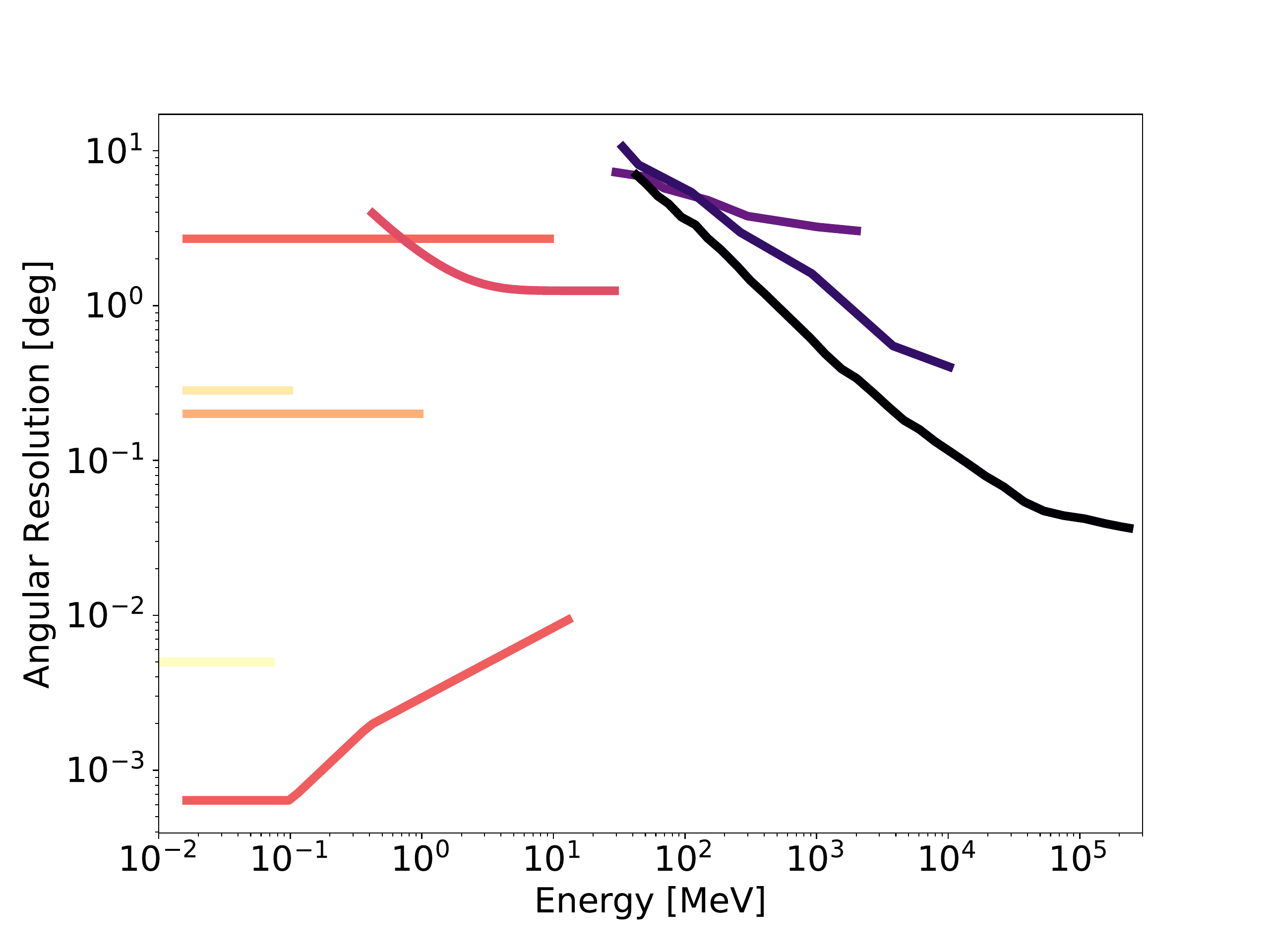}\\
    \includegraphics[width=0.49\textwidth,trim=0.2in 0.2in 0.9in 0.6in,clip=True]{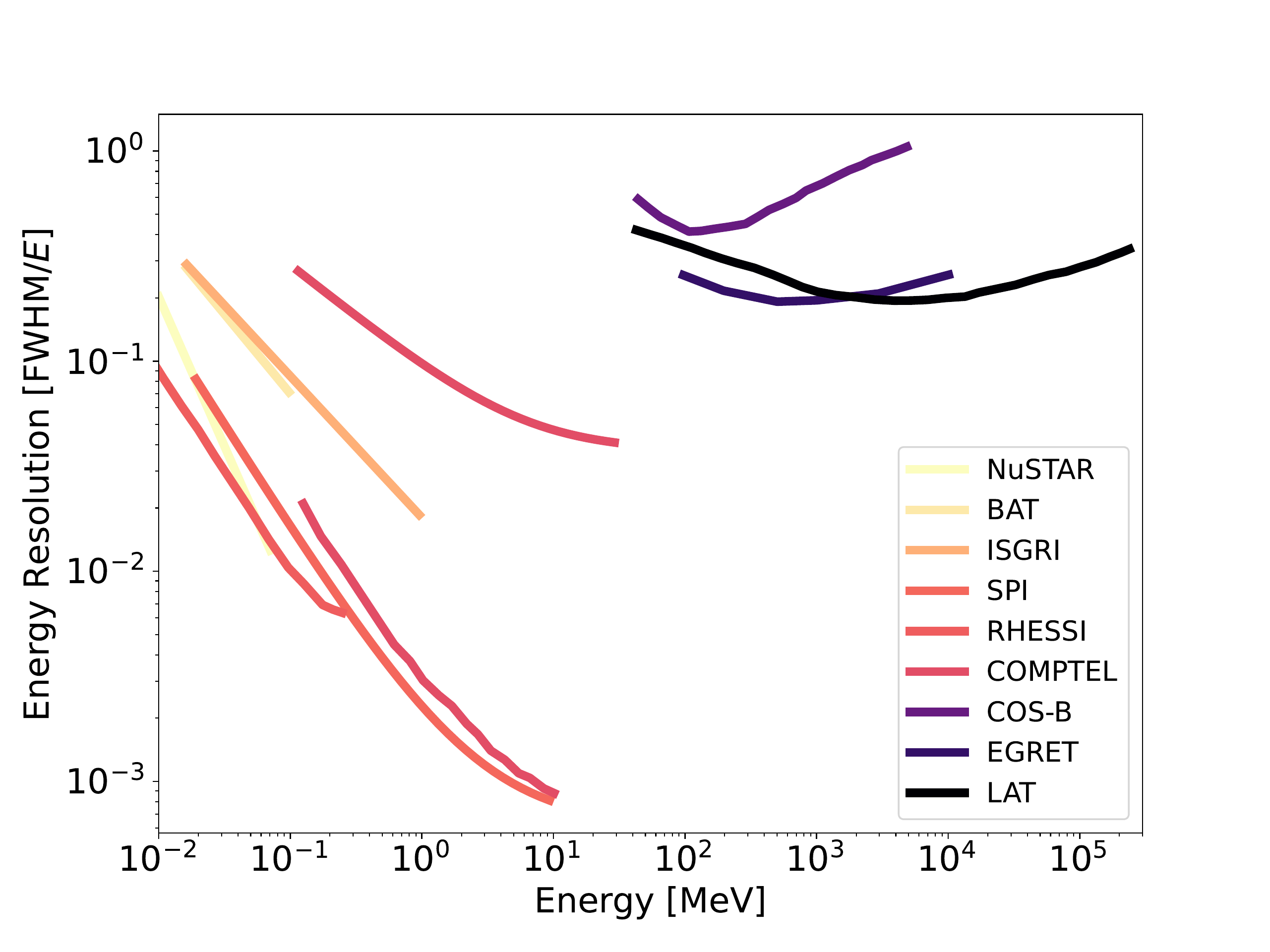}
    \includegraphics[width=0.49\textwidth,trim=0.2in 0.2in 0.9in 0.6in,clip=True]{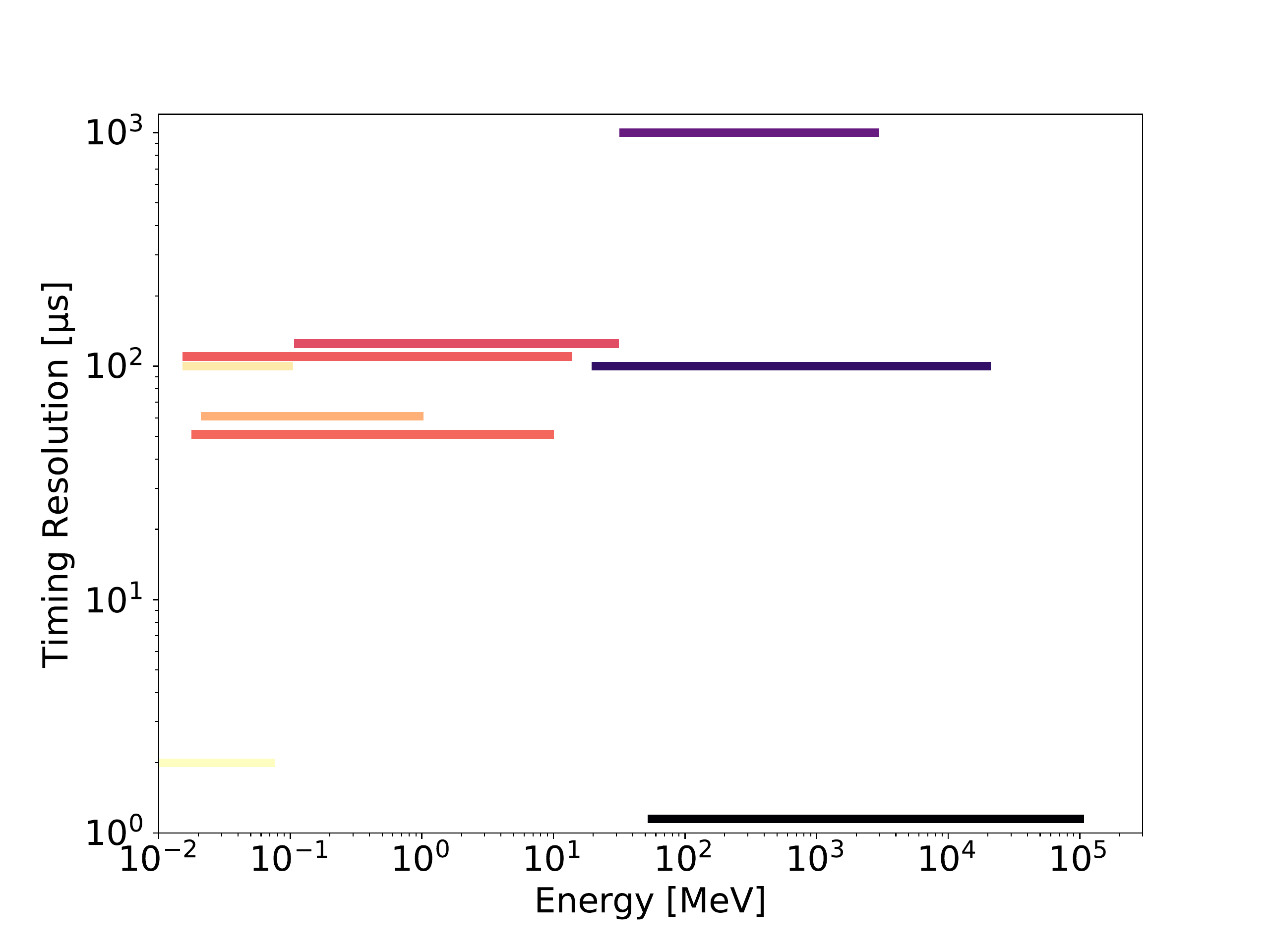}
    
    \caption{Characteristics of a selection of historic and current instruments as a function of photon energy. \textit{\textbf{Top left}}: Effective area. \textit{\textbf{Top Right}}: Angular resolution. \textit{\textbf{Bottom left}}: FWHM energy resolution. \textit{\textbf{Bottom right}}: Timing Accuracy.}
    \label{fig:astro_vs_instrument}
\end{figure}

A massive detector with limited positional but good energy resolution and deep enough to absorb most of the scattered photons can be used as calorimeter to measure spectra of incident $\gamma$ radiation.
Limited angular resolution can be achieved by fitting massive anti-coincidence wells around the detectors leaving an `acceptance angle' free, or by constraining the field of view with a passive or active collimator.
A modern high-resolution Ge spectrometer is the MeV spectrometer SPI on the INTEGRAL mission \citep{Winkler2003_INTEGRAL,Vedrenne2003_SPI}, which in addition to a massive anti-coincidence well encodes the incident $\gamma$-ray beam through a coded mask to enable imaging of radiation sources (Sec.\,\ref{sec:temporal_spatial_apertures}).

In order to detect a $\gamma$-ray via its pair-production interaction whilst extracting as much positional and energy information as possible, two main elements are required:
Firstly, the $\gamma$-ray must be made to interact, i.e. pair produce, in the detector.
In order to increase the probability of the $\gamma$-ray pair-producing, a high-Z material is required.
For the Energetic Gamma Ray Experiment Telescope (EGRET) detector aboard {\it{CGRO}}, this comprised tantalum foils \citep{1993ApJS...86..629T} while for both the Large Area Telescope (LAT), on board the {\it{Fermi}} satellite and the {\it{Gamma-Ray Imaging Detector}} ({\it{GRID}}) on the {\it{Astro-rivelatore Gamma a Immagini LEggero}} (AGILE) satellite, the high-Z converter material used is by tungsten \citep{2009ApJ...697.1071A, 2009A&A...502..995T}.
The resulting electron-positron pair must then be tracked as accurately as possible so that the direction of the incident $\gamma$-ray can be reconstructed.
This is done by measuring the passage of the electron/positron pair by the tracker.
For EGRET this was achieved by means of a multilevel spark chamber \citep{1993ApJS...86..629T}.
In both LAT and AGILE's {\it{GRID}}, the trajectory of the charged particles is recorded by layers of silicon strip detectors \citep{2009ApJ...697.1071A, 2009A&A...502..995T} (Sec.\,\ref{sec:pair_tracking_telescopes}).

To determine the energy of a $\gamma$-ray, it is desirable to stop the electron-positron pair in the detector via a calorimeter, where the total energy deposit is measured.
For EGRET a large NaI Total Absorption Shower Counter was the principal energy-measuring device while in LAT the calorimeter comprises 16 modules, each of which is composed of 96 CsI(T1) crystals.
The calorimeter on AGILE's {\it{GRID}} is also composed of CsI(T1), in this case 30 bars arranged in two planes \citep{2009A&A...502..995T}.
In addition to providing an energy measurement, a segmented calorimeter can also act as an anchor for the electromagnetic particle shower, providing further positional information to aid with pinpointing the direction of the incident $\gamma$-ray and to help with background discrimination (Sec.\,\ref{sec:data_cuts}).
The required elements of a $\gamma$-ray detector operating in the pair-production regime are, therefore, a tracker and a calorimeter. 

Not essential for the detection of the $\gamma$-ray but absolutely necessary so as to reject the overwhelming background of charged CRs that constantly bombard the instrument is an anticoincidence detector (ACD, Sec.\,\ref{sec:ACS}).
This allows the detector to self-veto upon the entry of a charged particle so it is essential that it have high detection efficiency for such particles.
The ACD of EGRET comprised a large scintillator which surrounded the spark chamber.
As is discussed in chapter about EGRET in this book, backsplash, whereby a charged particle generated inside the detector traversed the ACD and thus caused a false veto, became a problem above 10 GeV (Sec.\,\ref{sec:background_examples}).
To avoid backsplash, the ACDs of both the LAT and AGILE are segmented allowing only the segment adjacent to the incident photon candidate to be examined when searching for a veto.
This drastically reduces the effects of backsplash allowing for a much more efficient background rejection by the ACD.

All these considerations are summarised in the four basic parameters of any $\gamma$-ray telescope, most importantly the effective area, as well as the energy, angular and temporal resolution.
An overview of current and past $\gamma$-ray instruments in the MeV--GeV range is provided in Fig.\,\ref{fig:astro_vs_instrument}.
It is clear that the effective area is the reason why there is such a great loss in sensitivity in the MeV range ($\lesssim 100\,\mathrm{cm^2}$) compared to the keV or GeV range (both $\gtrsim 1000\,\mathrm{cm^2}$; Sec.\,\ref{sec:MeV_gap}).
However, because of Ge detectors, for example, the spectral resolution of MeV instruments (FWHM$/E \approx 10^{-3}$--$10^{-2}$) can supersede those of GeV instruments by two orders of magnitude.
The angular resolution of MeV telescopes can be similar to those of GeV telescopes, but only under specific circumstances, for example when observing the Sun in the case of {\it{RHESSI}} with a temporal modulation aperture (Sec.\,\ref{sec:temporal_spatial_apertures}).
Normally, Compton telescopes suffer from their inherently poor angular resolution on the order of degrees, whereas coded-mask telescopes could achieve arcminute resolutions or better (see Sec.\,\ref{sec:temporal_spatial_apertures}).
GeV telescopes can be considered almost direct imaging telescopes as the dispersion is only important for lower energies.
Because of the trackers, angular resolutions below the $0.1^{\circ}$-scale are possible.
Finally, the scarcity of $\gamma$-rays from celestial sources as well as their intrinsic temporal variability imposes a timing resolution requirement of approximately $100\,\mrm\mathrm{\mu s}$.
\subsection{Earth's Atmosphere and Space Environment}\label{sec:measurement_conditions}
\subsubsection{Atmospheric Effects}\label{sec:atmospheric_effects}
While at sea level the Earth's atmosphere blocks almost all low- and high-energy $\gamma$-rays to the extent that ground-based observations are impossible, high-altitude observations are still worth the effort and reduce the cost.
For instrument prototypes, in particular, balloon flights are often used to test new apertures and concepts.

\begin{figure}[ht]
    \centering
    \includegraphics[width=0.49\textwidth,trim=0.15in 0.25in 0.8in 0.8in,clip=True]{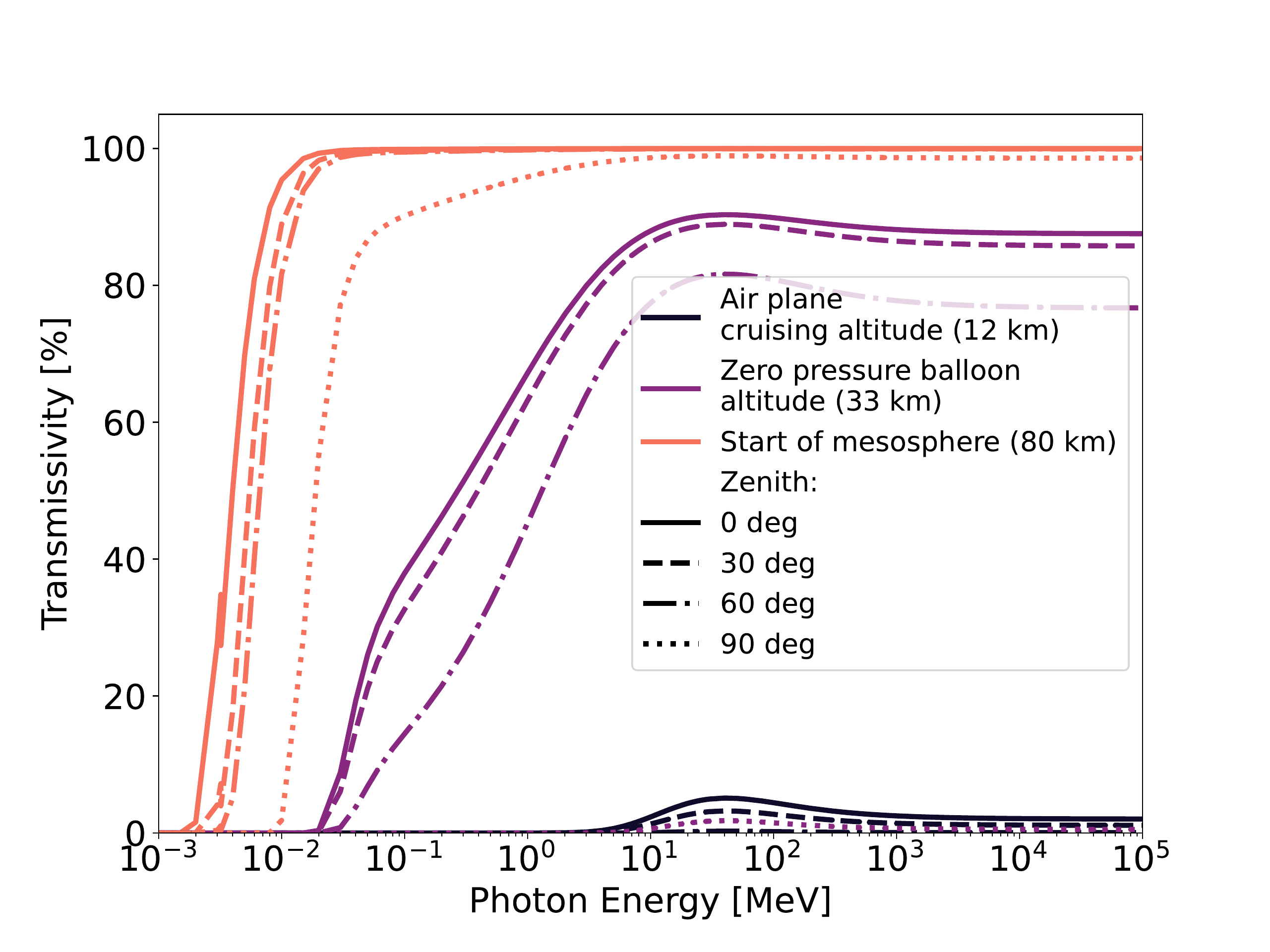}
    \includegraphics[width=0.49\textwidth,trim=0.15in 0.25in 0.8in 0.8in,clip=True]{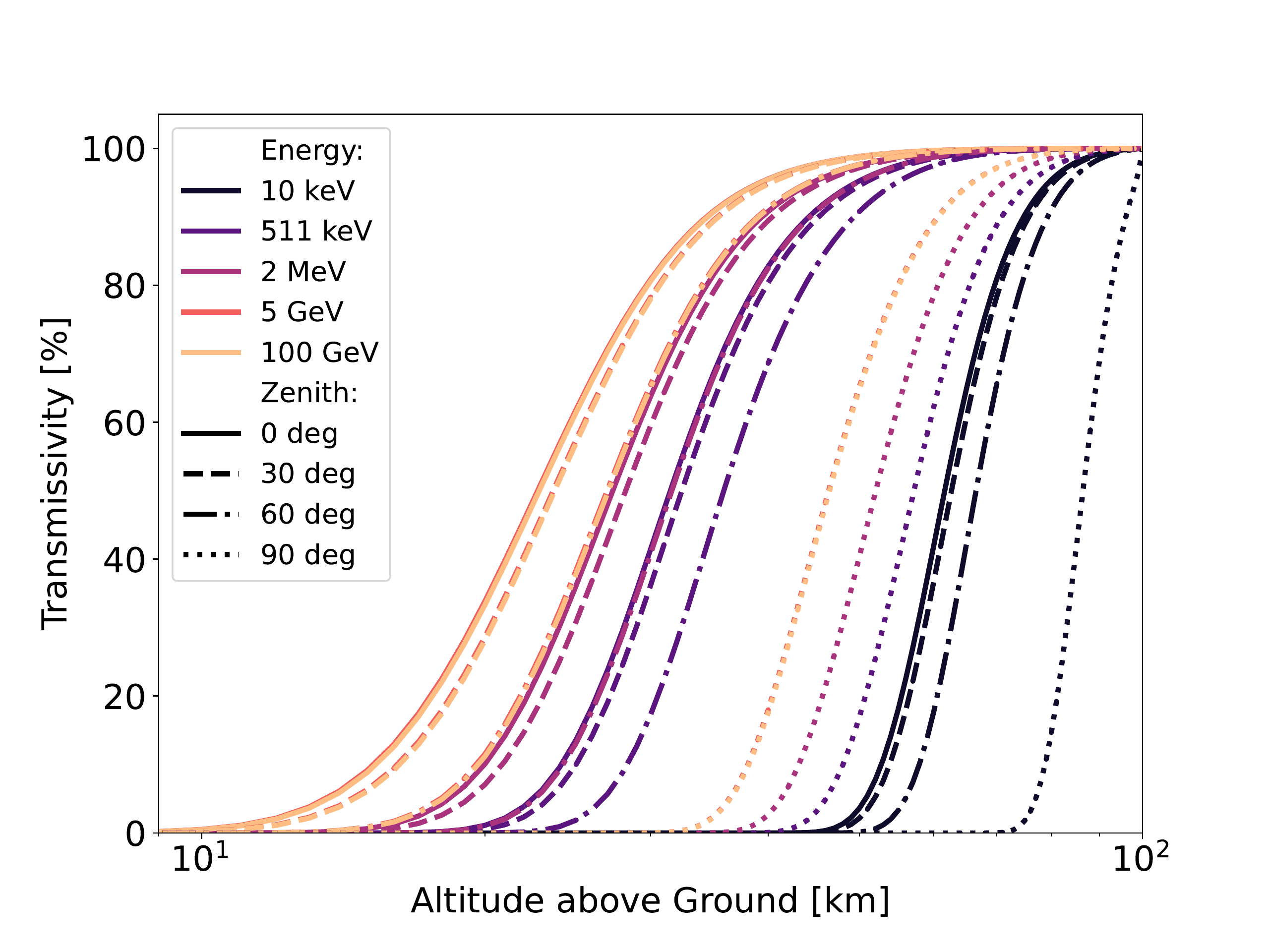}
    \caption{Transmissivity of Earth's atmosphere as a function of incoming photon energy (left) and observation altitude above surface (right) for different zenith angles.}
    \label{fig:transmissivity_atmosphere}
\end{figure}

In Fig.\,\ref{fig:transmissivity_atmosphere} the transmissivity of the Earth's atmosphere is shown for different photon energies, incidence angles (zenith), and altitudes above the surface \citep{Hubbell2004_NIST_DB,xcom}.
The transmissivity is defined as the probability for a photon to reach a certain altitude without previous interaction and therefore to be unabsorbed.
At aeroplane cruising altitudes (12\,km), for example, the chance for a 1\,MeV photon to pass through the upper layers of the atmosphere is $0.001\,\%$ at most (i.e. at zenith).
At this height, the transmissivity is maximised for 40\,MeV photons at about 5\,\%.
Because of the exponential decrease in the density of the atmosphere, the stratosphere layers of the atmosphere (up to 50\,km above ground) provide a useful environment for $\gamma$-rays telescopes.
At typical balloon flight altitudes of around 30\,km, the zenith transmissivity is already around 30\,\% for photon energies of 50\,keV.
Up to 40\,MeV, the transmissivity grows exponentially to about 85\,\% and slightly declines afterwards to flatten out at 80\,\% for GeV energies.
Clearly, with the beginning of the mesosphere at altitude of approximately 80\,km, essentially all $\gamma$-rays photons are directly measurable, and only soft and hard X-ray photons remain absorbed.
Beyond the von Karman line at around 100\,km, which conventionally defines the border between the atmosphere and space, all photons are readily detected as the transmissivity is nearly 100\,\% throughout the electromagnetic spectrum.

Most important for $\gamma$-ray observations at balloon altitudes, however, is the zenith angle dependence.
For the same photon energy and observation altitude, different zenith angles lead to vastly different transmissivities and therefore a much more drastic change in the effective area of the instrument (Sec.\,\ref{sec:instrument_designs}).
While response functions for balloon experiments take into account the zenith dependence of their effective area, the simulations to provide a reasonable measure of the atmosphere effects require different setups depending on the balloon position around the Earth.
This is the case because the local atmospheric conditions, including density and temperature for example, and in particular, the magnetic cutoff rigidity change with the Earth's latitude and longitude \citep{Smmart2005_cutoffrigidity}.

\subsubsection{In-Space Observations}\label{sec:in-space-observations}
Passing the 100\,km mark, $\gamma$-ray instruments experience the space environment which mainly concerns the distributions of charged particles.
In terms of onboard electronics, the instruments start to suffer more single event latch-ups and other effects.
These are short circuits caused by heavy ions or protons hitting the electronics and triggering semiconductor band transitions.
Apart from the latch-ups the detectors themselves are also more susceptible to incoming radiation.
This can be used as an advantage to measure the in-orbit particle spectrum, and therefore provide a measure for the instrumental background (Sec.\,\ref{sec:instrumental_background}).
Depending on the energy of the charged particles, they produce secondary particles when interacting with the instrument or satellite material.
The secondary particles compose most of the instrumental background for $\gamma$-ray measurements in space, especially in the MeV range.

\begin{figure*}
    \centering
    \includegraphics[width=0.45\textwidth]{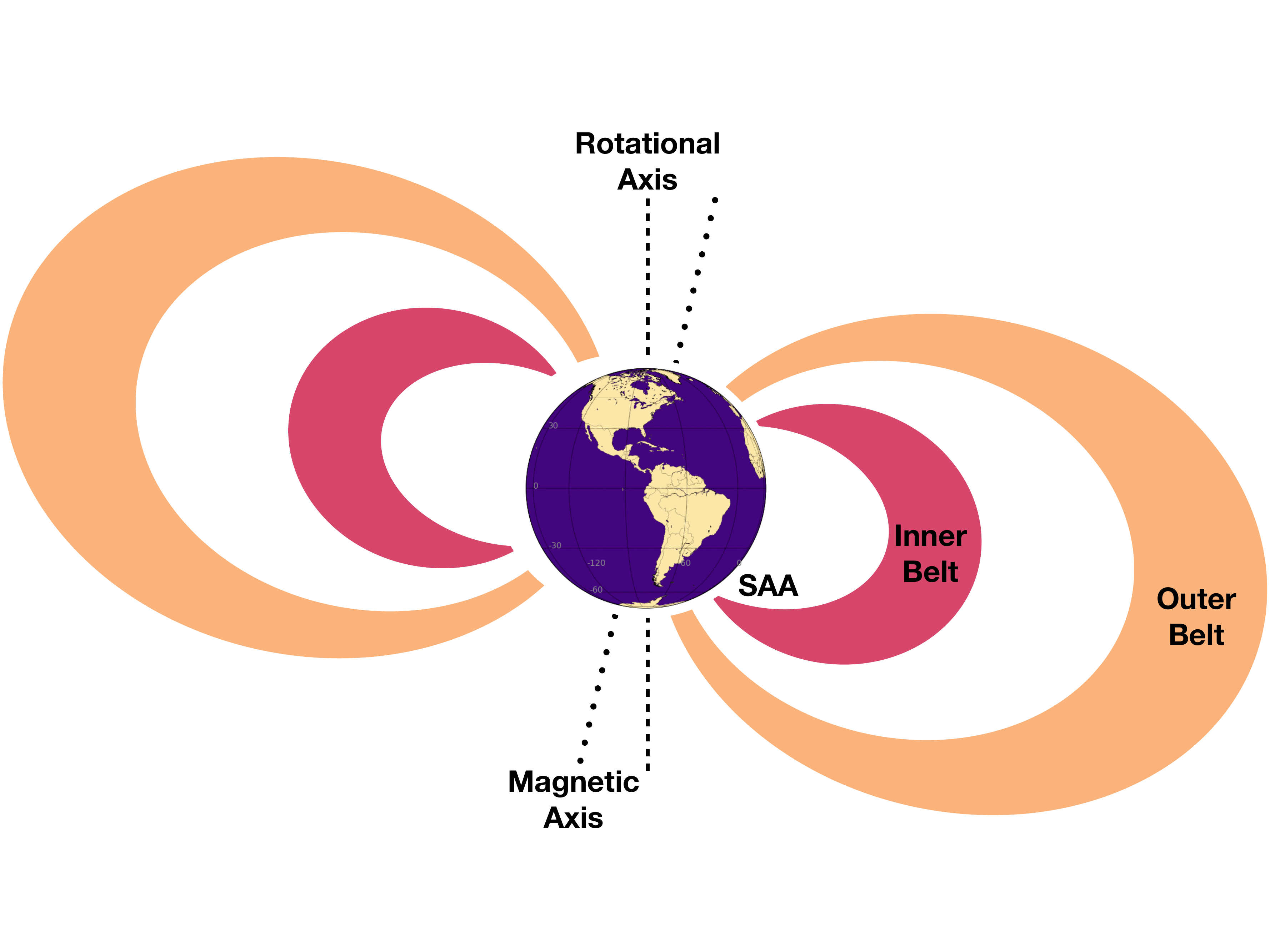}
    \includegraphics[width=0.54\textwidth,trim=0.6in 1.5in 1in 1.5in,clip=True]{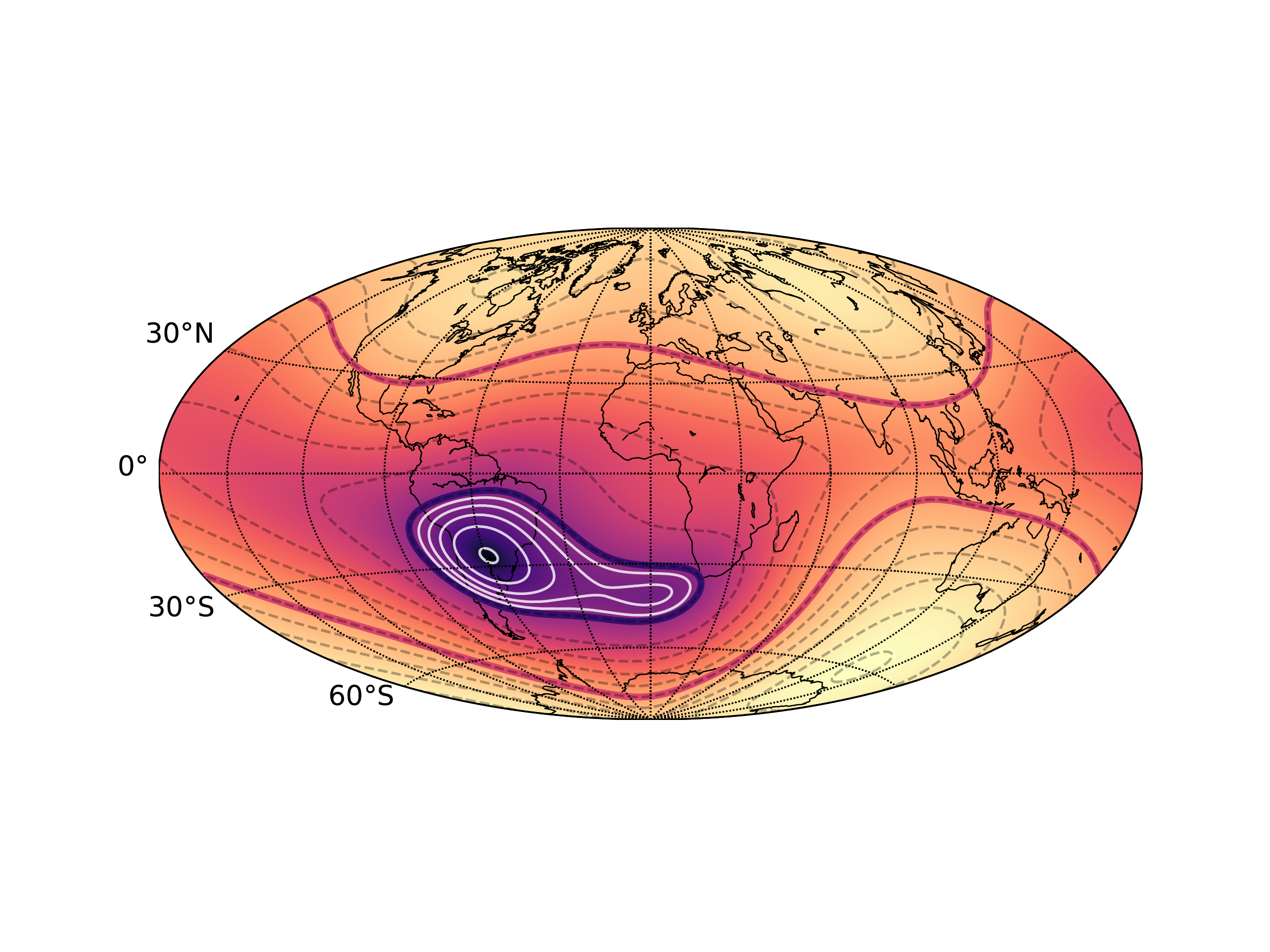}
    \caption{Van Allen Radiation Belts around Earth (\textit{\textbf{left}}) with Inner and Outer belts (to scale). Because the Earth's magnetic field is tilted with respect to its rotational axis (dashed line), the closest part of the Inner belt can reach to about 200\,km above the southern Atlantic. This is called the South Atlantic Anomaly (SAA, \textbf{\textit{right}}, adapted from \citep{Finlay2020_SAA}, and reproduced with permission). Shown is the difference to the mean magnetic field intensity of $45.8\,\mathrm{\mu T}$ (red solid line) in steps of $4.1\,\mathrm{\mu T}$ until the region defining the SAA (solid purple line). Inside the purple region, the steps are $0.6\,\mathrm{\mu T}$ for a minimum around Earth longitude and latitude of $60^{\circ}$W and $28^{\circ}$S.}
    \label{fig:SAA_van_Allen}
\end{figure*}

The concentration of charged particles around Earth is not homogeneous.
Because of the Earth's magnetic field, charged particles are trapped around the planet in torus-like accumulations (Fig.\,\ref{fig:SAA_van_Allen}).
These are known as the Van Allen Radiation Belts \citep[e.g.,][]{Ganushkina2011_VanAllenBelts}.
Two tori trap electrons and protons, and to a lesser extent $\alpha$-particles, reaching from 0.2 out to 2 Earth radii (Inner belt) and from 3 to 10 Earth radii (Outer belt), respectively.
While the Inner belt contains sub-relativistic electrons (few hundred keV) and protons ($\sim 100$\,MeV), the Outer belt also holds relativistic electrons (up to 10\,MeV).
The Outer belt is more easily influenced by the Sun and therefore more variable than the Inner belt.

Because the magnetic field of Earth is slightly tilted with respect to its rotational axis the centres of the belts are further shifted from Earth's centre, the Inner belt has an anomalously close approach at one specific region to the East of the South American continent.
This is called the South Atlantic Anomaly \citep[SAA, e.g.,][]{PavonCarrasco2016_SAA,Finlay2020_SAA}.
The anomaly represents an area in which the Earth's magnetic field is weakest relative to its surroundings (Fig.\,\ref{fig:SAA_van_Allen}, right).
In this region, the Inner belt approaches within 200\,km of the surface which results in higher abundances of energetic particles.
This leads to an enhanced instrumental background for satellite observatories (Sec.\,\ref{sec:instrumental_background}).

\subsubsection{Orbit Considerations}\label{sec:orbits}
There are options to alleviate the impact of the SAA and Van Allen Radiation Belts when the orbit of the satellite onboard which the instrument will be mounted is chosen.
However, not all instruments suffer from the effects of the increased radiation in the same way.
While MeV telescopes without major event selection capabilities (Sec.\,\ref{sec:data_cuts}) should avoid the SAA altogether, GeV instruments are typically placed in LEO, i.e. orbits between 200 and 2000\,km.
Most astronomical observatories in LEO are found between 450 and 600\,km.
Above an orbit of 1200\,km, the radiation belts would again lead to a much increased instrumental background.
For MeV transient observatories in particular, such as, for example {\it{Fermi}}-GBM, the enhanced particle flux at LEO is of only mediocre concern because the background for short time scales (on the order of seconds or less) can easily be determined from adjacent times.

\begin{figure}
    \centering
    \includegraphics[width=0.35\textwidth,trim=1in 1in 1in 1.15in,clip=True]{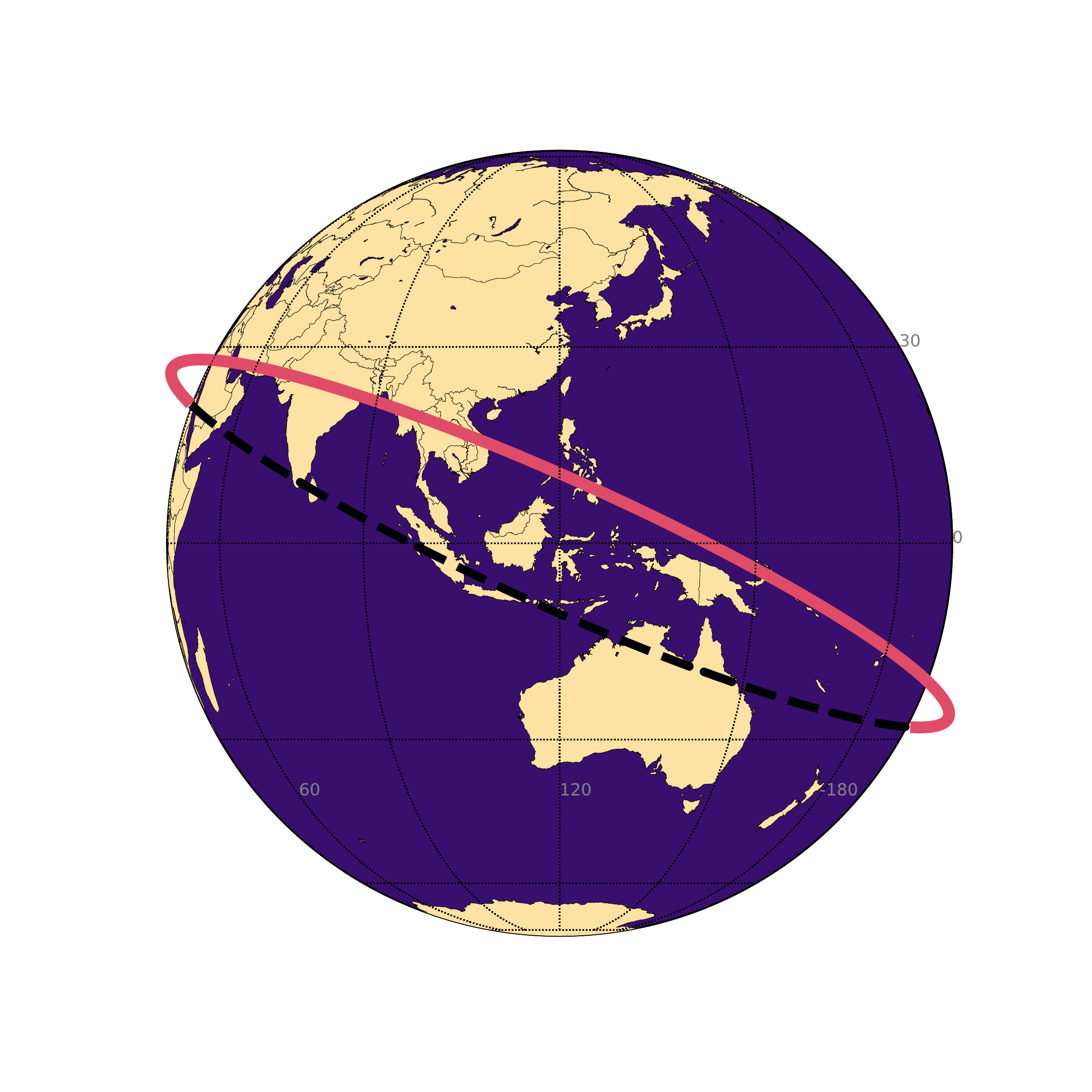}
    \includegraphics[width=0.27\textwidth,trim=2.5in 1in 2in 1in,clip=True]{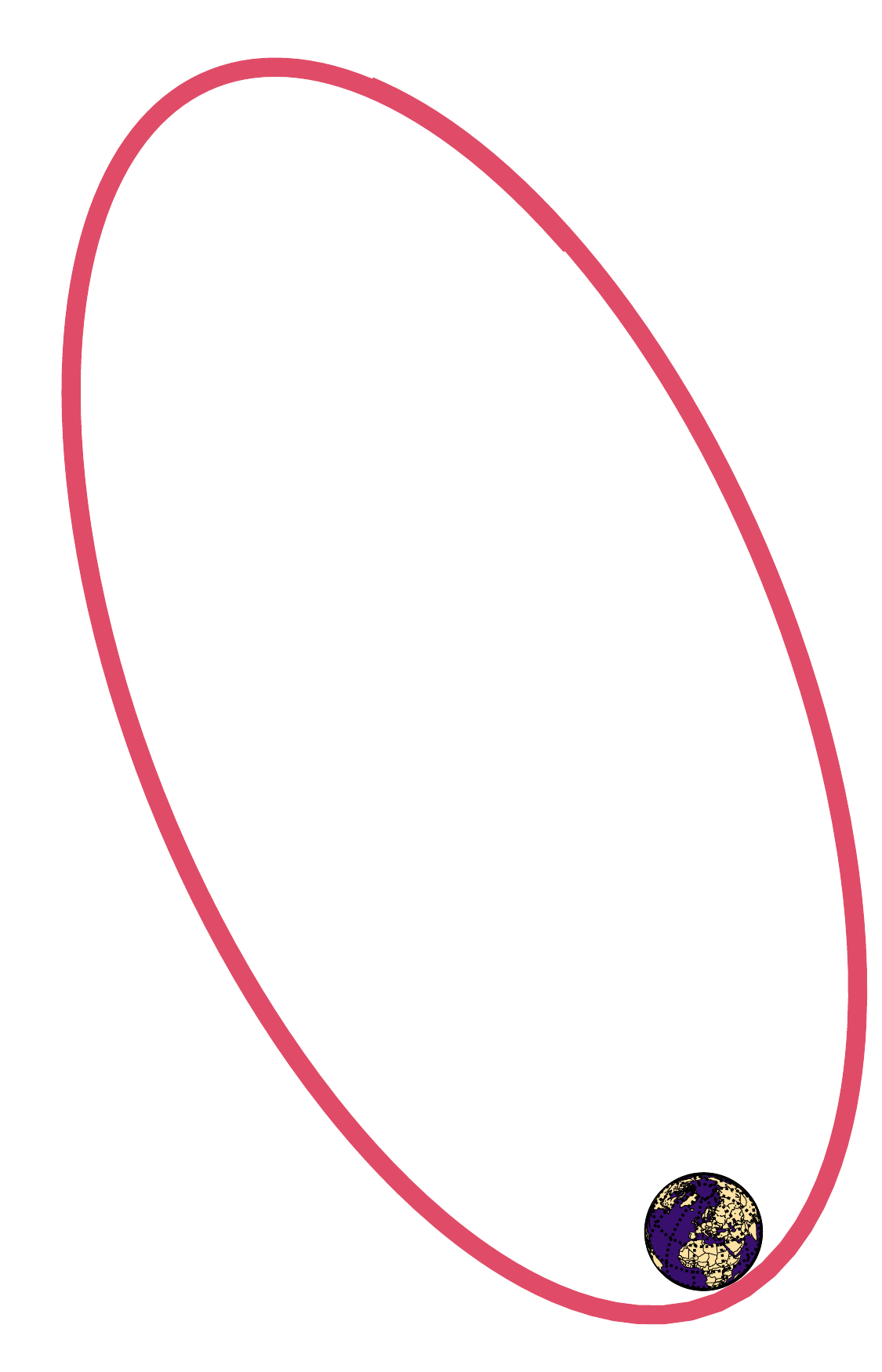}
    \includegraphics[width=0.35\textwidth,trim=1in 1in 1in 1.15in,clip=True]{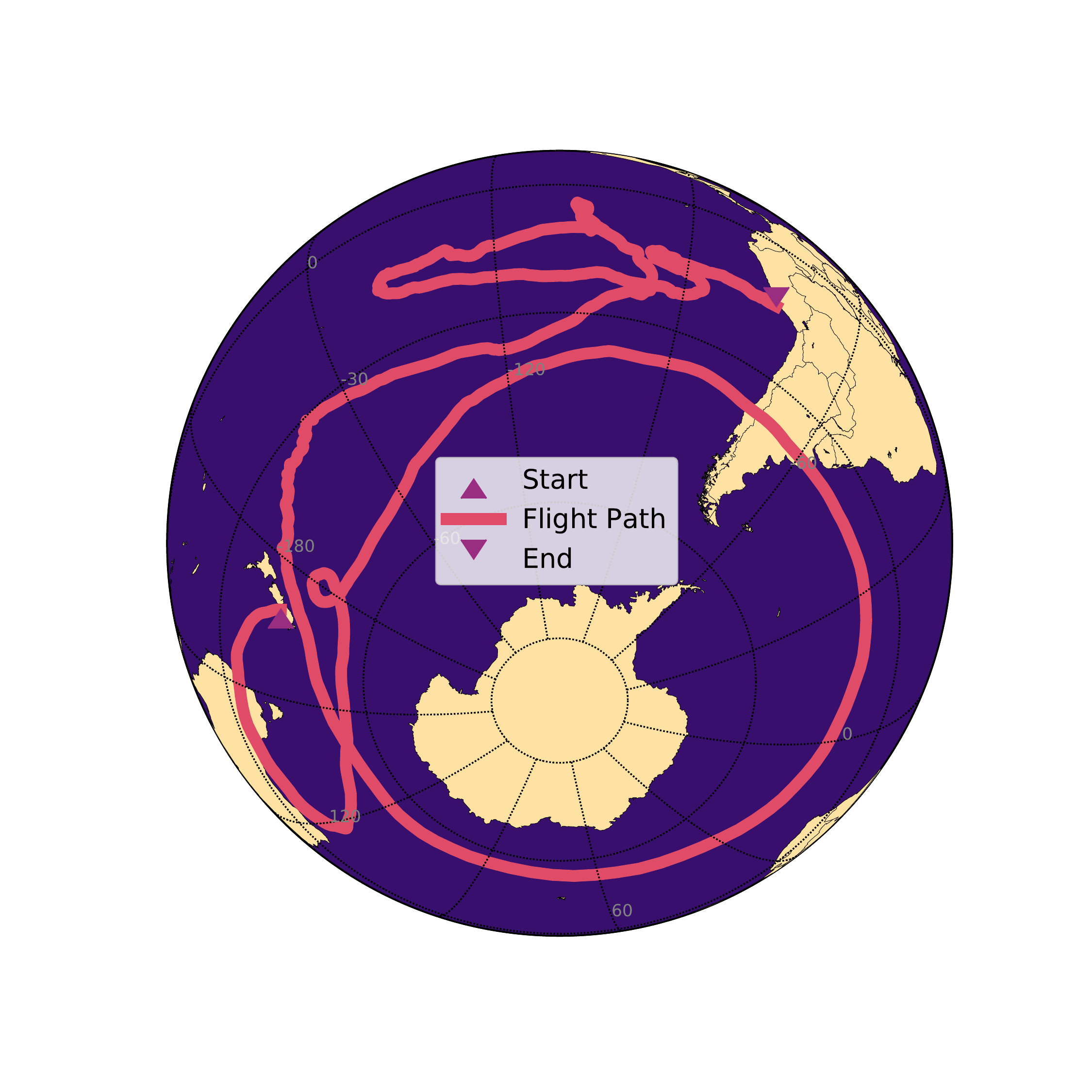}
    \caption{Satellite orbits and balloon flight paths. From left to right are shown the Fermi orbit, the INTEGRAL orbit, and the COSI balloon path from a campaign during 2016. The orbits in each panel are to scale.}
    \label{fig:satellite_orbits}
\end{figure}

For longer and targeted MeV observations, the radiation belts would lead to an insurmountable background rate which would heavily reduce the sensitivity of the instrument.
For this reason, MeV observatories like INTEGRAL chose high eccentricity and high inclination orbits to escape the radiation belts for a significant amount of their orbits.
The initial INTEGRAL orbit, for example, was a 72-hour orbit with an inclination of $52^{\circ}$ and an apogee and perigee of 154000 and 9000\,km, respectively.
Since the Outer belt is populated with charged particles to at most 10 Earth radii ($\sim 65000$\,km), most of the time spent in this orbit ($\sim 90$\,\%) is far away from the increased radiation.
However, the instruments onboard INTEGRAL have to be switched off every time it approaches Earth.

For special tasks which cannot (or can only inaccurately) be performed by single instruments, special orbits can be considered.
For example, some transient monitors have hardly any spatial resolution but can, however, be used in combination to provide highly accurate localisations (Sec.\,\ref{sec:IPN}).
The difference in the photon arrival times of transients can be used in triangulation to map overlapping annuli onto the sphere of the sky.
The larger the leverage arm, i.e. the larger the light travel distance between instruments, the better the localisation accuracy.
In particular, the Gamma-Ray Spectrometer onboard Mars Odyssey in a Mars orbit provides a valuable baseline for Earth-orbiting transient detectors.
This technique led to the term Interplanetary Network (IPN, \citep[e.g.,][]{Hurley2009_IPN}; Sec.\,\ref{sec:IPN}), for transient localisation with triangulation.
Another `orbit' of interest for $\gamma$-ray and other observatories is the Lagrange points, L2, of the Sun-Earth system (e.g., Wind, Spektr-RG), which also provide an excellent baseline for IPN measurements.

While satellites follow a specified path and can, most of the time, perform manoeuvres to correct their orbits (and to make sure that they re-enter the atmosphere when the mission is decommissioned), balloons have no or only little capability to adjust their flight paths.
Because of security concerns, among others, balloon flights are typically launched from remote areas, such as Antarctica, or those which are only sparsely populated.
After the launch, the balloons experience the natural Earth environment and float freely governed by wind (lower atmosphere), temperature (day and night cycle), and torque (rotation of the Earth).
Because the power generation has to be secured, which is mostly done with solar panels, the balloon gondolas are rotated towards the Sun during daylight.
This also holds the aspect angle of the instruments, which simplifies the analysis.
At night, the gondolas can again tumble freely and minuscule changes in the altitude can lead to extreme variations in the flight paths.
As examples, we show two satellite orbits as well as the long-duration balloon flight path of the COSI prototype in Fig.\,\ref{fig:satellite_orbits}.
For more details on orbital considerations, the reader is referred to the Chapter on orbits and background of $\gamma$-ray space instruments in this book.

\subsection{Instrumental Background}\label{sec:instrumental_background}
\subsubsection{Variations of the Background}\label{sec:background_variations}
The interaction of charged particles, i.e. in general CRs, with instrument and satellite material leads to several different components that are summarised under the term instrumental background.
These are all unwanted primary and secondary particles and photons which lead to enhanced count rates in the instruments, diluting the celestial signals of interest.
The interactions of CRs with matter lead to inverse Compton scattering, bremsstrahlung, nuclear excitation, spallation, radioactive build-up and decay, particle-antiparticle annihilation, and secondary particle production which can also undergo all of the previous interactions again.
This results in a cascade of interactions that, depending on the energy range of the instruments, are measured continuously.

\begin{figure}
    \centering
    \includegraphics[width=0.49\textwidth,trim=0.15in 0.2in 0.125in 0.1in,clip=True]{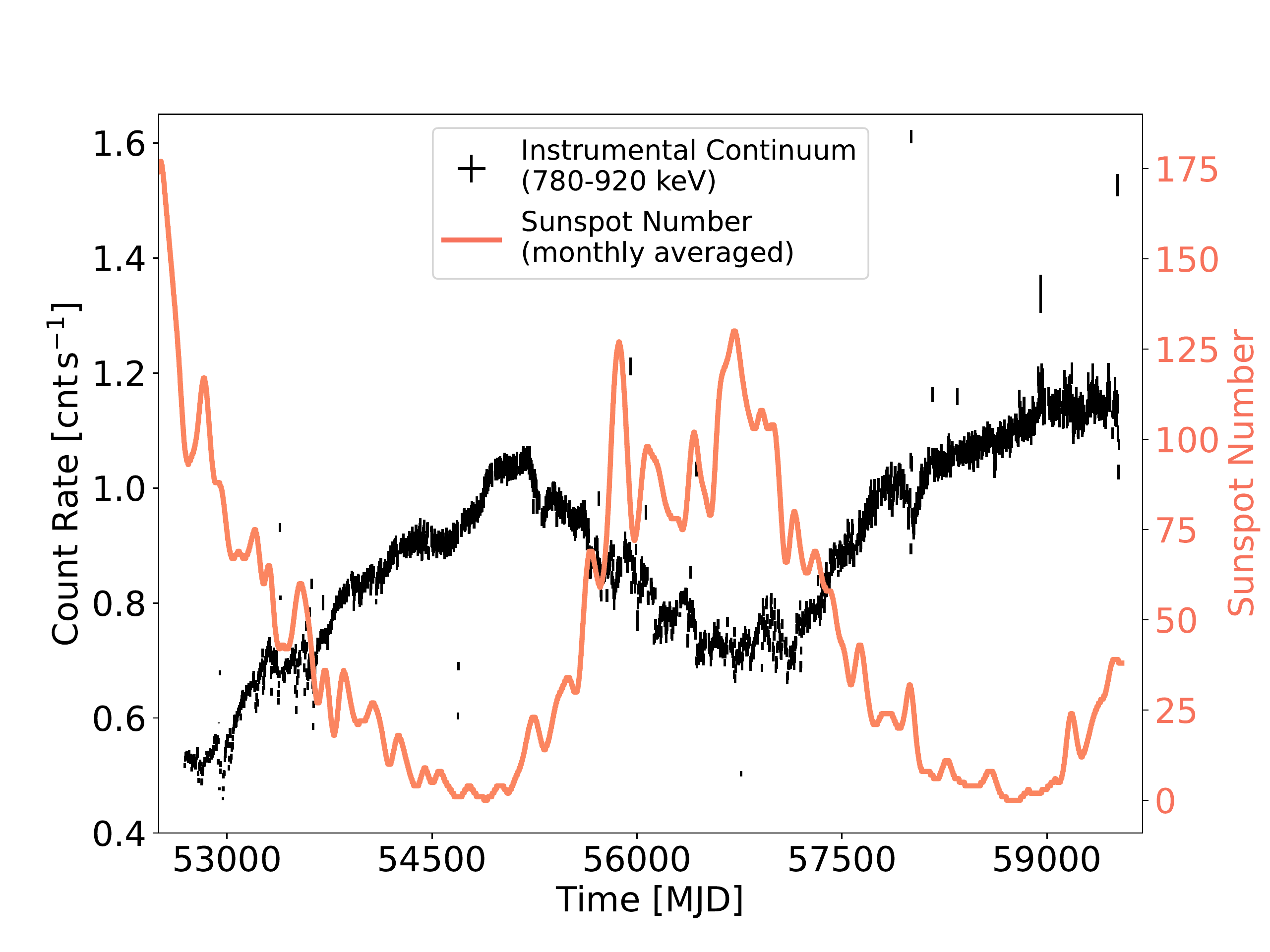}
    \includegraphics[width=0.49\textwidth,trim=0.15in 0.2in 0.125in 0.1in,clip=True]{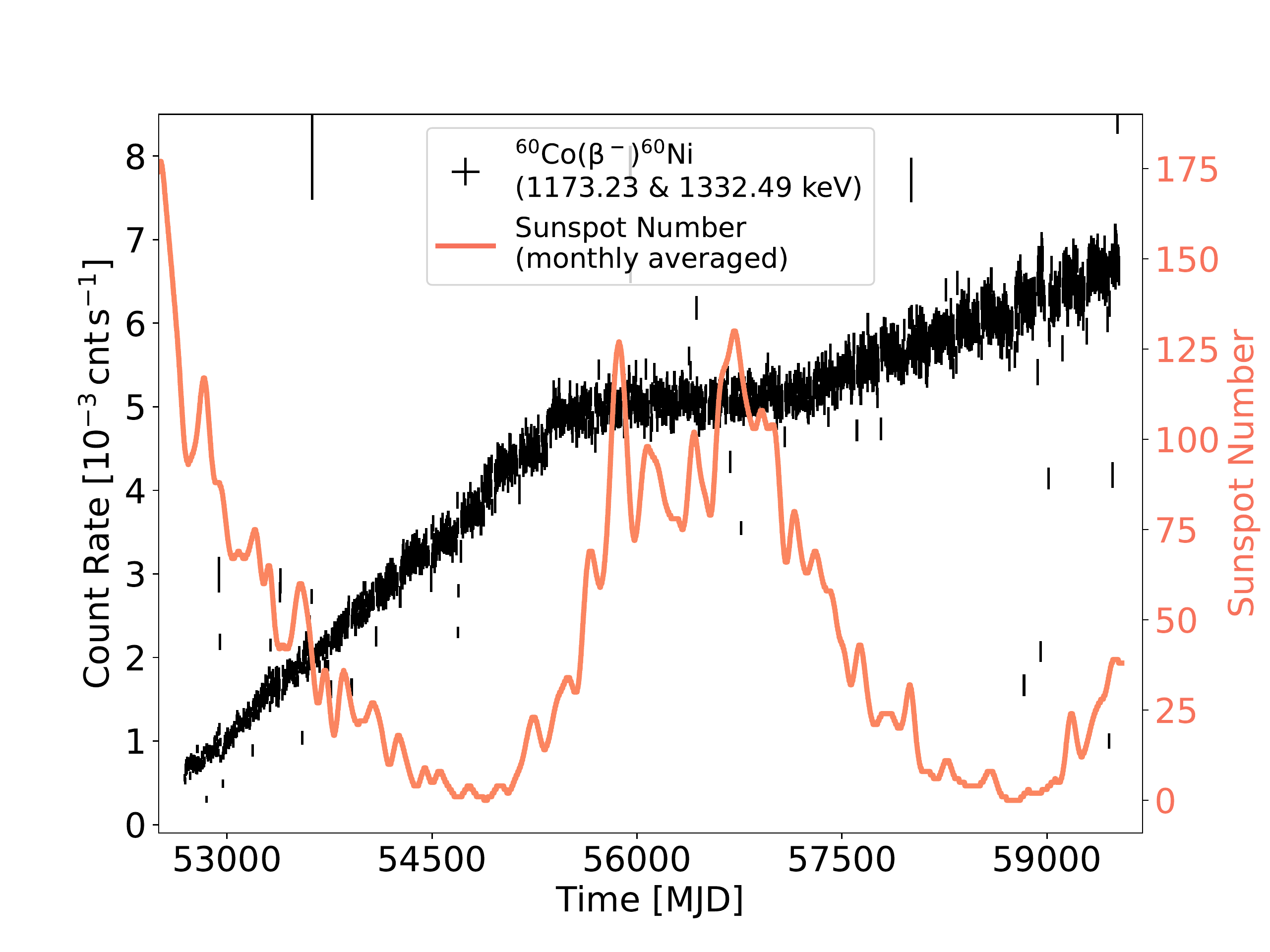}
    \includegraphics[width=0.49\textwidth,trim=0.15in 0.2in 0.125in 0.1in,clip=True]{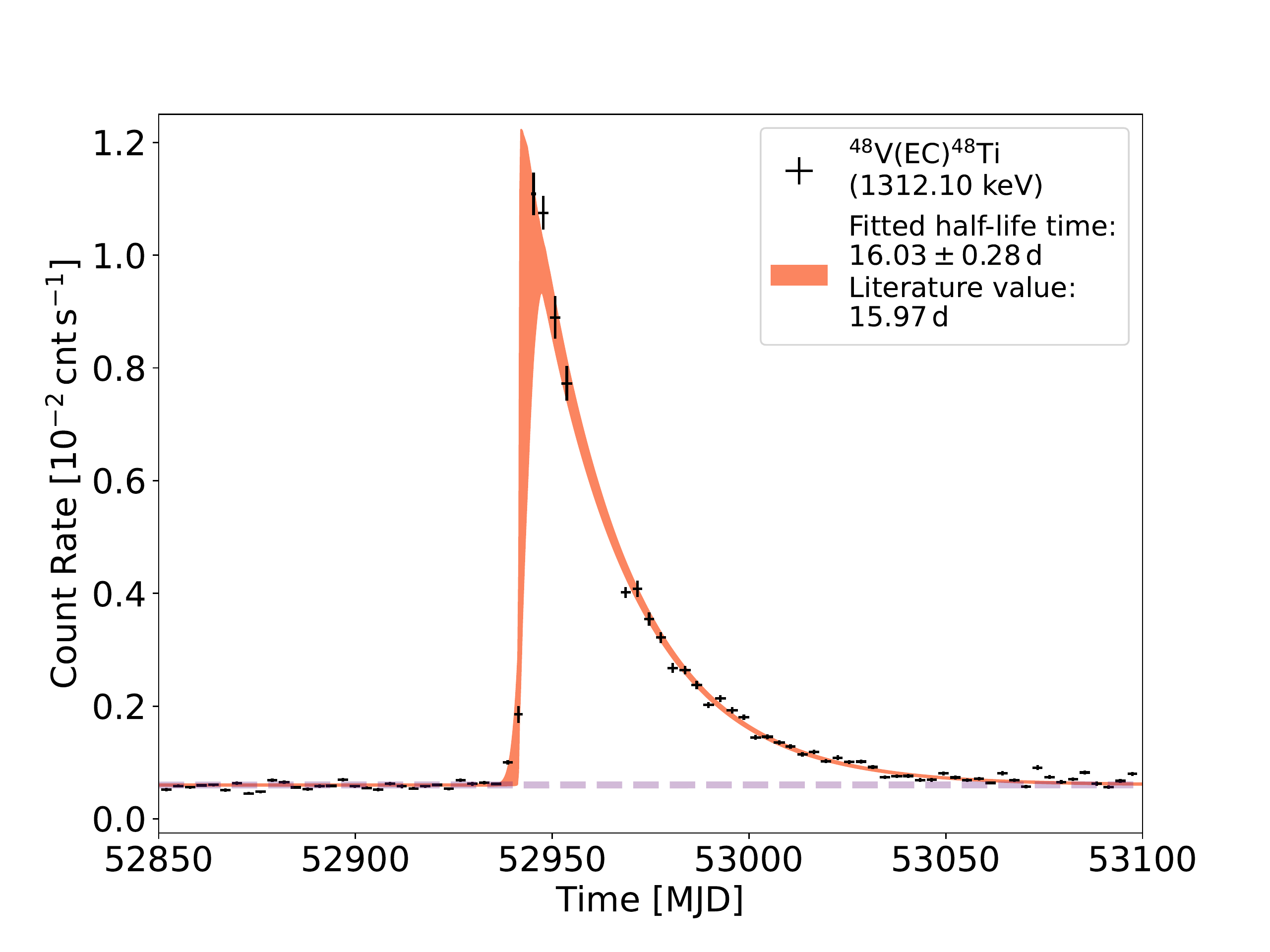}
    \includegraphics[width=0.49\textwidth,trim=0.15in 0.2in 0.125in 0.1in,clip=True]{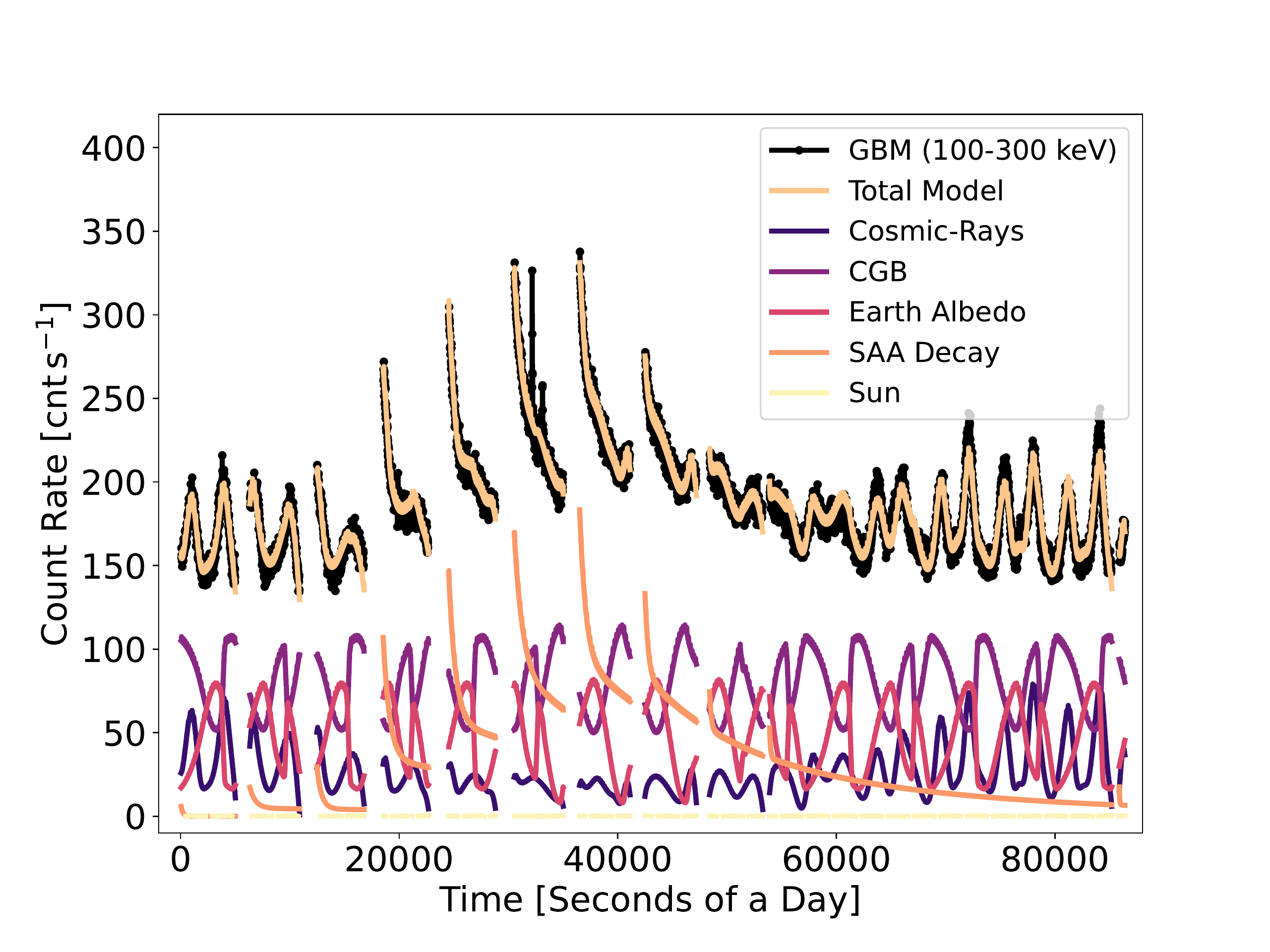}
    \caption{CR induced background rates for different processes from different origins. \textit{\textbf{Top left}}: Prompt MeV background is anticorrelated to the Sun-spot number, and thus with the solar magnetic field. \textit{\textbf{Top right}}: Radioactive build-up can occur when the lifetime of isotopes (here \nuc{Co}{60}) is much longer than the activation function (cosmic-ray flux, inverse proportional to Sun-spot number). \textit{\textbf{Bottom left}}: Solar flare events provide a large single dose of mainly protons during a short amount of time. Intermediate lifetime isotopes (here \nuc{V}{48}) are enhanced by a factor of ten and then decay according to their decay constants. \textit{\textbf{Bottom right}}: Equatorial LEO satellites pass the SAA every 90\,min, activating numerous short-lived isotopes which then decay during the next orbit. Adapted and updated from \citep{Diehl2018_BGRDB} and \citep{Biltzinger2020_GBM} - reproduced with permission.}
    \label{fig:activation_background}
\end{figure}

The largest impact on the amplitudes of these processes is given by the solar activity and the terrestrial and solar magnetic field.
Long-term trends in the instrumental background rate of MeV instruments are anti-correlated with the Sunspot number \citep[e.g.,][]{Clette2014_SunSpotNumber,Clette2016_SunSpotNumber}, which is a direct indicator of the solar magnetic activity cycle of eleven years (Fig.\,\ref{fig:activation_background}, top left).
The solar modulation of CRs is related to the intensity of the turbulent solar wind, which increases when the Sun's magnetic field is strong.
In other words this means that when there is a high number of Sunspots, the instruments are better shielded from CRs.
This leads to a reduction of the instrumental background rate which is why $\gamma$-ray missions are typically launched during or before the solar maximum.
One famous example for this is the Solar Maximum Mission (SMM) launched in 1980.
Two solar cycles later in 2002, INTEGRAL was launched -- also near the solar maximum.
Depending in addition on the chosen satellite orbit, the background rate can more than double between the solar maximum and minimum.
This effect is visible for prompt background phenomena such as continuous processes (e.g., bremsstrahlung; Fig.\,\ref{fig:activation_background} top left), nuclear excitation followed by fast deexcitation which typically happens on the order of nanoseconds, and particle production with fast decays from pions or $\beta$-unstable elements.

If the lifetime of the particles produced is (much) longer than the production time scale through CR bombardment, two other temporal evolutions of the background can be found.
For example, if the radiation dose hitting the satellite is drastically increased, such as during a solar flare event with a coronal mass ejection, the background rates from $\gamma$-rays of all isotopes in the satellite can rise by several orders of magnitude.
For isotopes produced during such events that are longer-lived, the background rate then stays at a high level even long after the initial dose.
In Fig.\,\ref{fig:activation_background}, bottom left, the rise of the background rate from the element \nuc{V}{48} is shown.
From a rather constant background rate of $\sim 6 \times 10^{-4}\,\mathrm{cnts\,s^{-1}}$ before the X-class solar flare on October 23rd 2003 (= MJD 52935), the \nuc{V}{48} rate rises to more than $1 \times 10^{-2}\,\mathrm{cnts\,s^{-1}}$.
Because \nuc{V}{48} has a half-life time of 16\,d, its rate decays only according to this decay time; the expected exponential decay is clearly seen.
Such nuclear reactions occur continuously, either converting stable satellite materiel to radioactive isotopes, which then decay promptly or with some delay, or directly exciting the nuclei of the instrument which then de-excite by the emission of $\gamma$-ray photons.
These $\gamma$-ray photons have specific energies so that individual isotopes and processes can be identified which helps in suppressing the instrumental background as a whole.

In the case of a regularly enhanced dose of radiation, for example by the passage through the SAA for LEO missions, the decays might not even go back to the base level because after about 90\,min, the next passage of enhanced radiation occurs.
This is shown in Fig.\,\ref{fig:activation_background}, bottom right, from a measurement of {\textit{Fermi}}-GBM over the course of one day.
Sixteen subsequent orbits and the different components making up the total measurement are shown.
While after the first three SAA passages, the corresponding levels go back to nearly zero, orbits 4--7 obtain a higher radiation dose so that until 40000 seconds, the background rate gradually builds up.
After orbit 8, only the very short-lived isotopes are seen in the data, while the build-up is still decaying on its own timescale.
Other components, such as the Earth albedo, the general CR activation rate outside the SAA, the cosmic $\gamma$-ray background, as well as the Sun as a $\gamma$-ray source itself, stay constant.
Only the change in orientation and aspect of the instrument with respect to the different astrophysical and background sources let the rates appear varying \citep{Biltzinger2020_GBM}.
If the radioactive decay time of isotopes is much longer than the activation function from CRs, more and more radioactivity is created inside the instruments.
In the case of \nuc{Co}{60}, for example, with a half-life time of 5.27\,yr, the decay rate is so small that over very long times, the background rate rises because there is radioactivity built up.
Fig.\,\ref{fig:activation_background}, top right, shows the rate of the two $\gamma$-ray lines at 1173 and 1332\,keV from the decay of \nuc{Co}{60} over 19 years of INTEGRAL/SPI measurements.
Clearly, as the Sunspot number goes down, i.e. the activation rate goes up, the \nuc{Co}{60} rate also rises.
Since the activation rate drops again after the solar minimum, but the material still decays, the background rate in these lines appears constant.
Then after the second maximum, the rate rises again.

\begin{figure}
    \centering
    \includegraphics[width=0.8\textwidth,trim=0.25in 2.5in 0.75in 2.5in,clip=True]{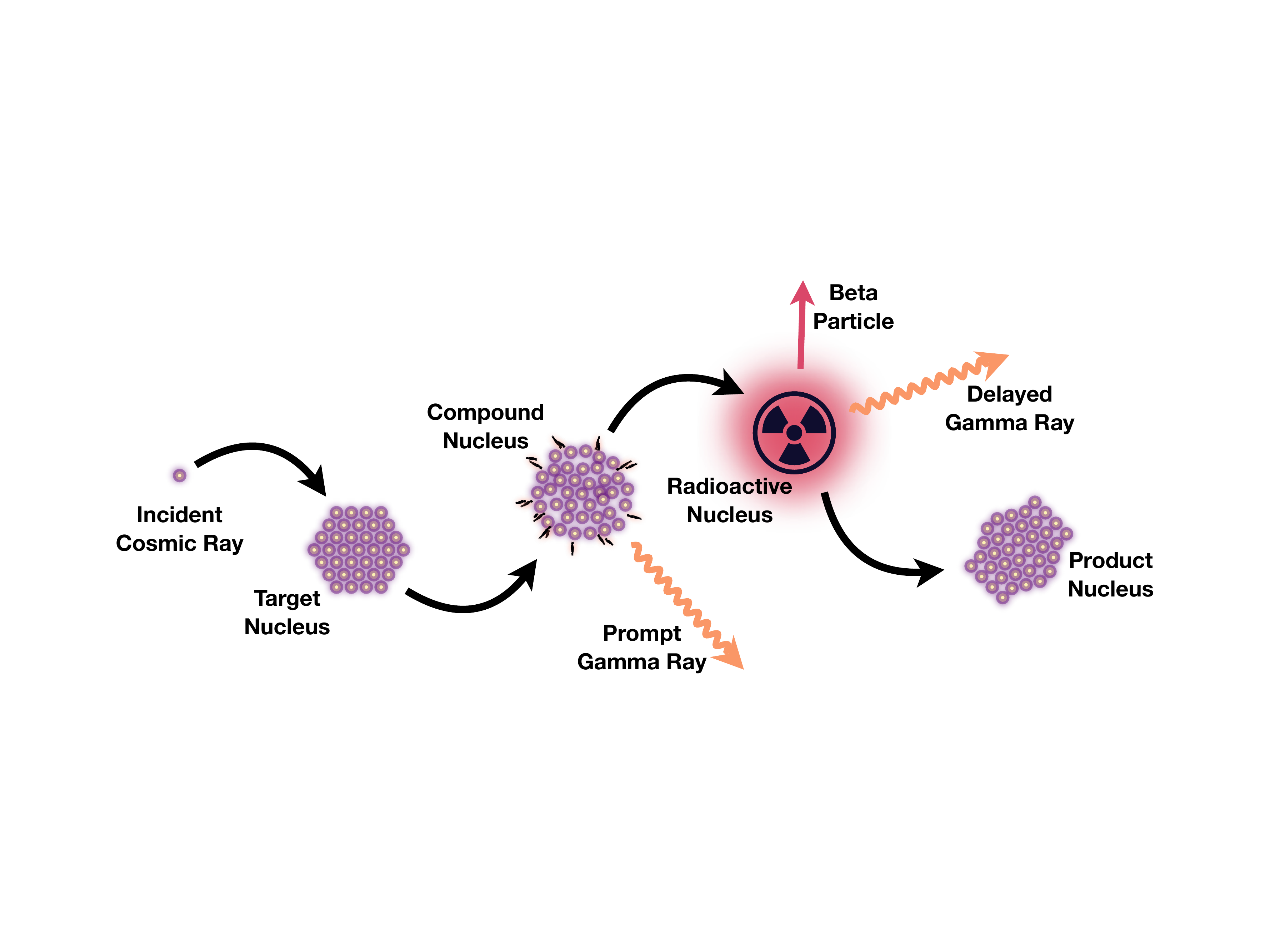}
    \includegraphics[width=0.35\textwidth,trim=4in 2.5in 4in 3.0in,clip=True]{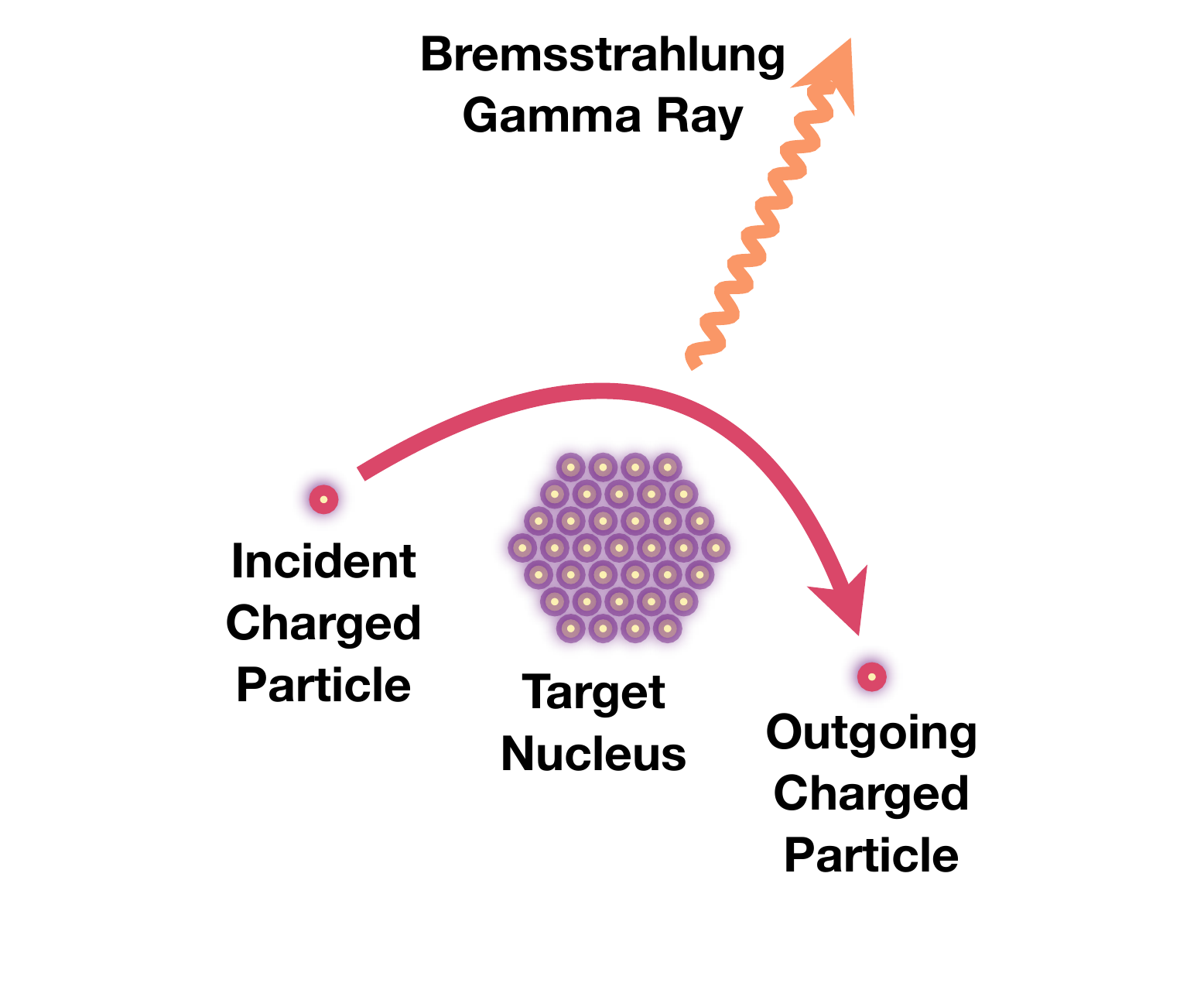}
    \includegraphics[width=0.44\textwidth,trim=3.0in 2.5in 3.0in 2.5in,clip=True]{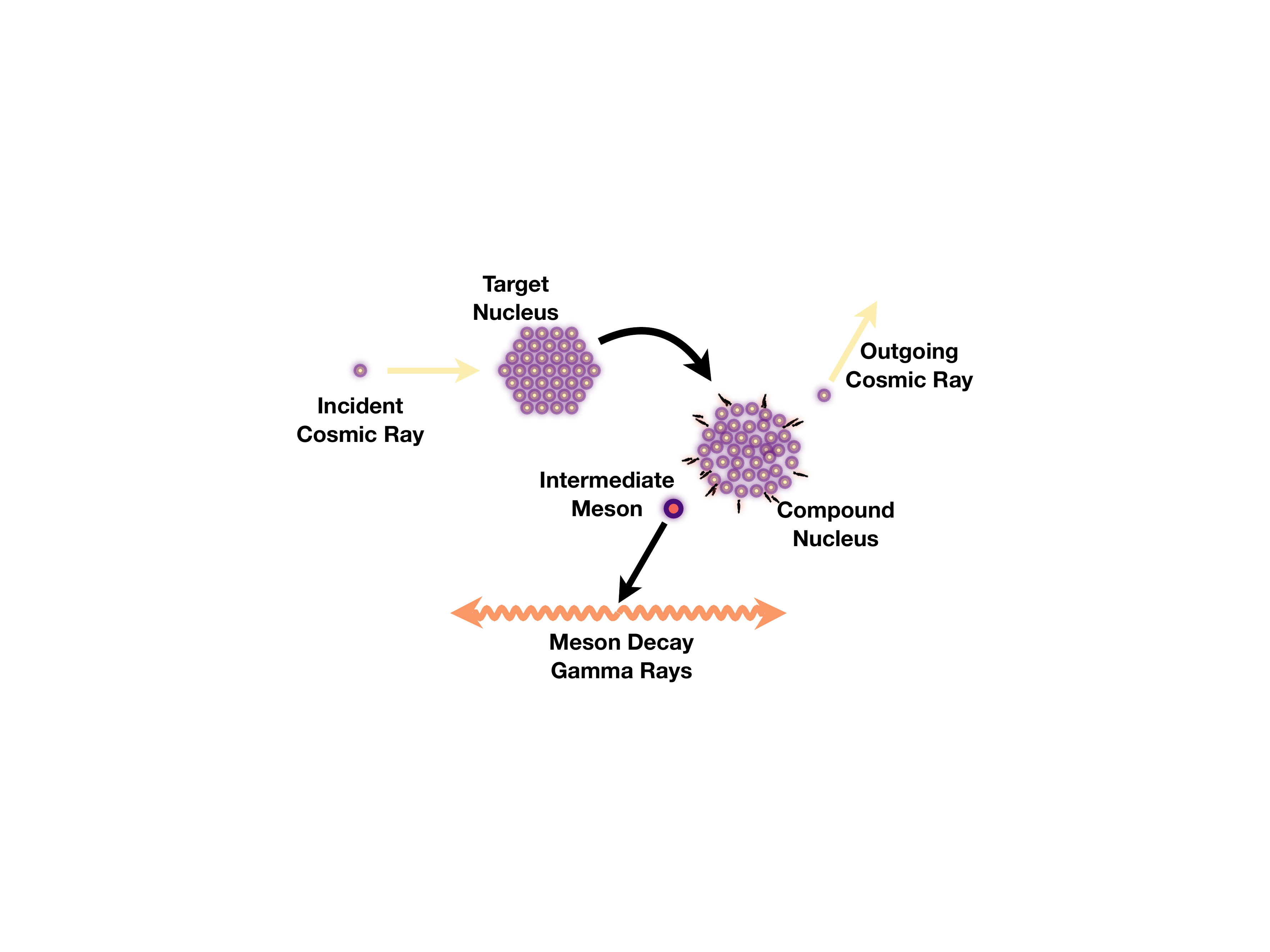}    
    \caption{Instrumental background processes. \textbf{\textit{Top}}: Nuclear excitation of instrument and satellite material by CR bombardment. An incident particle interacts with a nucleus from the instrument and forms a new isotope. This is formed in an excited state and de-excites by the emission of a prompt photon. The nucleus might still be left radioactive and decays (shown here as $\beta$-decay) toward a final nucleus, which may also involve the emission of a then delayed photon. \textbf{\textit{Bottom left}}: Bremsstrahlung of a charged particle moving in the field of a nucleus. A negatively charged electron approaches the positively charged electric field of a target nucleus. By a change of direction due to electrostatic attraction (or repulsion in the case of positrons), the electron is emitting bremsstrahlung photons equivalent to the change of its kinetic energy. \textbf{\textit{Bottom right}}: Particle production by relativistic CRs. If the incident CR is energetic enough, particle production can occur (similar to accelerator experiments). The thresholds to produce certain particles depends on the particles' rest masses and the interacting nuclei. In the case of mesons being produced, for example neutral pions ($\pi^0$), they decay on time scales of nanoseconds or less, and emit $\gamma$-ray photons.}
    \label{fig:background_processes}
\end{figure}

\subsubsection{Background as a Function of Energy}\label{sec:background_examples}
Since most $\gamma$-ray telescopes cover one or more decades of the electromagnetic spectrum, their measurements, and in particular their background, can appear quite different.
Depending on the spectral resolution, which, technologically, can be much higher at MeV energies compared to GeV energies, the general appearance changes.
At MeV energies, the background spectra are dominated by an electron bremsstrahlung continuum with a multitude of $\gamma$-ray lines on top (see Fig.\,\ref{fig:background_processes} for an overview of background processes).
Above $\sim 20$\,MeV, the decay and de-excitation lines from nuclei cease and the spectrum is a pure continuum up even to very high energies (TeV).
Pion production and decay (e.g., $p + p \rightarrow p + p + \pi^0$, followed by $\pi^0 \rightarrow \gamma\gamma$) describes the transition region from the MeV to the GeV background.
While these interactions would produce a spectrum peaking at 67.5\,MeV (half the rest mass of $\pi^0$) with a high-energy tail mimicking the incident proton spectrum, most of these interactions inside the instrument can be rejected due to their different signatures.

\begin{figure}
    \centering
    \includegraphics[width=0.98\textwidth,trim=0.0cm 4.0cm 0.0cm 0.0cm,clip=True]{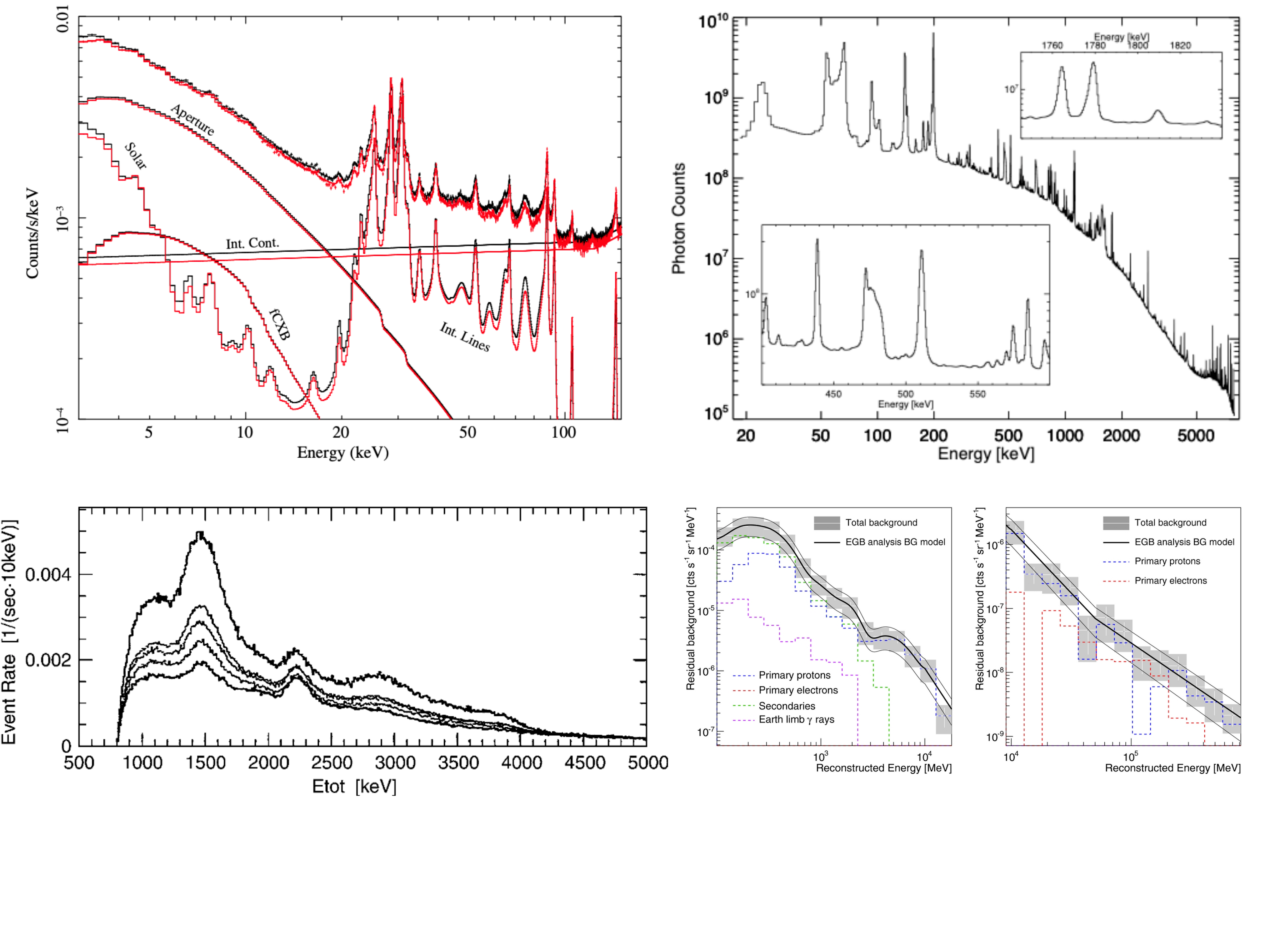}
    \caption{Example of measured (background) spectra. \textit{\textbf{Top left}}: {\it{NuSTAR}} from \citep{Wik2014_NuSTAR}. \textit{\textbf{Top right}}: INTEGRAL/SPI from \citep{Diehl2018_BGRDB}. \textit{\textbf{Bottom left}}: COMPTEL reproduced with permission from \citep{Schoenfelder1993_COMPTEL}. \textbf{\textit{Bottom right}}: {\textit{Fermi}}-LAT reproduced with permission from \citep{2015ApJ...799...86A}.
    }
    \label{fig:GammaRay_BG_spectrum}
\end{figure}

The nuclear lines directly reflect the elemental composition of the satellite, the instrument, and the Earth's atmosphere.
For example, shown in Fig.\,\ref{fig:GammaRay_BG_spectrum}, top, are the highly resolved {\it{NuSTAR}} and SPI background spectra.
Most of the lines below $\sim 100$\,keV are due to X-ray fluorescence of satellite material, i.e. atoms become ionized due to impinging radiation, which leads to an electronic transition from higher to lower shells, followed by the emission of a characteristic photon.
Depending on the element, these fluorescence photons can reach up to 115.6\,keV (uranium K-shell), formally being an X-ray photon due to its electronic nature, however falling into the `$\gamma$-ray' regime.
The strongest instrumental lines in {\it{NuSTAR}} are due to K-shell fluorescences of Cesium and Iodine at 28\,keV and 31\,keV, respectively.
Beyond the fluorescence lines, nuclear excitation lines, also appearing below 100\,keV, shape the background spectra up to $\sim 20$\,MeV.
Nuclear excitation is the interaction of an incoming particle with only the nucleus of an atom, therefore enhancing the energy scale of the process.
In instruments, either stable nuclei are excited directly by 1--100\,MeV particles, or nuclear reactions, such as proton or neutron capture, lead to new nuclei which are produced in an excited state and de-excite promptly.
For example, many of the strongest background lines in SPI are due to neutron captures and isomeric transitions of germanium isotopes.
Isomeric transitions are the spontaneous nuclear transitions of a meta-stable nuclear configuration to a less excited state by the emission of a characteristic $\gamma$-ray photon.
In SPI and other germanium detectors, multiple isotopes of germanium are naturally included in the crystals, so that multiple lines according to the different isotopes occur.
The SPI lines at $23.4$ and $175.0$\,keV are due to the second isomeric state of \nuc{Ge}{71m} ($T_{1/2} = 20$\,ms), and are coincidentally measured at $23.4 + 175.0$ ($T_{1/2} = 79$\,ns) $ = 198.4$\,keV to form its strongest background line \citep{Bunting1974_Ge198keV}.

Another strong line which always occurs in $\gamma$-ray measurements is the 511\,keV electron-positron-annihilation line.
Either $\beta^+$-unstable isotopes decay inside the satellite and produce a positron which quickly finds an electron to annihilate with, or CR bombardment leads to secondary positrons which slow down and also annihilate inside the satellite.

Compared to SPI, COMPTEL had poorer spectral resolution (Fig.\,\ref{fig:GammaRay_BG_spectrum}, bottom left), so that multiple lines overlapped and merged together as distinct line complexes, or weak lines were just smeared out and drowned in the continuum background.
A prominent line in the COMPTEL background was the neutron capture line on protons leading to a strong feature at 2.223\,MeV.
Most of these interactions occur for high accumulations of protons (hydrogen) which in COMPTEL was found either in its upper detector module filled with the liquid scintillator NE 213A (i.e. Xylene, $\mathrm{C_8H_{10}}$) or in {\it{CGRO}}'s fuel tanks filled with hydrazine ($\mathrm{N_2H_4}$) \citep{Schoenfelder1993_COMPTEL}.

\begin{figure}[t]
    \centering
    \includegraphics[width=0.8\textwidth,trim=0.25in 3in 0.25in 2.75in,clip=True]{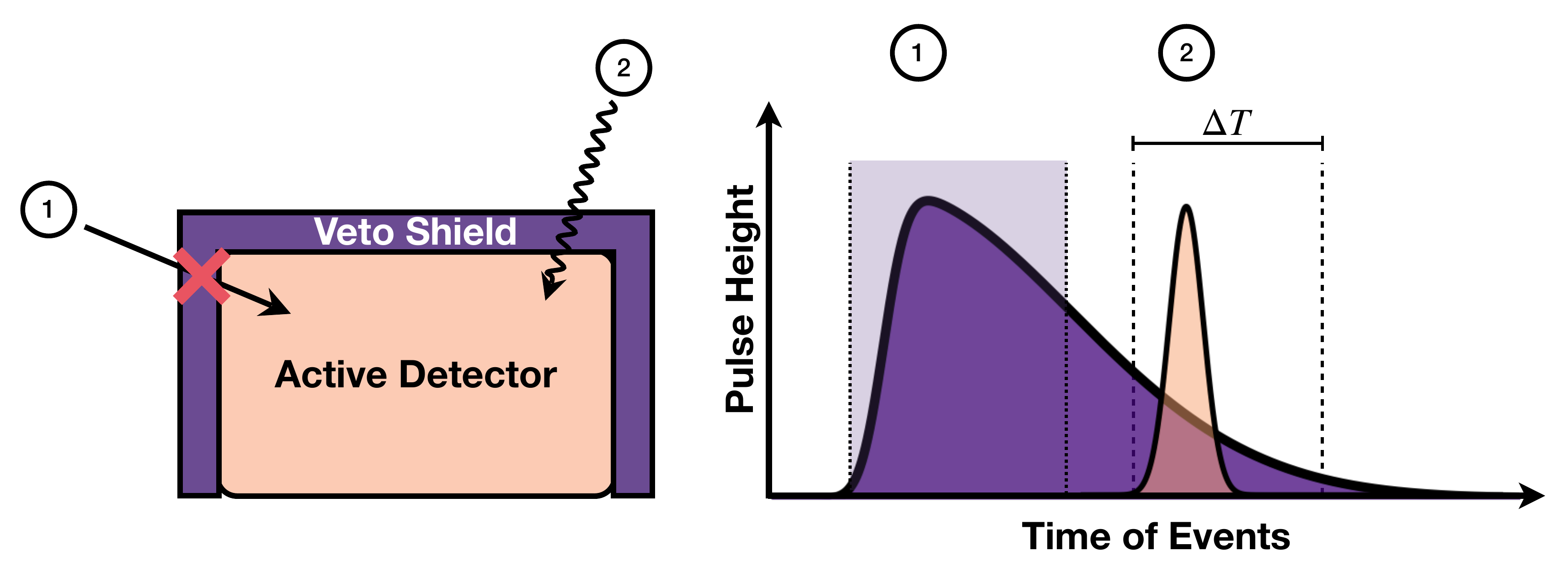}
    \caption{Schematic illustration of a ghost event. The remnants of electronic signals from the particles of a background event (1) that traversed the detector volume prior to the gamma ray (2) that triggered the instrument get read out along with the $\gamma$-ray signal.}
    \label{fig:ghost_events}
\end{figure}

\begin{figure}[b]
    \centering
    \includegraphics[width=0.47\textwidth]{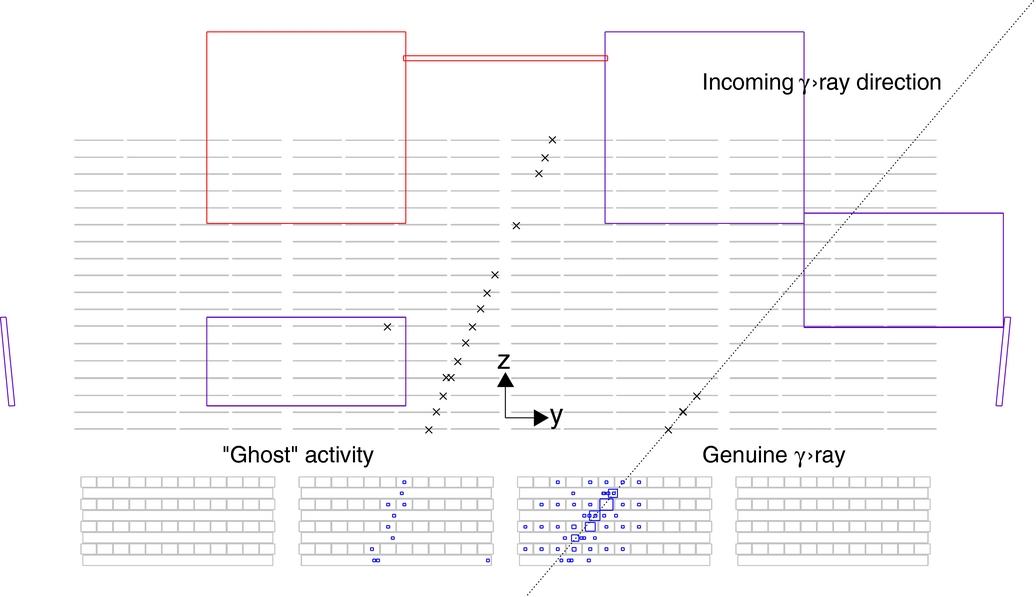}
    \includegraphics[width=0.31\textwidth]{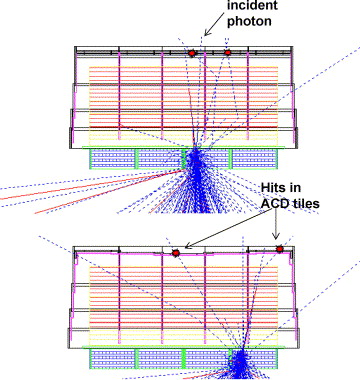}
    \caption{\textit{\textbf{Left}}: From \citep{2012ApJS..203....4A}, an example of ghost activity in the LAT. On the right of the figure is a genuine $\gamma$-ray whose reconstructed track is shown by the dashed line. The ghost activity is visible in the ACD, tracker and calorimeter. Only those ACD tiles with a signal are shown. \textit{\textbf{Right}}: Reproduced with permission from \citep{2007APh....27..339M} - an illustration of the effect of back-splash in the simulation of the LAT ACD. Red lines show the charged particles and blue dashed lines show the photons. The red dots show the signals in the ACD due to back-splash.}
    \label{fig:ghost_event_real}
\end{figure}

In the pair-production regime a reduction in the $\gamma$-ray detection efficiency can be due to a number of effects including instrumental pile-up, the incorrect vetoing of $\gamma$-rays and particle leakage into the detector.
One source of instrumental background is the residual signal that remains from the shower initiated by a charged particle, which has been vetoed, but whose decay time is such that traces still remain when a $\gamma$-ray enters the detector volume and causes a trigger \citep{2010arXiv1001.5005R}.
In this case, when the signals from the instrument are read out, there will be the signal due to the genuine $\gamma$-ray event but also the residual signal that remains from the previously vetoed event.
This can be seen schematically in Fig.\,\ref{fig:ghost_events}.
In {\textit{Fermi}}-LAT this residual signal is referred to as a `ghost' event and it can be present in the tracker, the calorimeter, the ACD or, indeed, in all three as is shown in Fig.\,\ref{fig:ghost_event_real}.
The effect has been modelled using simulations so its effects are well understood and are incorporated in the analysis of LAT data \citep{2012ApJS..203....4A}. 

Other sources of background that have been identified and effectively removed from the LAT data include non-interacting heavy ions and CR electrons that leak through the ribbons of the ACD \citep{2018arXiv181011394B}.
Improvements to the analysis and simulations post-launch have led to better particle-tracking algorithms \citep{2013arXiv1303.3514A} and $\gamma$-ray selections \citep{2018arXiv181011394B}.
Thus the effects of ghost events, leakage and non-interacting particles result in only a minor loss of efficiency in the LAT's $\gamma$-ray detection capabilities.

Another way in which an inefficiency is introduced for the detection of $\gamma$-rays at GeV energies is when a true $\gamma$-ray gets incorrectly vetoed.
In EGRET this was referred to as `back-splash' \citep{1993ApJS...86..629T}.
Although most of the particles in the electromagnetic shower travel along the direction of the incident $\gamma$-ray, a small fraction of them go in the backwards direction.
The low-energy photons in these showers Compton scatter electrons in the ACD and these charged particles can then cause a veto.
The effect became more pronounced at higher energies with EGRET's detection efficiency degraded by a factor of 2 at 10\,GeV compared to that at 1\,GeV \citep{2004APh....22..275M}.
The ACD for the {\it{Fermi}}-LAT was then optimised to avoid this issue \citep{2004APh....22..275M}.
An illustration of the effect of back-splash in the LAT ACD simulation model is shown in Fig.\,\ref{fig:ghost_event_real}, right.

\subsection{Background Suppression}\label{sec:background_suppression}
As shown in Fig.\,\ref{fig:GammaRay_BG_spectrum}, the measured detector rates from different instruments in the MeV to GeV range are on the order of $10^{-5}$--$10^1\,\mathrm{cnts\,s^{-1}\,keV^{-1}}$.
These rates are already reduced by different suppression mechanisms which decrease the rate of incoming particles and photons by several orders of magnitude.
Depending on the energy range and instrument again, the methods to reduce (instrumental) background  begin with the choice of the orbit (Sec.\,\ref{sec:orbits}).
However, most of the reduction in the MeV--GeV range is achieved by active anticoincidence shields, discrimination of signals in the readout electronics, and through case-specific data selections in the multi-dimensional data spaces of $\gamma$-ray telescopes.

\subsubsection{Anticoincidence Shields}\label{sec:ACS}
\begin{figure}[t]
    \centering
    \includegraphics[width=0.8\textwidth,trim=0.25in 2.5in 0.25in 2.25in,clip=True]{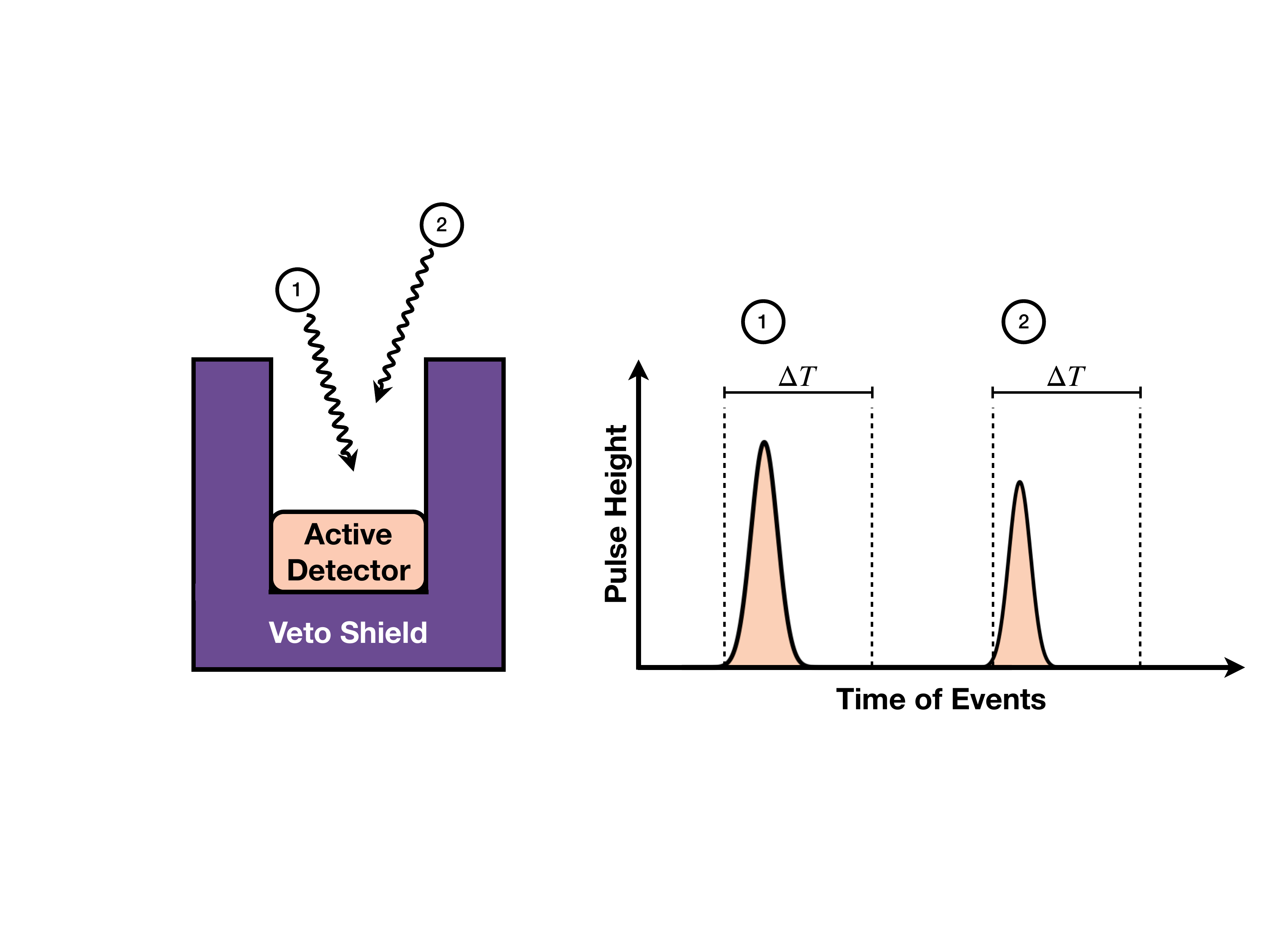}
    \includegraphics[width=0.8\textwidth,trim=0.25in 2.5in 0.25in 2.25in,clip=True]{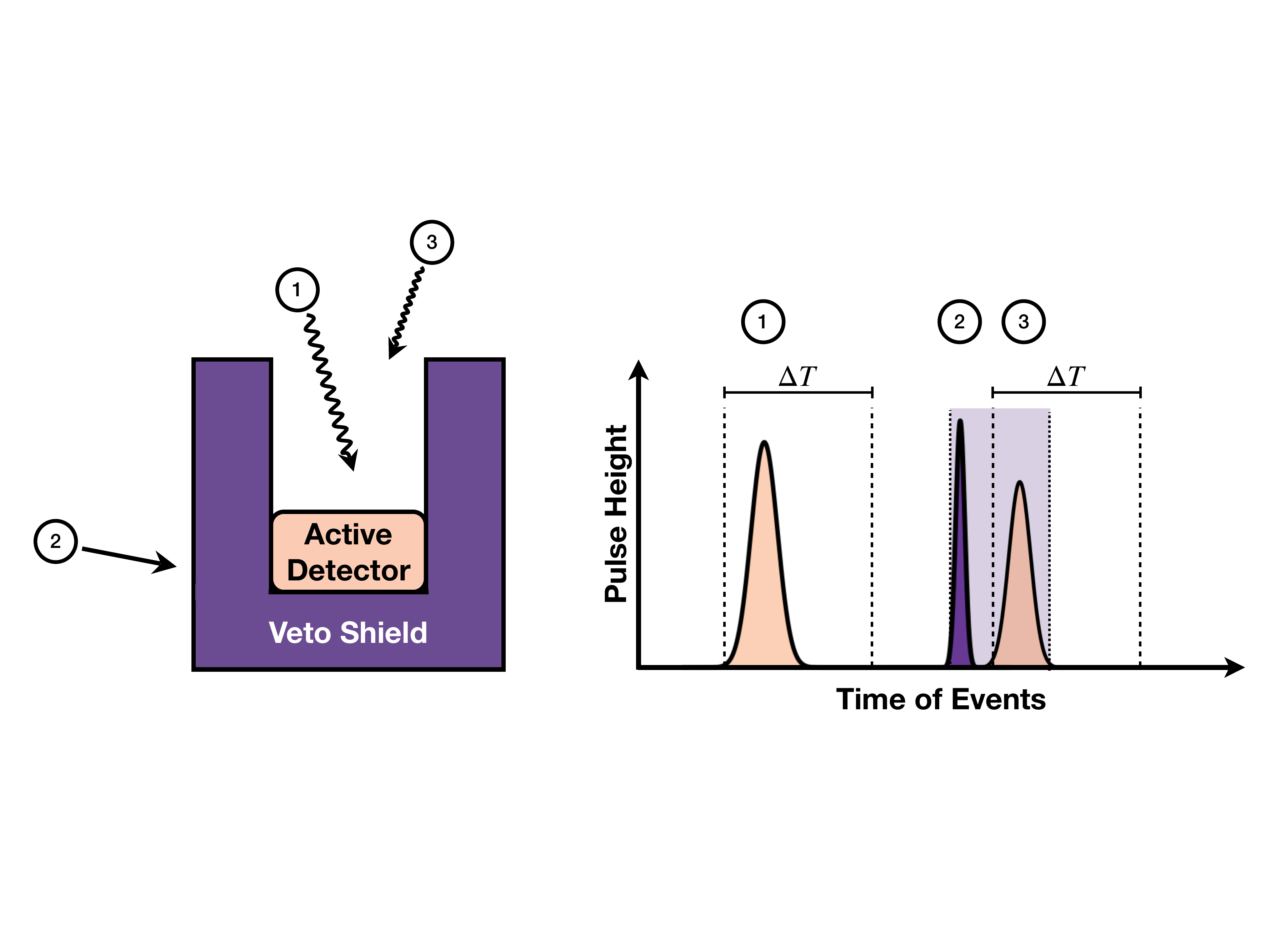}
    \caption{Anti-coincidence systems logics. \textit{\textbf{Top}}: Two photons, (1) and (2), arrive from inside the field of view of the instrument and deposit their energies in the active detector (left). In the electronics readout (schematic, right), a pulse is registered for each event, and with its start, the detecting system is unable to record any more events during a specified time $\Delta T$ (dead time). Either the pulse height or the integral over the entire pulse over the time of measurement $\Delta T$ converts the registered event into an electronic channel number, which will be associated with a photon energy after calibration. \textbf{\textit{Bottom}}: After photon event (1), a particle (2) hits the veto shield from the side. Shortly after, another photon (3) interacts with the detecting system. Because the veto shield triggers an anticoindicence (purple range, right), events (2) and (3) are both vetoed, and only event (1) is recorded.}
    \label{fig:anticoincidence_systems}
\end{figure}

The general idea of an anticoincidence shield is to veto unwanted particles and/or photons that would enter the detector.
This means the active detector is surrounded by another, sometimes U-shaped, active detector with a fast readout system.
In the case of a U-shaped detector, the effects are twofold: First, the inner detectors are shielded physically from all directions except for close to zenith (the size of the shield defines the field of view, Sec.\,\ref{sec:general_collimator}), and second, the inner detectors are shielded electronically from events that interact with the anticoincidence system.
In Fig.\,\ref{fig:anticoincidence_systems}, the normal, unvetoed observation case (top) and the vetoed observation case (bottom) with an anticoincidence signal triggered are shown.

Most modern MeV and GeV telescopes have veto systems made of scintillator crystals with a high light yield.
For example the veto shields of the IBIS and SPI telescopes onboard INTEGRAL are made of BGO, and show a typical count rate of up to $10^5\,\mathrm{cnts\,s^{-1}}$.
A considerable fraction of these counts would necessarily be measured in the main detectors and would heavily increase the average rate.

However, the veto shields also have a huge disadvantage: they are heavy and come with more mass than would actually be needed for the main detectors, effectively reducing their sensitivity.
More mass is equivalent to more instrumental background because CRs have more area to interact with.
That means that there is a trade-off between the increased mass and the background reduction where the latter typically gets precedence.

\begin{figure}[ht]
    \centering
    \includegraphics[width=0.80\textwidth,trim=0.75in 2.75in 1in 2.75in,clip=True]{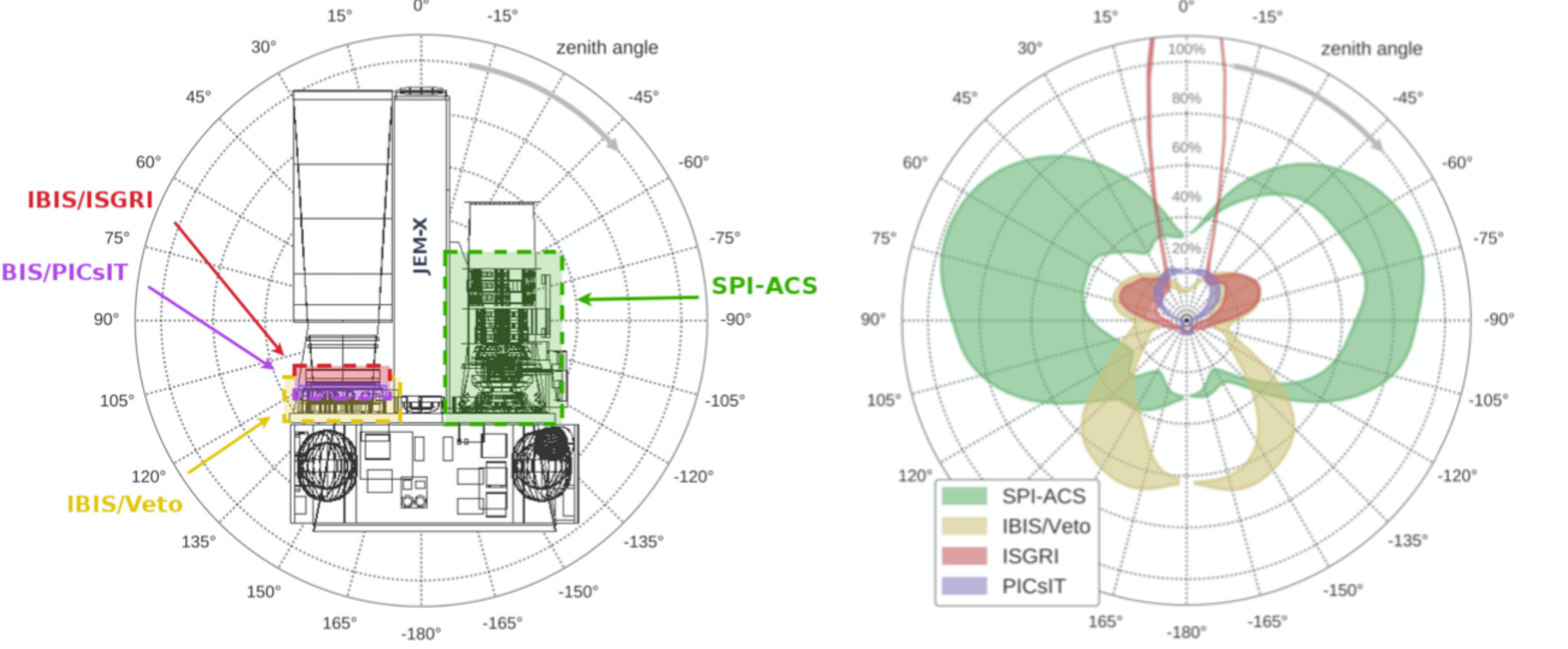}
    \caption{The INTEGRAL satellite as all-sky observatory. Shown are the different instruments and veto shields in the study of Savchenko et al. (2017, \citep{Savchenko2017_LVT151012}, \textit{\textbf{left}}) and the efficiency of the instruments with respect to the SPI-ACS (\textit{\textbf{right}}). For a given gamma-ray burst (GRB) spectrum and duration (here: Band function \citep{Band1993_BandFunction} with $\alpha = -1$, $E_p = 300$\,keV, and $\beta = -2.5$ for 8\,s), the different instruments would be expected to measure certain rates relative to each other. This describes a `$4\pi$ response'.}
    \label{fig:ACS_allsky}
\end{figure}

A major advantage of the massively increased photon collecting area of veto systems is their transient monitoring capabilities thanks to their quasi-all-sky fields of view.
While the main detectors of many instruments only observe in the zenith direction, the veto shields see the entire sky, unless blocked by the Earth.
If multiple instruments and shields onboard a satellite are combined, the sensitivity to transient events is largely enhanced, and the satellite functions as one big observatory.
Savchenko et al. (2017, \citep{Savchenko2017_LVT151012,Savchenko2017_GRB170817}) showed and used this for the INTEGRAL observatories, Fig.\,\ref{fig:ACS_allsky}.

Because the mass required to shield MeV detectors can comprise a considerable portion of the satellite payload, the effective collecting area supersedes that of the main camera by up to two orders of magnitude.
For example, the SPI veto shield ACS weighs 512\,kg and reaches a maximum effective area of $\sim 10^4\,\mathrm{cm^2}$ \citep[e.g.,][]{Savchenko2017_LVT151012}, however, without any spectral information (compared to the $10$--$10^2\,\mathrm{cm^2}$ of SPI).
The veto system on COSI, for example, made of CsI, weighs about 100\,kg \citep{Tomsick2019_COSI}.
This means that whenever an active veto system is installed, careful consideration should be given to whether spectral information can be added to its detection system so that a spectral analysis of transients can also be performed.

In the pair-production regime, where the background of charged CR particles outnumber the $\gamma$-ray events by a factor of $10^4$--$10^5$,  plastic scintillator tiles are mostly used nowadays for the ACD; this is the case for both LAT \citep{2007NIMPA.583..372M} and AGILE \citep{2009A&A...502..995T}.
These plastic tiles do not add too much weight and are a well-understood, efficient, reliable, and inexpensive technology \citep{2009ApJ...697.1071A}.
The ACD of LAT comprises a total of 89 scintillating plastic tiles, with varying surface areas (between 561 and 2650\,cm$^2$) and thicknesses (between 10 and 12\,mm), 16 of them on each of the four sides and 25 on the top of the instrument \citep{2012ApJS..203....4A}.
The ACD of AGILE comprises 13 independent charged particle detectors \citep{2006NIMPA.556..228P}.
As discussed in Sec.\,\ref{sec:background_suppression}, the segmentation of the ACDs of both LAT and AGILE allows for a localisation of the veto signal to avoid false vetoes due to back-splash.
To cover the gaps between ACD tiles in the X- and Y-axis, the LAT also has eight flexible scintillating ribbons.
In LAT the signal generated upon the passage of a charged particle is transmitted to the 194 PMTs (two for each ACD tile and two for each of the ribbons) via wavelength shifting fibres and clear fibres so that the veto can be registered.
The signals from the AGILE plastic scintillators are read out via optical fibres connected to 16 subminiature PMTs.
The total mass of the ACD on LAT is 284\,kg (the combined mass of LAT is 2,789\,kg) while that of AGILE is 22.5\,kg (the combined mass of the AGILE scientific instruments is $\sim 100$\,kg).
%

\subsubsection{Pulse Shape Discrimination}\label{sec:PSD}
Another useful technique to filter out particle events, for example in MeV telescopes, can be achieved by measuring the shape of the incident pulse in the electronics.
These pulse shape discriminators (PSD) include templates of rise times to peak and fall times to base level for different particle types so that unwanted particles can efficiently be ignored.
The templates depend on the interaction locations in the detectors as well as on the charge carrier mobility as a result of the electric fields and applied voltages \citep{Philhour1998_PSD}.
The general idea to distinguish, for example, $\beta$-particles from photons interacting with the detectors, is that the particles mostly interact in one particular site to deposit parts of their kinetic energy, whereas photons show deposits in multiple sites.
This means that single-site events from electrons could potentially be rejected, which enhances the sensitivity of the instrument whenever the photon energies imply a high probability of scattering within the detector volume.
%
%

\begin{figure}[h]
    \centering
    \includegraphics[width=0.8\textwidth,trim=0.25in 2.5in 0.25in 2.25in,clip=True]{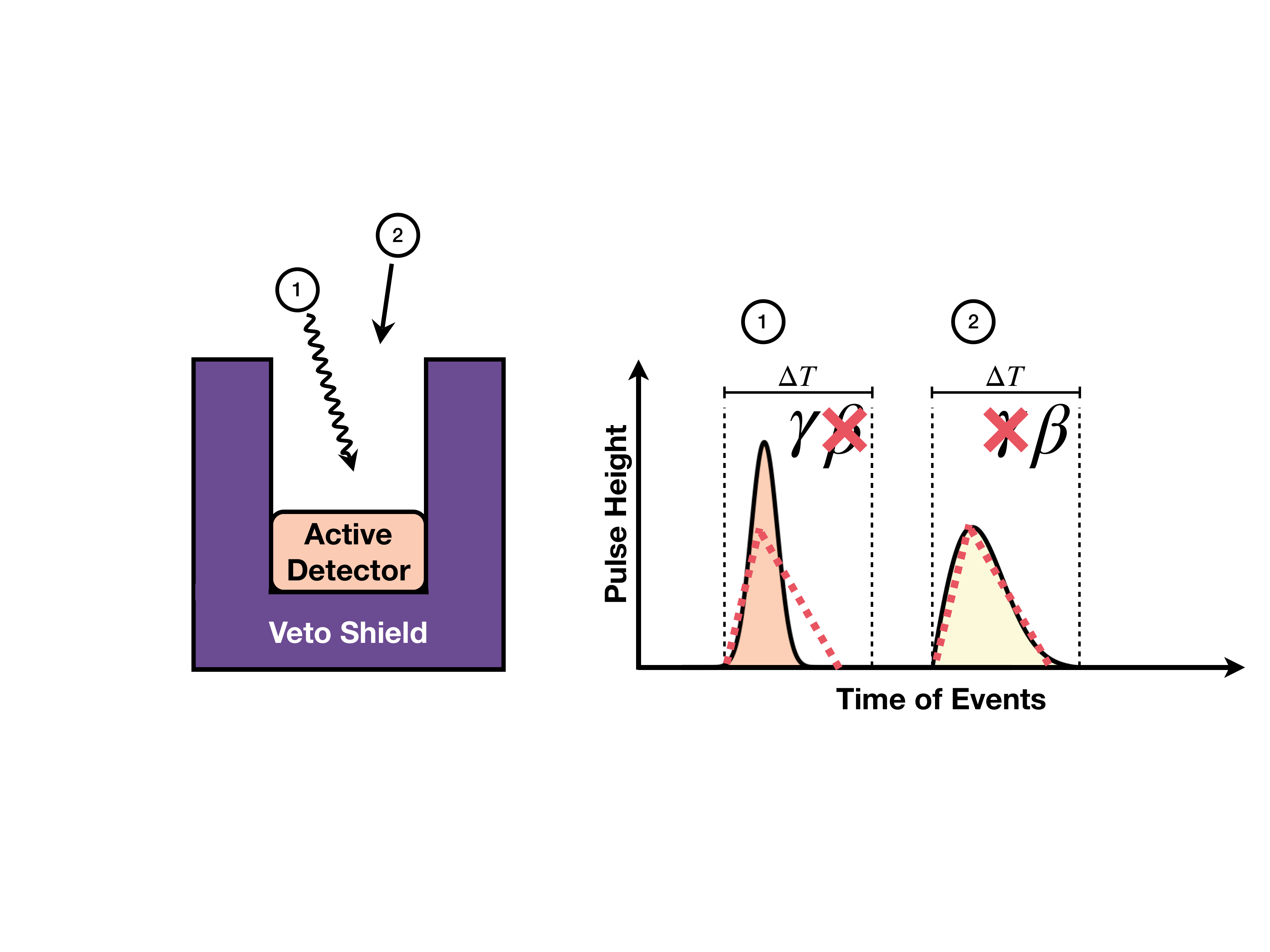}
    \caption{Working principle of a pulse shape discriminator. Events (1) (photon) and (2) (electron) are recorded by the detector. Their pulse shapes are compared to a template (dotted red). If the pulse shape is similar to a $\beta$-particle template, it is recognised and filtered out.}
    \label{fig:PSD_sketch}
\end{figure}

Because photons can also be directly absorbed in only one interaction, the energy threshold for a PSD should be set around the turnover from photo-electric absorption to Compton scattering (Fig.\,\ref{fig:cross_sections}), which depends on the material and geometry of the instruments.
In Fig.\,\ref{fig:PSD_sketch} a sketch of pulse-shape-discriminated particles compared to photons is shown.
PSD electronics have been employed, for example, in INTEGRAL/SPI, and are also used to suppress electronic noise which arises from the saturation of its Analog Front-End Electronics \citep{Roques2019_SPI}.

\subsubsection{Tailored Data Selections}\label{sec:data_cuts}
The data that can be sent from the satellite to ground stations for further analysis and diagnostics is limited and so, those that are downloaded must be considered and selected carefully.
In the case of the LAT for example, the on-board trigger is designed to pre-scale the volume of each particular event class that is downloaded so that a maximum of $\gamma$-ray candidates can be kept whilst also sampling a sufficient quantity of particular background and periodic trigger events to help characterise and keep track of the conditions under which the signal is detected.
Once downloaded, the data can be subjected to different sets of analysis cuts, each designed with particular scientific goals in mind.
For LAT, these are known as {\textit{event classes}} and they are optimised to address different science cases including, for example, transients, steady point sources or diffuse backgrounds.
The quality and efficiency of the cuts is different for each class. 
Similar data selections can apply for the event selections in Compton telescopes, for example, utilising the Compton Data Space \citep{Schoenfelder1993_COMPTEL} to distinguish background and sky photons.
%

\section{Astrophysical Sources of Gamma Rays: Not one fits all}\label{sec:science_cases}\label{sec:astrophysical_sources}
\begin{figure}[ht]
    \centering
    \includegraphics[width=0.98\textwidth]{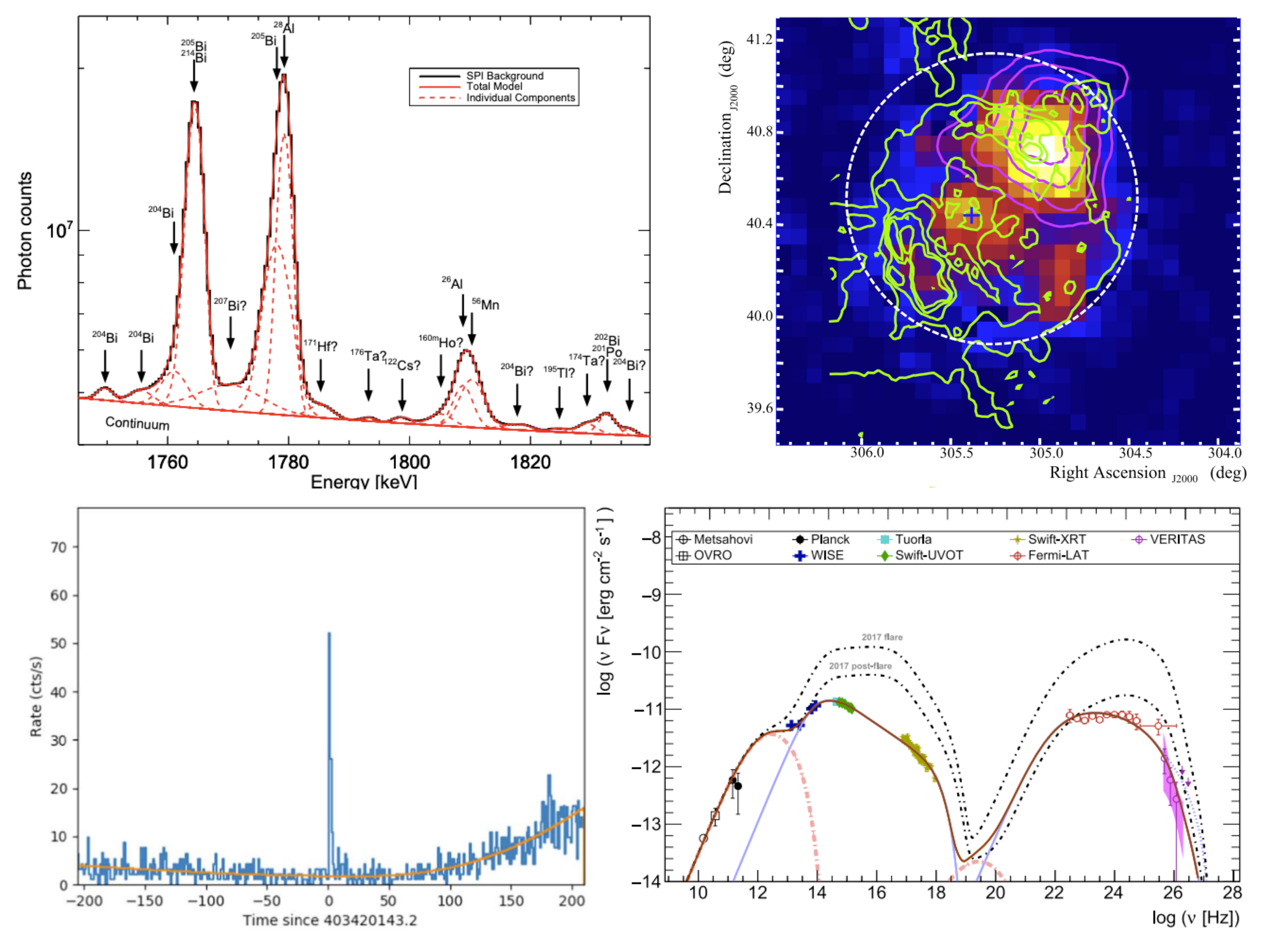}
    \caption{\textit{\textbf{Top left}}: Gamma-ray spectroscopy using INTEGRAL/SPI. The spectral decomposition into lines is shown for near the $^{26}$Al line \citep{Diehl2018_BGRDB}. \textit{\textbf{Top right}}: A significance map showing the $\gamma$-ray sky above 15\,GeV around the supernova remnant $\gamma$-Cygni (G78.2+2.1 / VER\,J2019+407) reproduced with permission from \citep{2016ApJ...826...31F}. Shown are the 1420\,MHz observation from the Canadian Galactic Plane Survey at brightness temperatures from 22 to 60\,K (green), the VHE source, VER J2019+407, smoothed photon excess contours (magenta), and the location of the $\gamma$-ray pulsar, PSR\,J2021+4026 (blue cross). The boundary of the extended LAT source 3FGL\,J2021.0+4031e is indicated by the white dashed circle. \textit{\textbf{Bottom left}}: The light curve of GRB131014 in the 0.03--1\,GeV energy range. The polynomial fit to the background is shown by a red line. The data are analysed using the LAT low-energy technique, designed to optimise the study of bright transient events below $\sim 1$\,GeV \citep{2019ApJ...878...52A}. \textit{\textbf{Bottom right}}: The spectral energy distribution of the blazar 1ES\,1215+304 from \citep{2020ApJ...891..170V}. The data and model are from the source when it was found to be in a low state. Shown are the blob synchrotron and synchrotron self-Compton (SSC) contributions (pale blue), the jet synchrotron and SSC emission (dotted–dashed pink), the intrinsic SSC emission without absorption from the extra-galactic background light (dotted blue), and the sums of all components (thick brown and thick black dotted–dashed).}
    \label{fig:science_panorama}
\end{figure}

\noindent Depending on the scientific goal of the observations being undertaken, the $\gamma$-ray instruments look very different because they are designed for specific tasks.
Figure\,\ref{fig:science_panorama} shows a selection of images which highlight the diversity of the science that can be studied at $\gamma$-ray energies.
The instrument capabilities need to be optimised according to both the energy range of the $\gamma$-rays being sought and the science case under study.
An in-depth description of both Galactic and extragalactic $\gamma$-ray science can be found in Volume 3 of this Handbook.

Once the instrumental background has been taken into account in the data analysis, the signal that remains is that due to astrophysical $\gamma$-rays.
Depending upon the energy range being investigated and on the pointing direction on the sky, this could be a superposition of a number of different components.
Each of these components needs to be modelled and understood in order to study the $\gamma$-ray emission detected.
Gamma-ray sources can appear point-like or extended, depending upon the combination of their intrinsic nature and on the angular resolution and exposure time of the instrument.
The $\gamma$-ray emission from resolved sources will lie on top of that from the diffuse $\gamma$-ray background, itself a combination of unresolved point sources, the isotropic diffuse background and possibly containing so-called exotic components such as contributions from dark matter annihilation and axions.
Many solar-system objects, for example the Sun and the Moon, are $\gamma$-ray emitters and, in addition to being studied in their own right, constitute a foreground source that has to be accounted for when they pass between the $\gamma$-ray telescope and more distant sources for those instruments who can operate in their presence. 
The spatial, spectral and temporal nature of the $\gamma$-ray sources being investigated are important considerations when designing an instrument and optimising the observational and analysis strategy.

Some $\gamma$-ray sources, certain supernova remnants or radio galaxies, for example, are extended and can have multiple emission components or exhibit different spectral features at different locations.
The identification of a position-dependent photon index can help map out the underlying structure of the source and thus, high angular and spectral resolution is a requirement.

Sources can be steady emitters, meaning that they emit a flux that does not vary significantly with time.
Often, extended sources of $\gamma$-rays have been found to belong to the class of steady emitters.
The $\gamma$-ray emission from point-like sources can be variable over many different timescales from minutes (some AGN) to years (e.g., some binary systems) or, indeed, it can be periodic (pulsars and binaries) quasi-periodic (some AGN) or episodic (AGN, some pulsar wind systems).
This variable signal may well sit on top of a more steady component.
Other sources of $\gamma$-rays, such as GRBs and perhaps fast radio bursts, are one-off events meaning that their detection is dependent upon having a large enough field of view.

The spectral properties of the $\gamma$-rays being studied should also be considered.
Many $\gamma$-ray sources have continuous spectra that follow a power law, due to the non-thermal nature of their emission.
The spectra of sources whose $\gamma$-ray emission is due to nuclear transitions will have a line nature.
Many dark matter models also predict mono-energetic $\gamma$-ray signals meaning that spectral lines at an energy corresponding to the mass of the annihilating or decaying particle are sought.
Similarly, the annihilation signature of neutral pions (an indicator of hadronic processes at work in the $\gamma$-ray source) will exhibit a characteristic bump.
Many $\gamma$-ray sources also exhibit spectral breaks and cutoffs so, depending upon the importance of accurately measuring these spectral features, the energy resolution of the instrument is an important consideration.

\section{Instrument Designs}\label{sec:instrument_designs}
Gamma-ray measurements in the MeV and GeV range classically rely on the modulation of one or more data space dimensions.
Because single photons are counted in individual detector units, such a variation can appear minuscule and still lead to a significant change if treated properly by statistical means.
The recognition of one or zero counts in the complex data spaces over a longer period of time leads to almost unique inferences when the full instrument response is applied.
The instrument response is, in general, a kernel function that converts an (astro)physical model, such as a point-like or extended source with a certain spectral shape with physical units, into the native data space of the instrument, always counting photons per detector, time, energy (electronic read-out channel), or other entities, as a function of its intrinsic coordinates given as zenith and azimuth angle.
The instrument's geometrical detecting area $A_{\rm geom}$ is therefore reduced to an effective area $A_{\rm eff}$ which depends not only on the incident photon energy $E_{\rm inc}$, time $T$, zenith and azimuth angle ($Z,A$) of the source, but also on the entire structure of the instrument, environmental conditions (temperatures, voltages, etc.), and on the satellite mountings and orbit.
Even for the simplest of all instrument designs, collimators (Sec.\,\ref{sec:general_collimator}), it holds true that
\begin{equation}
    A_{\rm eff}(E_{\rm inc},Z,A,T,\dots) \leq A_{\rm geom}\mathrm{.}
    \label{eq:effective_area_vs_geom}
\end{equation}
For example, the geometrical detecting area of INTEGRAL/SPI's 19 Ge detectors is $508\,\mathrm{cm^2}$ while the maximum effective area for $Z=0^{\circ}$ is $125$ and $65\,\mathrm{cm^{2}}$ for 0.1 and 1.0\,MeV, respectively \citep{Vedrenne2003_SPI,Sturner2003_SPI,Attie2003_SPI}.

In what follows, different instrument designs, i.e. different aspects of modulation in various data spaces, are outlined briefly.
The reader is referred to the subsequent Chapters in which each of the $\gamma$-ray telescopes apertures are explained in more detail.

\subsection{General Considerations: A Gamma-Ray Collimator}\label{sec:general_collimator}
The basic modulation categories are summarised into temporal, spatial, energetic, and other apertures, as well as combinations thereof.
As described earlier in this Chapter, the history of low-energy $\gamma$-ray detectors started with collimators which should be considered a temporal and spatial modulator by the classification above.
They are described as the basic principle from which other designs can be derived in the following.

\begin{figure}[!h]
    \centering
    \includegraphics[width=0.40\textwidth,trim=4in 2in 4in 2in,clip=True]{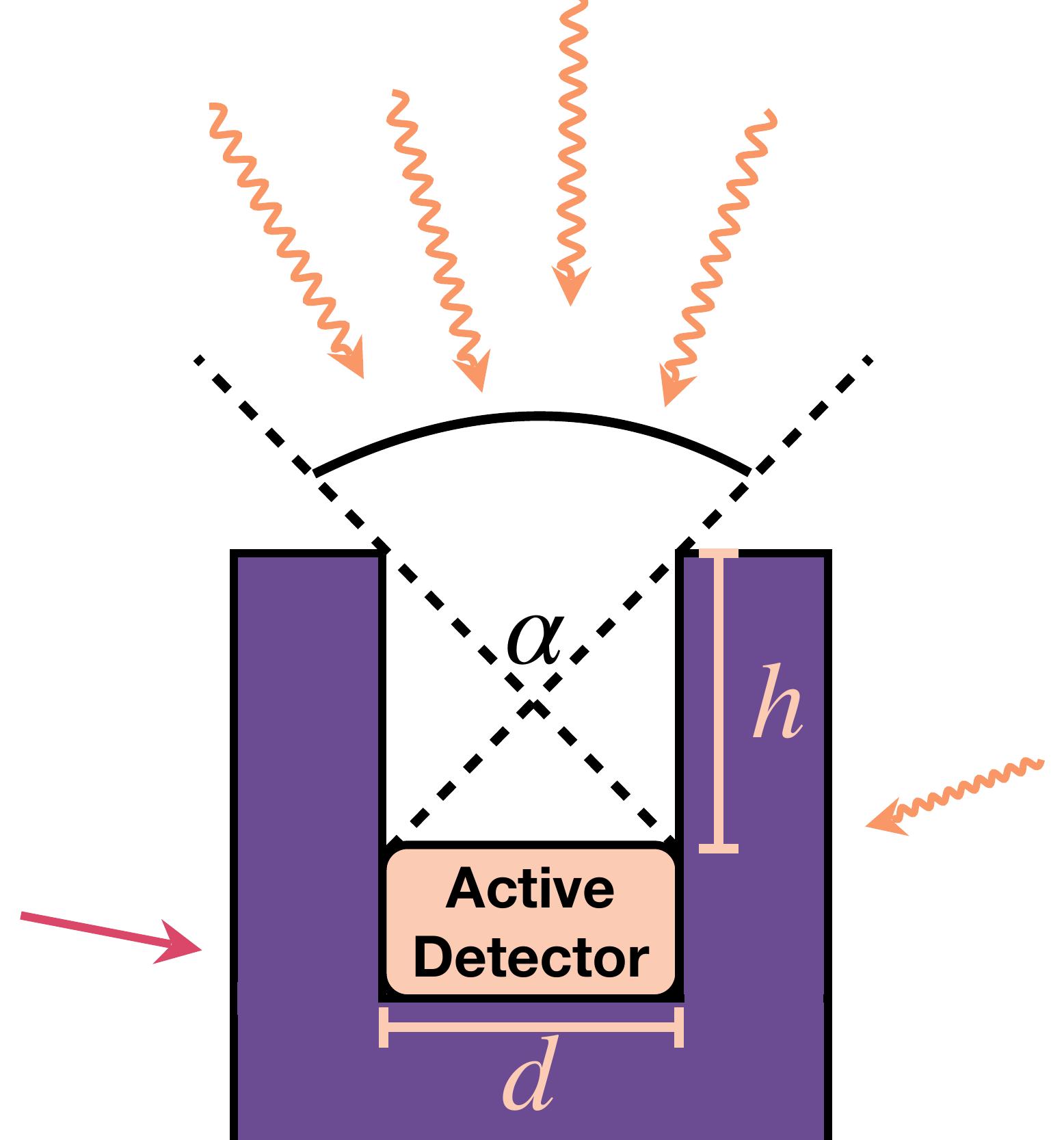}
    \includegraphics[width=0.40\textwidth,trim=4in 2in 4in 2in,clip=True]{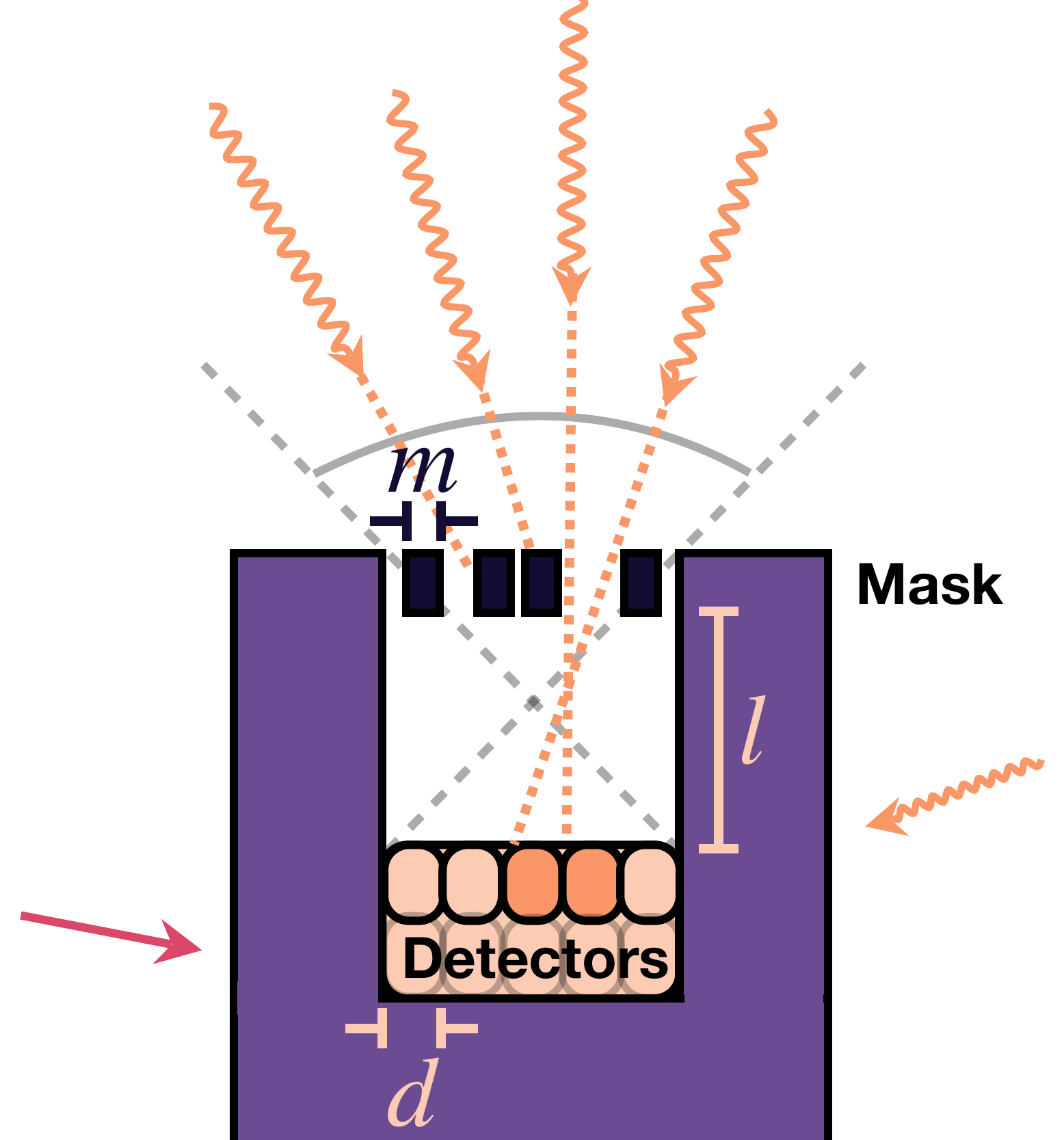}
    \caption{Sketches for collimator and coded mask telescopes. \textbf{\textit{Left}}: A collimator is built from an active detector that is surrounded by passive or active material to block photons and particles from the side. Only photons within the field of view, defined by the opening angle, $\alpha$, are recorded. \textbf{\textit{Right}}: A coded mask telescope adds opaque and transparent mask elements at the opening of the collimator tube. If the active detector is pixellated, this encodes incoming $\gamma$-ray photons spatially, and their origin can be inferred.}
    \label{fig:collimator_coded_mask_sketch}
\end{figure}

Collimator apertures are designed as large, often cylindrical, tubes with a detector unit (the camera) at the base of the tube (Fig.\,\ref{fig:collimator_coded_mask_sketch}, left).
The tube itself shields photons and particles from the side and the back, most of the time being itself an active $\gamma$-ray detector to veto those unwanted events.
In this way the central camera only observes in the zenith direction with a field of view given by the measurements of the tube.
As an example, if we take a cylindrical camera with diameter $d$, placed in a collimating tube with height $h$, the field of view, defined by the opening angle $\alpha$ of the aperture, would be given by
\begin{equation}
    \alpha = 2 \arctan(d/h)\mathrm{.}
    \label{eq:field_of_view}
\end{equation}
If material opacities are ignored for the moment, the angular response of the collimator with one detector unit as the camera can be expressed analytically for a plane-parallel beam of light as $A_{\rm eff}(Z,A) = A_{\rm geom} \left(\Theta(Z+\alpha/2) - \Theta(Z-\alpha/2)\right)$, where $\Theta(x)$ is the Heaviside step function and $Z$ and $A$ are the zenith and azimuth angle, respectively.
This means that if the source is inside the field of view, the camera can detect all of the emitted photons, while it sees zero counts when the source's aspect angle is greater than half the opening angle.
Departing from this ideal view, for example if the central camera consists of more than one detector and is therefore pixellated, the rise of a source with respect to the camera (decreasing zenith) now leads to a gradual increase of the effective area until the source is directly above the camera.
For small fields of view, this results in an effective area of approximately $A_{\rm eff}(Z,A) =  \frac{A_{\rm geom}}{\pi} ( \arccos(\tau) - \sin(2\arccos(\tau)) )$, where $\tau = \tan(Z)/\tan(\alpha/2)$.
This is strongly simplified and only serves as a means to describe the zenith-dependence of a collimator-type instrument.

These ideal treatments are erroneous once a real instrument is considered:
The field of view is not a sharply defined region as described above, but depends on energy.
The higher the photon energy, the higher the probability that the photon is not absorbed by the collimating material so that it may be detected even from `outside' the field of view.
If the instrument is not perfectly cylindrical, for example if it is hexagonal or octogonal, the effective area gains an azimuth dependence.
Finally, and probably most importantly for the analysis of $\gamma$-ray data, the incident photon energy $E_{\rm inc}$ is not necessarily the measured photon energy $E_{\rm meas}$:
Because of Compton scattering, escape peaks, and instrumental spectral resolution, the measured photon energy is related to the incident photon energy only by a known but non-invertible redistribution matrix (Sec.\,\ref{sec:measurements}).
This means that a measured spectrum is never representative of the source spectrum so that the latter must be inferred by forward modelling.
The forward modelling then requires complete knowledge of the instrument, which is condensed in the response, often separated into an effective area contribution plus an energy redistribution, and which assumes a certain source model.
The responses of $\gamma$-ray telescopes are typically determined by particle physics simulations using GEANT (Sec.\,\ref{sec:simulations}), which are then validated by calibration measurements on Earth using either radioactive sources or particle accelerator beams (Sec.\,\ref{sec:calibration}).
The source model can be versatile but requires the basic parameters of the object of interest, such as position, spatial extent, spectral shape, and temporal behaviour.
If the spectrum of a point source is being analysed, the first two properties are typically fixed to known values.
With more elaborate techniques, however, all unknown parameters of the observed target can be inferred in a single inference step.

Different collimators have already been flown on balloon experiments between the 1960's and 1990's.
The most successful collimator aperture was OSSE on {\it{CGRO}} \citep{Johnson1993_OSSE}.
It consisted of four independent, single-axis orientable, and actively shielded NaI(Tl)-CsI(Na) detectors, each surrounded by a tungsten shield.
The fields of view of the detectors were $3.8^{\circ} \times 11.0^{\circ}$ and sensitive in the 0.05--10\,MeV photon range.
Due to its four independent units, OSSE could measure the instrumental background by simultaneously pointing away from and at the source of interest.
This has the advantage that only the instruments themselves are moved and not the entire satellite.
This technique was then further used to perform the first temporal and spatial modulated measurement which led to the first image reconstruction ever of the Galactic diffuse 511\,keV emission \citep{Purcell1993_511,Purcell1997_511}.
Because the rise into and away from the fields of view of the four detectors changes uniquely with time over several years, an image could be reconstructed by singular value decomposition.
It was shown for the first time that the 511\,keV emission from the centre of the Galaxy was not point-like and variable, but extended and constant.
\subsection{Temporal and Spatial Modulation Apertures, Geometry Optics: Coded Mask Telescopes}\label{sec:temporal_spatial_apertures}
Collimators have no spatial, i.e. angular, resolution.
The camera is pointed at a source to detect photons, and is then moved away so that a background estimate can be provided.
This is the simplest form of a temporal (or spatial) modulation: on--off observations.
However, if more than one source is in the field of view, they might be difficult to analyse separately, especially if the field of view is large.
The apertures are therefore changed to include more information.

One way to improve the angular resolution is to place a mask on the top of the collimator's shielding tube which encodes the incoming light beam to cast shadows onto the detecting area, so-called `shadowgrams'.
The first mentioning of coding $\gamma$-rays appears in Mertz \& Yound (1962, \citep{Mertz1962_fresnel_coding}) in the context of Fresnel transformations of images.
This mask consists of opaque and transparent elements so that a fraction of the incoming light is blocked and only certain parts of the camera are illuminated.
Much finer variations in the aspect angle change between source and telescope can be recorded with such a coded mask.
The improvement in angular resolution then depends on the mask element size $m$, the size of the detector pixels $d$, and the separation between mask and camera $l$.
In order to separate shadowgrams from different source positions inside the field of view, the detector plane must therefore be pixelated.
For technical reasons, the detector size is adjusted to the science case and in particular the photon energy.
While in the MeV range this means that one detector (one `pixel') is several cm in size, which ultimately limits the angular resolution, the pixels can be much smaller (few mm) in the case of 100\,keV detectors.
This originates from the attenuation lengths required to stop a 1\,MeV photon (e.g. in tungsten $\mu^{-1} \approx 1.6$\,cm) compared to a 100\,keV photon ($\mu^{-1} \approx 250\,\mathrm{\mu m}$).
The mask elements should be as small as possible for the angular resolution to be maximised.
However, the sensitivity of the instrument suffers when the mask element size is smaller then the detector size.
The angular resolution of a coded mask telescope is approximately given by
\begin{equation}
    \delta\Theta = \sqrt{ (m/l)^2 + (d/l)^2 }\mathrm{,}
    \label{eq:coded_mask_resolution}
\end{equation}
and the positioning accuracy by $\delta\alpha \approx (S/N)^{-1} \delta\Theta$ with $S/N$ being the signal-to-noise ratio of the source given a suitable background estimate.
This means that the localisation of a coded mask telescope naturally supersedes its angular resolution when the source is strong.
The optimal trade-off between angular resolution, localisation accuracy, and sensitivity is provided when $m \approx d$ \citep{Skinner2008_coded}.

In order to remove ambiguities in the mask patterns that can emerge if a certain degree of symmetry is involved in the instrument design, targeting coded mask telescopes follow a particular observing strategy.
Given the angular resolution and specific geometry of the instrument (symmetries, field of view), an observation pattern can be performed instead of staring at the source of interest for a long time.
For example, INTEGRAL performs a rectangular $5 \times 5$-pattern around the source of interest, called dithering, to optimally sample the different shadowgrams of the mask onto the camera.
This can also provide a measure of the unknown instrumental background during this observation because the shadowgrams of the sources inside the field of view smear out over longer periods of time \citep{Siegert2019_SPIBG}.

Present and past coded-mask telescopes are ISGRI (0.03--0.4\,MeV, \citep{Ubertini2003_IBIS}) and SPI (0.02--8\,MeV, \citep{Vedrenne2003_SPI}) on board INTEGRAL, {\it{Swift}}-BAT (0.015--0.15\,MeV, \citep{Krimm2013_SwiftBAT}), the CZT Imager onboard Astrosat (0.01--0.15\,MeV), the All Sky Monitor (ASM) onboard RXTE (0.002--0.012\,MeV), and the wide field camera (WFC) onboard {\it{Beppo}}SAX (0.002--0.030\,MeV).
Details about coded-mask telescopes are provided in the respective Chapter of this book.

Another possibility to remedy the need for spatial variation can be achieved if parts of the instrument itself are movable.
With several sub-collimators, which had been realised in the {\it{RHESSI}} imaging system \citep{Hurford2002_RHESSI,Smith2004_RHESSI}, for example, an arcsec angular resolution had been achieved in $\gamma$-ray observations of the Sun.
In the case of {\it{RHESSI}}, a pair of separated but parallel grids (opaque slats and transparent slit-like elements) inside each of its nine collimator tubes is rotated with respect to the detector plane at 15 revolutions per minute.
This leads to the effect that a change in aspect angle produces a modulation of the transmission of the grid pair in time.
The rotating shadow of the slats in the top grid then falls on the slits or slats of the read grid which results in a time-modulated transmission from zero to 50\,\% and back.

Similar techniques have been applied for solar flare observations onboard the {\it{Hinotori}} mission with its rotating modulation collimator (RMC, 0.02--0.04\,MeV; \citep{Sakurai1991_Honitori}), HXT onboard Yohkoh (0.02--0.1\,MeV; \citep{Acton1992_Yohkoh}), and the balloon-borne HEIDI (High Energy Imaging Device; \citep{Crannell1992_HEIDI}) solar telescope with two RMCs.
A currently active temporal modulation telescopes is Insight-HXMT (0.02--0.25\,MeV; \citep{Zhang2018_HXMT}).

\subsection{Quantum Optics in the MeV: Compton Telescopes}\label{sec:compton_telescopes}
In the 1920's, A. H. Compton introduced the classical concept of elastic scattering for the interaction of photons with matter.
He showed that, in the energy range from $\sim 100$\,keV to $\sim10$\,MeV, this scattering takes place between the incoming photon and an atomic electron.
This results in an energised recoil electron and a deflected photon of reduced energy: in order to characterise the interaction, both secondary components must be measured.

Elastic scattering conserves energy and momentum 
$\ \   E_{0} = E_{\rm scat} + E_e $ , 
$\ \  \vec{p_{0}} = \vec {p_{\rm scat}} + \vec{p_e}$  , where 
$ \left| \vec{p_{0}} \right| = h \nu_{0} / c\ ,\ 
 \left| \vec{p_{\rm scat}} \right| = {h \nu_{\rm scat} / c} $,   and 
$ \left| \vec{p_e} \right| = m_{e} v \gamma \ \ $ 
 with 
 $\ \gamma = {1 / {\sqrt{1-\beta ^2}} }$ with $\beta = v/c$, which leads to the so called `Compton equation':
\begin{equation}\label{eq:Compton_wavelength}
    \lambda_{\rm scat} - \lambda _{0} = {h\over {m_{e} c}} (1-\cos \varphi) 
\end{equation}
where $h$, $m_{e}$, and $c$ are Planck's constant, the electron rest mass, and the speed of light, respectively.
The fraction ${h/ {m_{e} c}} = 2.426 \times 10^{-12}$\,m is often called the Compton wavelength, which is the wavelength shift for a $90^\circ$ scattering.
It is important to note that, in a Compton scattering interaction, the incident photon can never lose all of its energy even if it is completely backscattered.

\begin{figure}[b]
    \centering
    \includegraphics[width=0.40\textwidth,trim=3.2in 2in 3.2in 2.7in,clip=True]{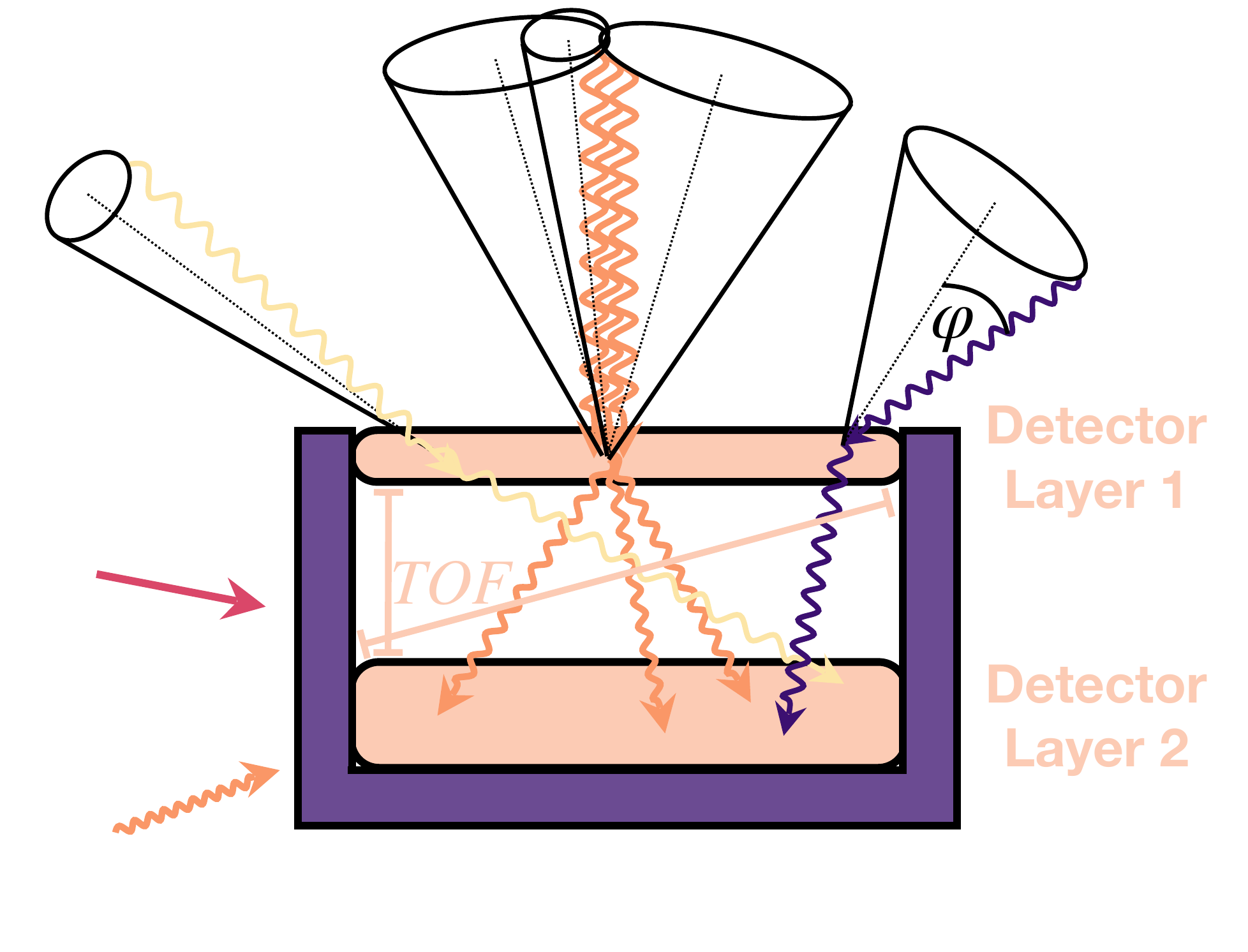}
    \includegraphics[width=0.40\textwidth,trim=3.2in 2in 3.2in 2.7in,clip=True]{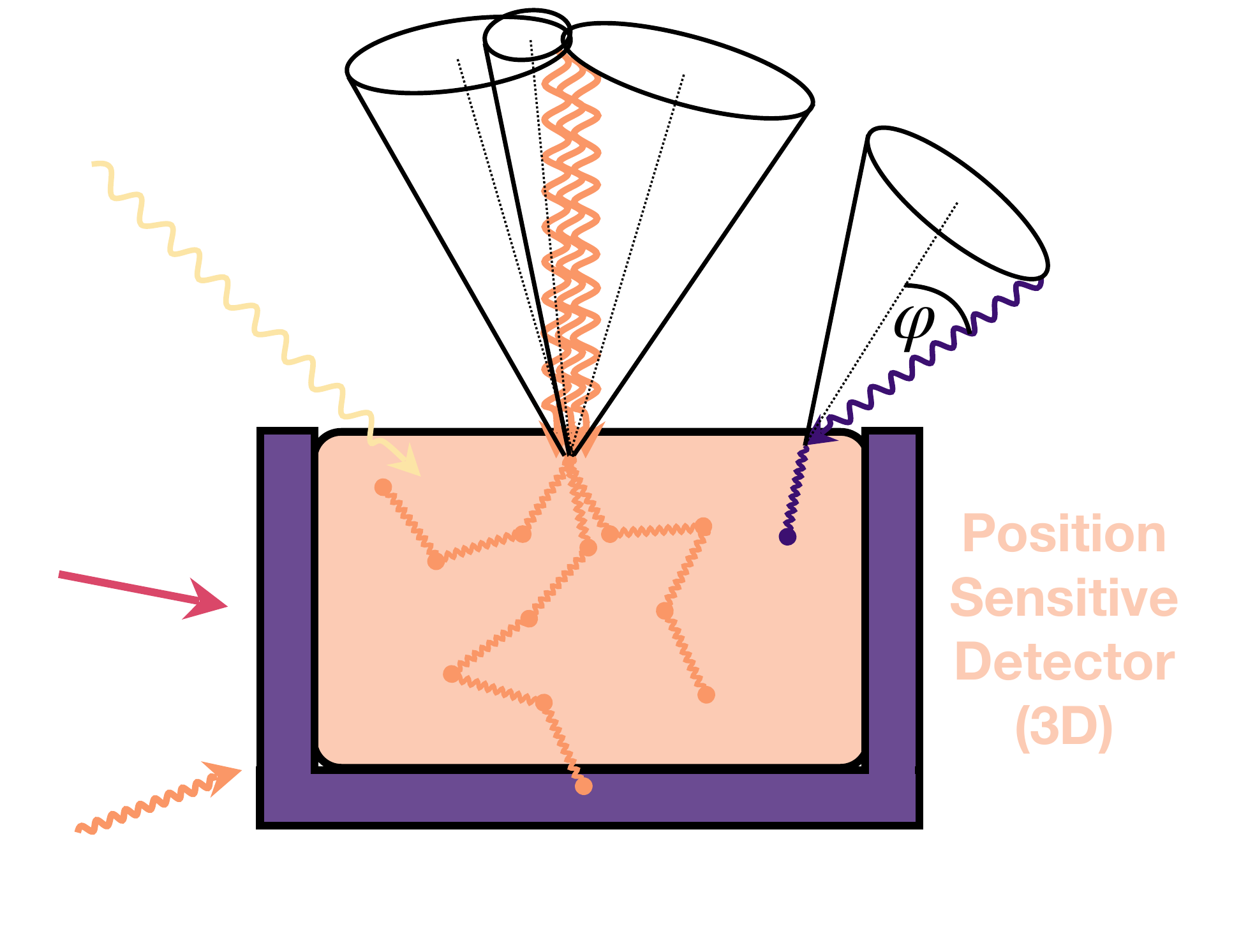}
    \caption{Compton telescope designs. \textbf{\textit{Left}}: Classic two-layer Compton telescope. Photons scatter in the upper detector layer 1 via Compton scattering and are absorbed in the lower detector layer 2. Given the Compton scattering Eq.\,(\ref{eq:Compton_wavelength}), each photon can be associated with a circle in the sky that forms a cone with the interaction point in the upper layer. The opening angle of this cone is given by the Compton scattering angle $\varphi$. Individual astrophysical sources can be identified by intersections of rings (orange photons). The time-of-flight ($TOF$) between the two layers is indicated with its minimal value (perpendicular path between layers) and maximal value (define by the opposite edges of the layers). \textit{\textbf{Right}}: Instead of two layers, a position sensitive detector, for example made of several detector strips or with gas, allows the photons potentially to scatter more often (zig-zag paths), always according to Compton scattering. The first interaction inside the detector volume again defines the Compton scattering angle. Three or more scatters help to identify the photon origins more clearly. In the case of the yellow photon on the right, only photo-absorption occurs, so that no Compton event reconstruction is possible.}
    \label{fig:compton_telescope_designs}
\end{figure}

Since, in this energy range, the photon energy is much higher than the binding energy of atomic electrons, the target electrons are taken to be free and non-interacting.
This is a good approximation at these energies.
For low-energy photons interacting with inner-shell atomic electrons however, `Doppler broadening' of the angular response occurs \citep{Zoglauer2003_DopplerCompton}.
The total cross section (or absorption coefficient) of Compton scattering in any target material depends directly on the electron density, and therefore on the nuclear charge, $Z$, of the detector material (Sec.\,\ref{sec:cross_sections}).

\noindent Equation\,(\ref{eq:Compton_wavelength}) can be solved for $\varphi$ and the wavelengths converted to energy,
\begin{equation}\label{eq:Compton_scatter_angle}
\varphi = \arccos \left[ 1 - m_{e} c^2 \left(
{1\over E _{\rm scat}} - {1\over {{E _{0}}}} \right)
\right],
\end{equation}
where the energy of the scattered photon is
\begin{equation} \label{eq:Photon_scattered}
 E_{\rm scat} = h \nu _{\rm scat} = {E_{0} \over {1+ {E_{0}\over m_{e} c^2} {(1- \cos \varphi) }}}\mathrm{,} 
\end{equation}
and the kinetic energy of the recoil electron is
\begin{equation} \label{eq:Electron_scattered_1}  
K_e = E_{0} - E_{\rm scat} = {{E_{0} (1-\cos \varphi )} \over { {m_{e} c^2}+E_{0} (1-\cos \varphi)}}\mathrm{.}
\end{equation}
$K_e$ can also be expressed in terms of the angle $\Theta $ between the incident photon and the direction of the recoil electron, making use of the relation  $\cot (\Theta) = (1+\alpha)  \tan (\varphi /2)$ where $\alpha = {E_{0} / {m_{e} c^2}}$, such that
\begin{equation} \label{eq:Electron_scattered_2}
 K_e = {{2 E_{0} \alpha \cos^2 \Theta} \over {(1+\alpha)^2 - \alpha ^2 \cos^2 \Theta}}\mathrm{.}
\end{equation}

\noindent Equations\,(\ref{eq:Compton_wavelength})--(\ref{eq:Electron_scattered_2}) are directly based on the kinematics of the elastic Compton scattering process and are the basis for various realisations of Compton telescopes.
In the `classical' two separated detector design (e.g., COMPTEL; Fig.\,\ref{fig:compton_telescope_designs}, left) a scattering detector D1 and an absorbing detector D2 trigger on a time-of-flight delayed coincidence signal and measure the positions and energy deposits of the two interactions.
The positions indicate the path of the scattered photon between D1 and D2.
Assuming the primary photon's energy is the sum of both energy deposits (i.e. no undetected energy leakage occurred), the scattering angle is given by Eq.\,(\ref{eq:Compton_scatter_angle}).
The direction of the incident photon will then be somewhere on a cone around the scattered photon trajectory.
If many photons from a distant point source are registered their individual `event cones' all intersect at the direction to this source.

For compact Compton telescopes (Fig.\,\ref{fig:compton_telescope_designs}, right), the basic principle is the same, however instead of two interaction layers, a position sensitive detector volume is used.
This decreases the size of the telescope because no time-of-flight information is used and instead event reconstruction techniques are applied to identify the possible paths of scattering $\gamma$-rays inside the instrument.
If the pixelation and threshold sensitivity allows recording the track of a Compton recoil electron, its initial direction (before Molière scattering disturbs it) can be used to reduce the event circle to an arc-length.
This can improve the overall sensitivity considerably.

The distribution of closest offsets between the event-cones and the true source direction is called the angular-resolution-measure (ARM).
The width and `lopsidedness' of the ARM distribution is caused by uncertainties in the position and energy measurements and by possible energy leakage from the system.
\subsection{Quantum Optics for Higher Energies: Pair Tracking Telescopes}\label{sec:pair_tracking_telescopes}
\begin{figure}[t]
    \centering
    \includegraphics[width=0.70\textwidth,trim=3.0in 2.7in 3.0in 3.0in,clip=True]{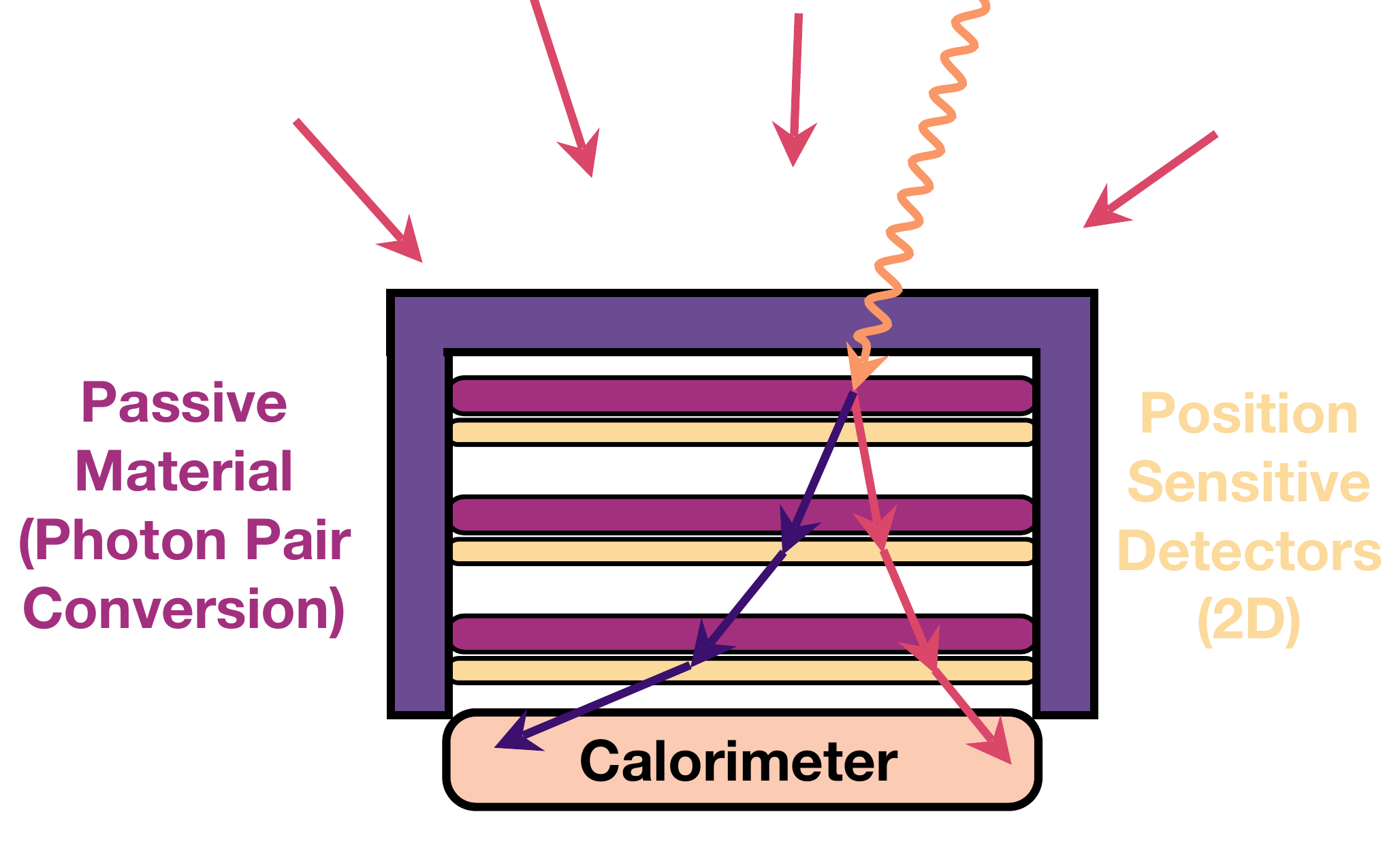}
    \caption{Pair-tracking telescope design. Incoming high-energy photons are converted into electron-positron pairs in passive material. With position sensitive detectors, the electrons and positrons are tracked until they deposit all their energy in a calorimeter. The tracks and final energy deposit are used to determine the total energy and direction of the incoming photons. In contrast to MeV telescopes, the background is predominantly from CRs, so that the anticoincidence system behaves like a shield above the tracker and calorimeter.}
    \label{fig:pair_tracking_telescope_designs}
\end{figure}

The measurement of the energy of a $\gamma$-ray in the pair-production regime is done by measuring the secondary products: electrons, positrons, and recoils on the target nucleus or electron.
The latter recoils are not easily measurable, but they are of minor importance for higher $\gamma$-ray energies.
The energies of the pair particles can be characterised by their scattering behaviour (Molière scattering for low-energy electrons) or they can be totally absorbed in a deep calorimeter, where the initiated shower at high energies gives additional information for the total energy.
Thus, to first order in most cases, the energy of the $\gamma$-ray is calculated by summing the energy deposited in each of the crystals of the calorimeter in the case of the calorimeter on LAT or AGILE or by means of the pulse-height analyzers (PHAs) for the Total Absorption Shower Counter (TASC) NaI(T1) calorimeter on EGRET.
Of course, some fraction of the energy of the incident $\gamma$-ray will have been deposited elsewhere in the detector prior to the shower's arrival in the calorimeter.
So, this energy must be estimated and added on to that deposited in the calorimeter.
The tracking information deduced from the tracker and the shower's position or trajectory into the calorimeter can be used to estimate its path through the detector and, combined with the energy measured in the calorimeter, can be used to estimate the energy deposited elsewhere in the detector (Fig.\,\ref{fig:pair_tracking_telescope_designs}).

The 3D passage of the electrons and positrons until they reach the calorimeter is reconstructed using a track finding algorithm.
For {\it{Fermi}}-LAT, for example, the algorithm starts by generating a track hypothesis, i.e. a proposed trajectory made of locations and directions, that is accepted or rejected given the detector signals \citep{Atwood2009_FermiLAT}.
One possibility to fit these tracks and therefore to identify origin of the initial photon on the celestial sphere is by pattern recognition:
The algorithm starts by assuming an $(x,y)$-position in the top layer, which is compared to a subsequent hit as well as the energy deposited in the calorimeter.
If these positions are close in the multidimensional dataspace, a candidate track is created by Kalman fitting.
This process is repeated into the next layers, always taking into account the covariance matrices of the previous steps, until an adequate figure-of-merit is found.
This ensures the correct propagation of uncertainties when scattering in different materials.
In the Calorimeter-Seeded Pattern Recognition for {\it{Fermi}}-LAT, this process provides a $\chi^2$ goodness-of-fit criterion for the Kalman fit, the number of hits in the tracker and the number of gaps (layers not hit or unrecognised).
A quality parameter is derived from a combination of these values, sequencing the possible candidate tracks from `best' to `worst'.
In addition, the fitted values also return an error ellipse to each photon's position in the sky.

For the higher-energy $\gamma$-rays, the shower will not be completely absorbed by the calorimeter so this missing energy must also be estimated.
For low-energy $\gamma$-rays a significant fraction of their energy can be deposited in the tracker.
In these cases the tracker is considered to be a sampling calorimeter \citep{2009ApJ...697.1071A} and, in the case of the LAT, the number of silicon strips that had hits in them is used to estimate the energy deposited therein.
In the case of EGRET, by the time the shower induced by a low-energy $\gamma$-ray reached the TASC, there may not have been sufficient energy deposit to trigger one or both of the PHAs \citep{1993ApJS...86..629T}.
These events were assigned a different class and the energy was estimated by an alternative means.
Another case in which the calorimeter cannot be used to estimate the energy of the incident $\gamma$-ray is when the direction is such that the shower misses the calorimeter.
In these cases, again, the energy deposited in the tracker can be used to provide an estimate of the energy albeit one with a larger uncertainty.
In all cases, energy loss due to leakage must be taken into account.
This leakage can occur out the sides and back of the calorimeter, in the various materials that comprise the detector volume and, in the case of LAT and AGILE, between the internal gaps in the calorimeter modules.

\subsection{Scattering Information: Gamma-Ray Polarimeters}\label{sec:polarimeters}
All basic interactions of light with matter from Sec.\,\ref{sec:cross_sections} are intrinsically sensitive to the polarisation properties of the photons.
In the case of longer wavelengths, filters can be used to distinguish between different polarisation angles $\eta$ and the polarisation degree $\Pi$.
In the case of hard X-rays to high-energy, GeV photons, such filters are impossible to construct so that the polarisation parameters of astrophysical sources are inferred from the distribution of secondary particles in the instruments.
The differential cross sections for photoelectric effect, Compton scattering, and pair production show an asymmetry with respect to the incoming photon's polarisation angle, whose amplitude is proportional to the polarisation degree (also called polarisation amplitude).
This asymmetry can be measured if position sensitive detectors for photons and resulting particles are employed.
It is important to note that the total cross sections for the three main interactions are unchanged by photon polarisation, and that with these techniques, only linear polarisation can be measured.
Conceptually, the polarisation degree can be defined as the maximum variation in the azimuthal scattering probability \citep{Novick1975_XrayPolarimetry,Lei1997_ComptonPolarimetry} as
\begin{equation}
    \Pi = \frac{d\sigma_{\perp} - d\sigma_{\parallel}}{d\sigma_{\perp} + d\sigma_{\parallel}}\mathrm{,}
    \label{eq:polarisation_degree}
\end{equation}
\noindent which is to be compared to the actual possible modulation capabilities of the instruments.
In Eq.\,(\ref{eq:polarisation_degree}), $d\sigma_{\perp}$ and $d\sigma_{\parallel}$ are the scattering cross sections for photons perpendicular and parallel to the emission plane.

Photoelectric absorption produces an electron whose angular distribution depends on the polarisation of the incident photon.
In the classical sense, the electrons accelerate in the direction of the electric field of the incident photon.
However, the exact distribution of photoelectrons also depends on the microscopic properties of the absorbing material, for example the electronic band structure in solids, which makes an exact derivation of the differential cross section as a function of polarisation angle cumbersome.
In the Born approximation \citep{Sauter1931_PE,Scofield1989_PE,Costa2001_PE,Sabbatucci2016_PE}, the differential cross section can be expressed as
\begin{eqnarray}
    \frac{d\sigma_{\rm PE}(E_{\gamma},\eta)}{d\Omega} = r_e^2 Z^5\alpha^4 \left( \frac{m_ec^2}{E_{\gamma}} \right)^{7/2} \frac{4\sqrt{2}\sin^2\theta\cos^2\eta}{(1-\beta\cos\theta)^4}
    \label{eq:photoeffect_polarisation}
\end{eqnarray}
\noindent where $E_{\gamma}$ is the total energy of incoming photon, $r_e$ is the classical electron radius, $Z$ the material's charge number, and $\alpha$ the fine-structure constant.

The initial derivation of the Compton scattering cross section already included the polarisation angle of incident photons \citep{Klein1929_comptoncs},
\begin{equation}
    \frac{d\sigma_{\rm CE}(E_{\gamma},\eta)}{d\Omega} = \frac{1}{2}r_e^2 \left( \frac{\lambda}{\lambda'} \right)^2 \left[\frac{\lambda}{\lambda'} + \frac{\lambda'}{\lambda} - 2\sin^2\theta\cos^2\eta \right]\mathrm{,}
    \label{eq:comptonscattering_polarisation}
\end{equation}
\noindent with $\lambda/\lambda' = (1 + (E_{\gamma}/(m_ec^2))(1-\cos\theta))^{-1}$.
In this case, the scattered photon obtains a preferred direction with respect to the incident polarisation angle.

The differential cross sections for pair production with polarised photons have been calculated and studied in \citep[e.g.,][]{Maximon1962_PP,Motz1969_PP,Depaola1998_PP,Bakmaev2008_PP}, among others, for numerous cases including form factors, partial screening of the nucleus charge, pair creation in the electron field, among others.
The equations resulting from these calculations are too long to be useful in this summary, however they all have one factor in common: they depend on the polarisation angle $\eta$ as $d\sigma_{\rm PP}(E_{\gamma},\eta)/d\Omega \propto A(1 + B\cos^2\eta)$.
The factor $\cos^2\eta$ appears in all the differential cross sections for the basic interactions of polarised light with matter.
Making use of the resulting azimuthal distributions of either scattered photons (Compton scattering) or produced particles (photoelectrons, pairs) will infer the polarisation of the incoming photons.

Dedicated instruments that use these techniques are, for example, IXPE \citep{Weisskopf2016_IXPE} in the photon range 2--8\,keV (photoelectric effect), and POLAR in the energy range 50--500\,keV (Compton scattering; \citep{Produit2005_POLAR}).
Other instruments can measure polarisation of low- and high-energy $\gamma$-rays, however have not been initially designed for this task.
These include INTEGRAL/SPI \citep[e.g.,][]{Kalemci2004_polarisation,Chauvin2013_polarisation} by scatterings between its detectors, {\it{CGRO}}-COMPTEL \citep[e.g.,][]{McConnell2016_COMPTEL_polarisation}, the COSI balloon \citep{Lowell2017_COSI_polarisation}, all in the photon range 0.1--10\,MeV, and {\textit{Fermi}}-LAT \citep[e.g.,][]{Giomi2017_LAT_polarisation}.

\subsection{Other Apertures: Combinations and Wave Optics}\label{sec:other_apertures}
There are more $\gamma$-ray telescope concepts, some of which have, to date, never flown, and some of which are not feasible technically without major advances in space flight, for example.
In the following, an overview of other such apertures is given.

\subsubsection{Coded-Mask Compton Telescopes}\label{sec:coded_mask_compton_telescopes}
Using the angular resolution from coded mask telescopes, Eq.\,(\ref{eq:coded_mask_resolution}), it is clear that the separation between the detector and the mask impacts the resolution as $\delta\Theta \propto l^{-1}$.
Therefore, increasing the collimator tube (anti-coincidence shield) length will result in better angular resolution, however this comes with the problem of enhanced mass, and consequently higher background and narrower field of view.
One possibility to alleviate this problem is to use a combination of a coded aperture mask with a Compton telescope:
In this way, the anti-coincidence shield does not necessarily need to fill the gap between the detectors and the mask, but only needs to cover the position sensitive detectors.
Thus, there are two fields of views, one covering a small region from the mask to the detector, and one defined by the veto shield that surrounds the detectors.
A deployable mast (see also Sec.\,\ref{sec:reflective_optics}) could place the mask several tens of meters above the camera, which narrows the field of view to $\alpha \propto 2l^{-1}$, while at the same time improving the angular resolution by a similar factor $\delta\Theta \propto l^{-1}$.
The Compton telescope part can then be used to only select photons that passed through the mask which results in considerable background rejection in addition to the veto shield.

The imager IBIS aboard INTEGRAL is composed of two layers below a coding aperture, ISGRI and PICsiT \citep{Lubinski2009_PICsITanalysis}.
This may be considered a coded-mask Compton telescope, however it only works up to $\approx 3$\,MeV because the mask becomes too transparent at higher energies.
The separation between mask and camera is 3.2\,m and the Compton telescope layers are 9\,cm apart, however everything is still shielded by the IBIS Veto system.
This results in an angular resolution of 12\,arcmin within a field of view of $9^{\circ}$.
A proposed instrument that would extend these capabilities is GECCO, the Galactic Explorer with a Coded aperture mask Compton telescope \citep{Orlando2021_GECCO}.
GECCO's mask-detector separation would be about 20\,m, resulting in a $4^{\circ}$ field of view with an angular resolution of 1\,arcmin.
The pixellated camera would be made of CZT, resulting in a spectral resolution of $\approx 1\,\%$.

\subsubsection{Reflective Optics for Gamma-Rays}\label{sec:reflective_optics}
X- and $\gamma$-rays that approach any material perpendicular to its surface will either be absorbed, undergo Compton scattering, or produce pairs, so that refraction of high-energy photons onto a focal plane -- the typical case for optical photons -- is difficult.
The refractive index for most materials in X- and $\gamma$-ray is close to $1.0$ (or smaller) so that refractive optics (classical lenses) cannot be used for high-energy photons (see, however, Sec.\,\ref{sec:diffractive_optics}).
Therefore, the only way to `focus' high-energy photons is to use grazing incident optics that rely on reflection off mirrors in an X-ray Wolter-type telescope \citep{Wolter1952_WolterTelescope,Wolter1952_WolterTelescope2}, for example.
For incident angles smaller than some critical value that depends on the photon energy and refraction index of the material, the photons undergo total reflection and thereby avoid photoelectric absorption.
In general, the critical angle is proportional to $\sqrt{\rho}/E_{\gamma}$, where $\rho$ is the density of the material.
Thus, for a fixed incidence angle, photons can only be totally reflected up to a cutoff energy that is related to the K-edge of the reflecting material.
For photons above 10\,keV, the critical angles approach values that are too small to be used in Wolter telescopes unless the reflective coatings are very thin or the focal length very long:
For example, platinum at an incidence angle of $0.07^{\circ}$ shows a reflectivity of more than 90\,\% up to 68\,keV, and sharply drops below 20\,\% at higher energies.

Wolter telescopes have an effective area that is approximately
\begin{equation}
    A_{\rm eff}(E_{\gamma}) \approx 8 \pi f L \theta^2 R^2(E_{\gamma})\mathrm{,}
    \label{eq:wolter_aeff}
\end{equation}
\noindent where $f$ is the instrument's focal length, $L$ is the mirror length, $\theta$ is the incidence angle, and $R(E_{\gamma})$ is the reflectivity as a function of photon energy $E_{\gamma}$ \citep{Pareschi2021_WolterIOptics}.
Clearly, for the highest possible effective area, the focal length should be maximised as the material parameters $R(E)$ and $L$ are naturally limited.

One instrument in which this focal length maximisation, together with Pt/C multilayer coatings to effectively reflect photons below the Pt K-absorption edge at 78.4\,keV, is employed is {\it{NuSTAR}} \citep{Harrison2013_NuSTAR}.
{\it{NuSTAR}} is the first focusing hard X-ray telescope ever launched into orbit.
It employs 133 nested grazing-incidence shells in a conical approximation to a Wolter telescope to focus photons onto a focal plane made of a pixellated CZT detector.
{\it{NuSTAR}}'s focal length of 10\,m is achieved by a deployable mast that was extended after the satellite was launched into its orbit

Because reflective optics are ultimately limited by Eq.\,(\ref{eq:wolter_aeff}), i.e. focusing higher photon energies would require smaller incidence angles ($\propto \theta^2$) and a much larger focal lengths ($\propto f$) to accommodate a large effective area, these types of apertures are probably not suited beyond photon energies of $\sim 200$\,keV.
A proposed instrument that is based on reflective optics is PheniX \citep{Roques2021_PheniX}, which would have a focal length of 40\,m, also obtained by an extendable mast.

\subsubsection{Diffractive Optics}\label{sec:diffractive_optics}
Beyond total reflection, there are the possibilities for refraction and diffraction of $\gamma$-ray photons.
Yang (1993, \citep{Yang1993_FresnelLenses}) discussed refractive optics for photon energies up to 1\,MeV, however found that absorption and scattering severely limits the ability to form an efficient telescope with large effective area.
Therefore, only diffraction permits improvement upon the classic non-focusing $\gamma$-ray instruments.
In general, the diffraction limit defines the spatial resolution $s_d = 1.22 \lambda f / d$, with $f$ being the focal length, $\lambda$ the photon wavelength, and $d$ lens diameter, and provides the fundamental limit to the achievable angular resolution, $\theta_d = s_d / f$.
Thus, for photon energies around 1\,MeV ($\lambda \approx 1.24$\,pm), the angular resolution approaches microarcseconds ($\mathrm{\mu ''}$).
This is possible by the use of Phase Fresnel Lenses \citep{Miyamoto1961_PhaseFresnelLense} in which concentric rings of precisely placed crystals diffract the high-energy photons onto a detector at the focal point.
The distance of the focal point where the $\gamma$-ray detector is placed, however, is related to the lens diameter $d$, the photon energy $E$, and the pitch size of the Fresnel lens $p$ by
\begin{equation}
    f = 0.4 \times 10^{6} \left(\frac{p}{\mathrm{1\,\mrm{mm}}}\right) \left(\frac{d}{\mathrm{1\,m}}\right) \left(\frac{E}{\mathrm{1\,MeV}}\right)\,\mathrm{km,}
    \label{eq:focal_length_fresnel}
\end{equation}
which makes the realisation of a telescope on a single spacecraft almost impossible \citep{Skinner2001_GammaRayLenses}.
It should be noted that Fresnel lenses suffer severely from chromatic abberation, effectively worsening the achievable angular resolution by factors of a few, and distorting the received spectrum.

Nevertheless, concepts to build $\gamma$-ray-focusing (concentrating) telescopes exist and are discussed in this book.
An important part of these concepts is the formation flight of two or more spacecrafts in sync.
Another possibility to achieve a much enhanced sensitivity would also to use a deployable boom, such as used in the ASTENA proposal \citep{Frontera2019_ASTENA,Frontera2021_ASTENA511}.
Here, the goal is not to approach the diffraction limit, but to construct a design feasible with current technologies.
With a 20\,m focal length, an angular resolution of 30\,arcmin could be achieved in the bandpass between 50 and 600\,keV, reaching a continuum sensitivity of more than two orders of magnitude better than INTEGRAL/SPI thanks to ASTENA's effective area of more than $7\,\mrm{m^2}$ \citep{Frontera2019_ASTENA}.

\subsubsection{Interplanetary Network}\label{sec:IPN}
In fact, several spacecrafts are already used in combination in the so-called `Interplanetary Network' \citep[IPN; e.g.][]{Hurley2009_IPN}.
While the spacecrafts are not flying in formation, their absolute distances to each other can be used for triangulation of celestial burst-like signals, such as GRBs or soft gamma repeaters.
In particular the localisation is performed by a comparison of the arrival times from the different $\gamma$-ray instruments wherein the precision is given by the distances of the spacecrafts and the absolute number of detected photons.
The further the instruments are separated, i.e. the larger the baseline of potentially several hundred millions of kilometers, the more accurate the localisation will be.

The triangulation technique is explained in \citep{Hurley2013_IPN}, for example, and depicted here briefly:
A transient event is measured with a delay time $\delta T$ at two different spacecrafts.
Given the separation $D$ of the spacecrafts, the transient is localised onto an annulus on the celestial sphere with half-angle $\Theta$ as
\begin{equation}
    \cos(\Theta) = \frac{c \delta T}{D}\mrm{,}
    \label{eq:IPN_equation}
\end{equation}
with $c$ being the speed of light.
The `error box' or width of the annulus is provided by the uncertainty of the time delay as $\sigma_{\Theta} = c \sigma_{\delta T} / (D\sin(\Theta))$.

Burgess et al. (2021, \citep{Burgess2021_nazgul}) showed that this classical triangulation method has weaknesses because the choice of uncertainties may be ill-defined.
The authors developed a novel method that can robustly estimate the position of a transient via a hierarchical Bayesian model.
In particular, they forward-fold the unknown temporal signal evolution, described by random Fourier features, and fit this model to the time series data of each instrument.
This takes into account the appropriate Poisson likelihood and consequently the uncertainties generated by the method are more robust and in many cases more precise compared to the classical method.

The IPN started in 1977; its third version, IPN3, was operating with Ulysses, {\it{CGRO}}, Pioneers Venus Orbiter, Mars Observer, and {\it{Beppo}}SAX.
Currently, the IPN localisations come from Konus-{\it{Wind}}, Mars Odyssey, INTEGRAL, {\it{Swift}}, AGILE, BepiColombo, and {\it{Fermi}}.
In total, more than 32 spacecrafts have been involved in the IPN so far.

\subsection{Gamma-Ray Detectors}\label{sec:detectors}
Most $\gamma$-ray emission processes are continuum-like. 
Instrumental resolution is, therefore, not too important except for when one wants to do line spectroscopy.
Since lines only appear up to the MeV range ($\lesssim 20$\,MeV), spectral resolution is less of an instrument design driving factor above $\sim 20$\,MeV. 
When choosing the materials that compose the target for the incident $\gamma$-ray signal, a tradeoff between instrumental resolution, weight, sensitivity and power is always at play.

The materials that are used to detect the by products (charged particles) of the incident $\gamma$-ray at these energies broadly comprise two main categories, solid state detectors and scintillators.
Solid state detectors are discussed in detail in several Chapters of this book. 
They are lightweight, compact and more tolerant to a space environment than vacuum or gas-based detectors. 
In general, they comprise a semi-conductor material (such a silicon, germanium or CZT) which is reverse-biased so that the electrons and holes can move freely in the so-called depletion region. 
When a charged particle (electron or positron) enters this sensitive area of the crystal, ionisation is produced. 
This signal is then transferred via a charge-sensitive preamplifier where it is converted to a voltage pulse proportional to the strength of the ionisation signal.
For example, the spectrometer SPI uses an array of 19 high purity cooled geranium detectors to perform high-resolution spectral measurements between 18\,keV and 8\,MeV.

\begin{figure}[b]
    \centering
    \includegraphics[width=0.6\textwidth,trim=0in 0in 0in 0in,clip=True]{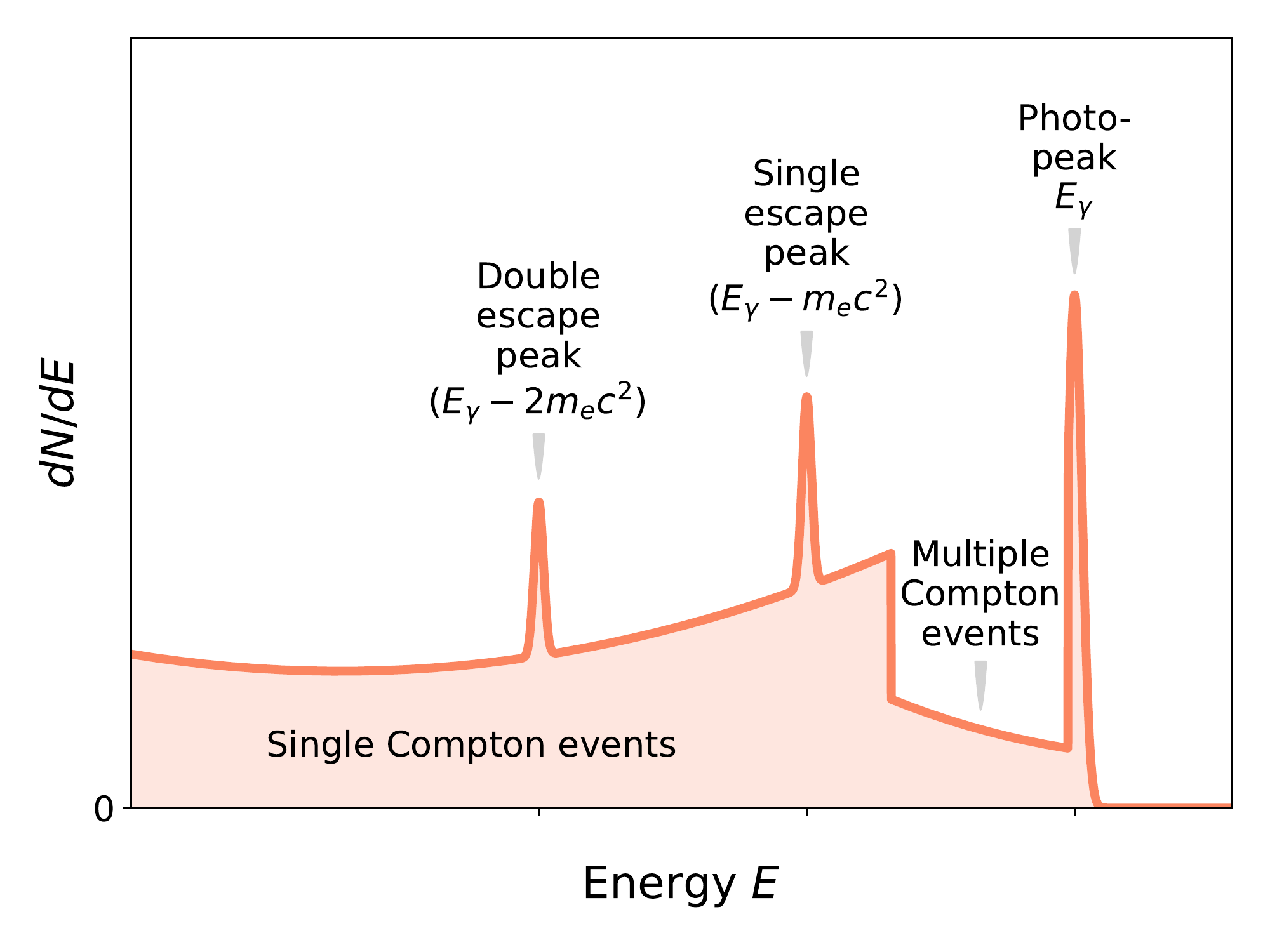}
    \caption{General MeV $\gamma$-ray measurement features as observed in detectors. A beam of photons with energies $E_{\gamma}$ create a photo-peak at this energy. Compton scattering inside the detector leads to a characteristic continuum below the photo-peak. If the incoming photon energy is greater than $2m_ec^2 = 1.022$\,MeV, single and double escape peaks can emerge.}
    \label{fig:gamma-ray_measurements}
\end{figure}

Silicon-strip detectors are used to detect the passage of the electron-positron pairs for both the LAT on {\textit{Fermi}} \citep{2009ApJ...697.1071A} and for the GRID on AGILE \citep{2006NIMPA.556..228P}.
Scintillation detectors comprise a material that produces light when it is traversed by a charged particle. 
The scintillation light is recorded by a photodetector (often a photomultiplier tube or photodiode) so that the passage of the charged particle and, in certain cases, its energy can be measured. 
Scintillator materials can be organic or inorganic in nature. 
Inorganic scintillators, including NaI and BGO, are usually chosen for calorimeter systems due to their high density and effective atomic number which means that they have a high stopping power. 
Inorganic scintillators comprise four main categories: plastic, glass, single crystal and liquid. 
Plastic scintillators are those most commonly used in $\gamma$-ray applications due to their light weight, low cost and robustness. 
The light yield from inorganic scintillators is typically higher than those from organic scintillators so that it is typically used as a material for veto shields.

\subsection{Understanding Gamma-Ray Measurements}\label{sec:measurements}
The nature of $\gamma$-ray measurements can be understood as the recording of photons into complex data spaces due to the vastly different apertures.
These data spaces are typically shown in the form of back-projections, such as the shadow pattern of coded-mask instruments or the rings from Compton telescopes.
In the case of GeV instruments, this `imaging' is only weakly influenced by dispersion which is why we focus more on the MeV instruments here.
Nevertheless, the concepts here apply to all dispersed measurements.

\begin{figure}[b]
    \centering
    \includegraphics[width=0.24\textwidth]{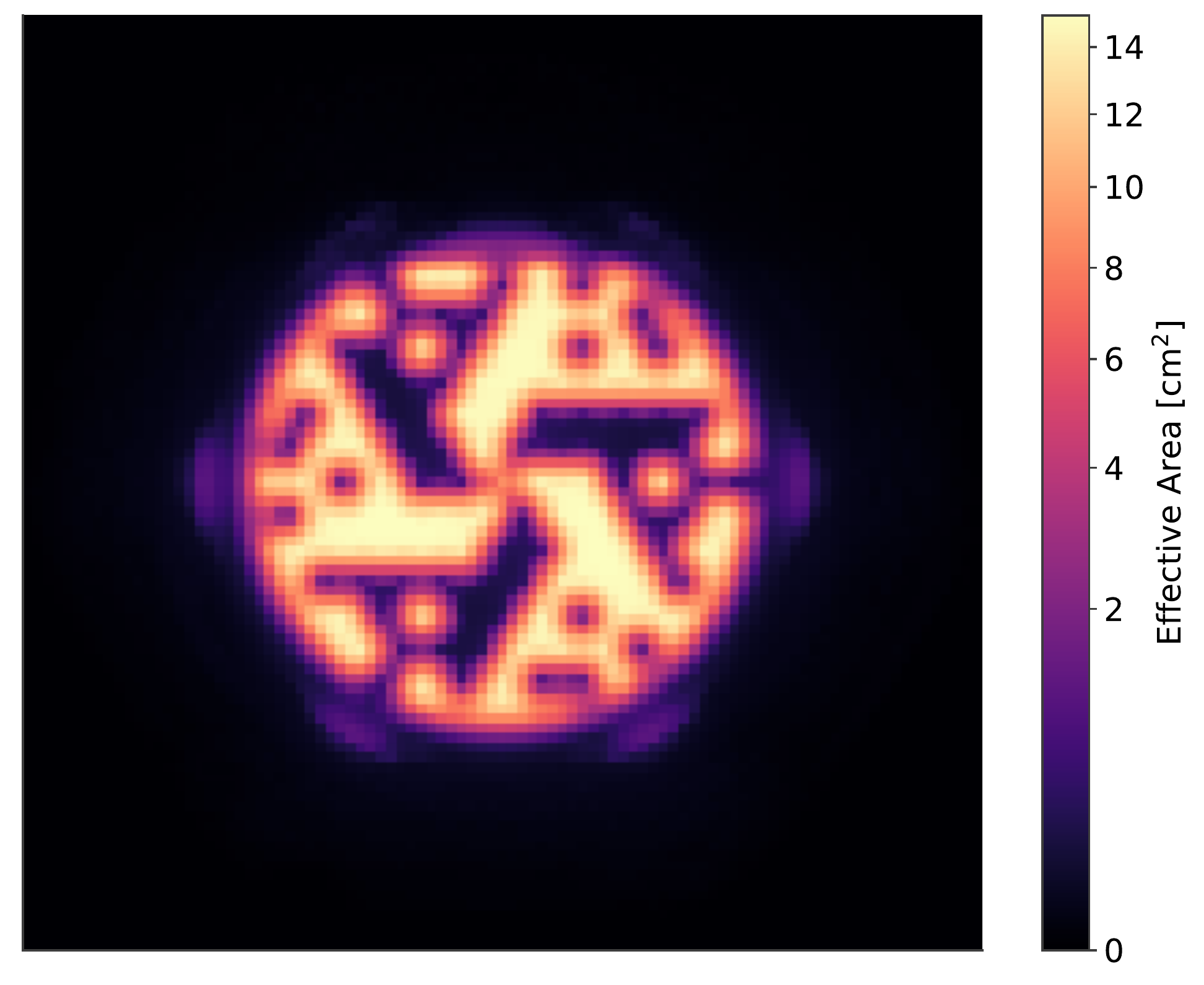}
    \includegraphics[width=0.24\textwidth]{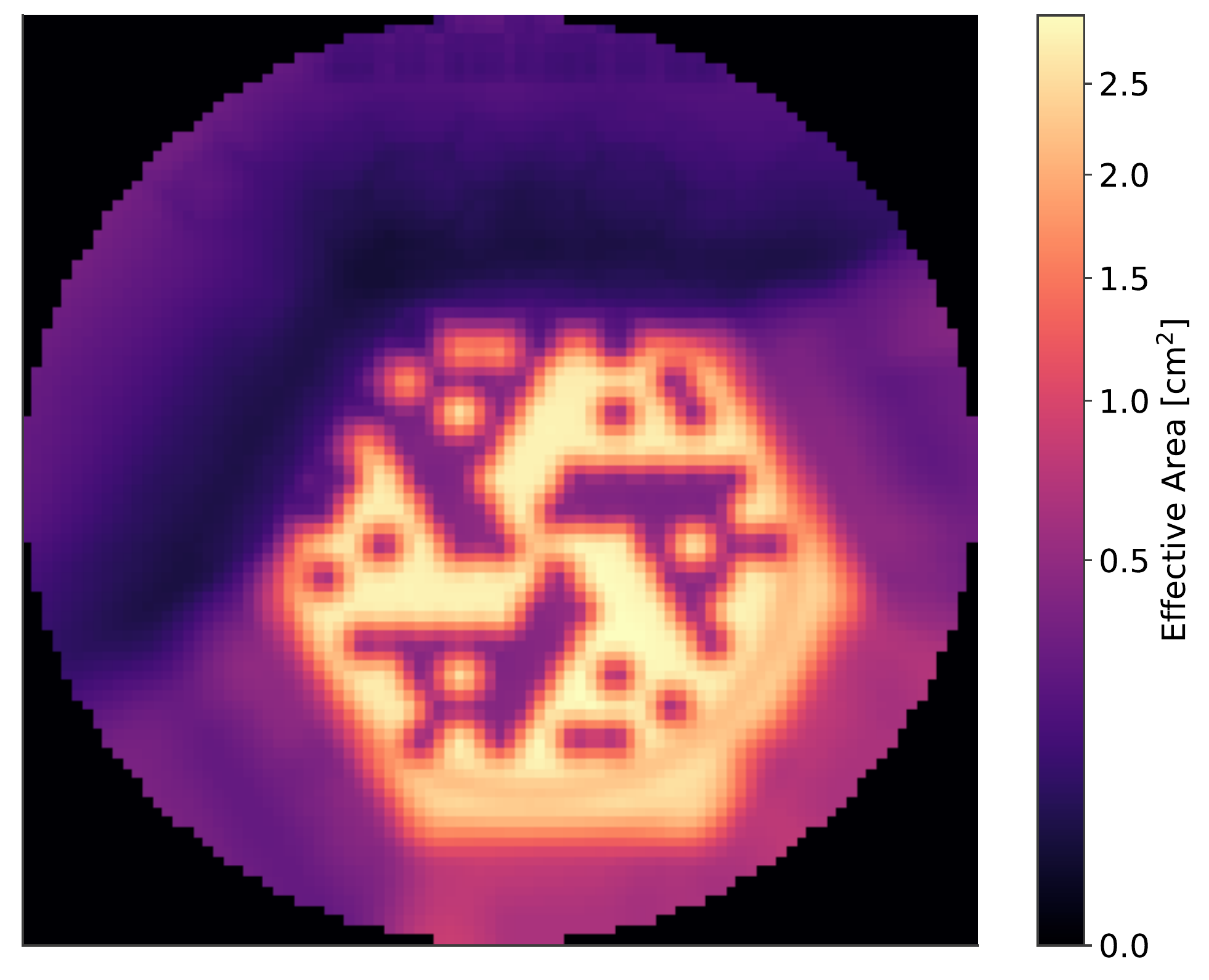}
    \includegraphics[width=0.24\textwidth]{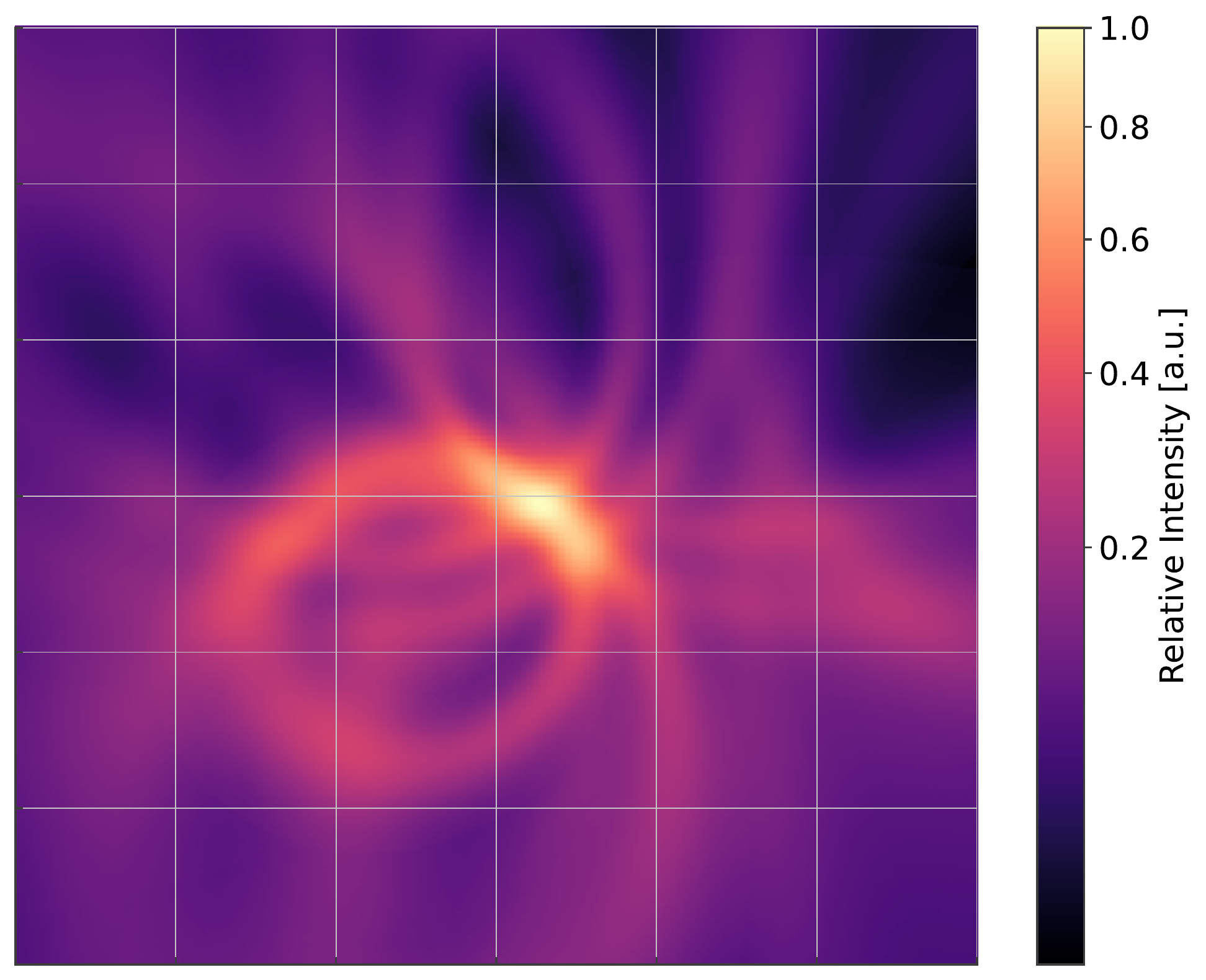}
    \includegraphics[width=0.24\textwidth]{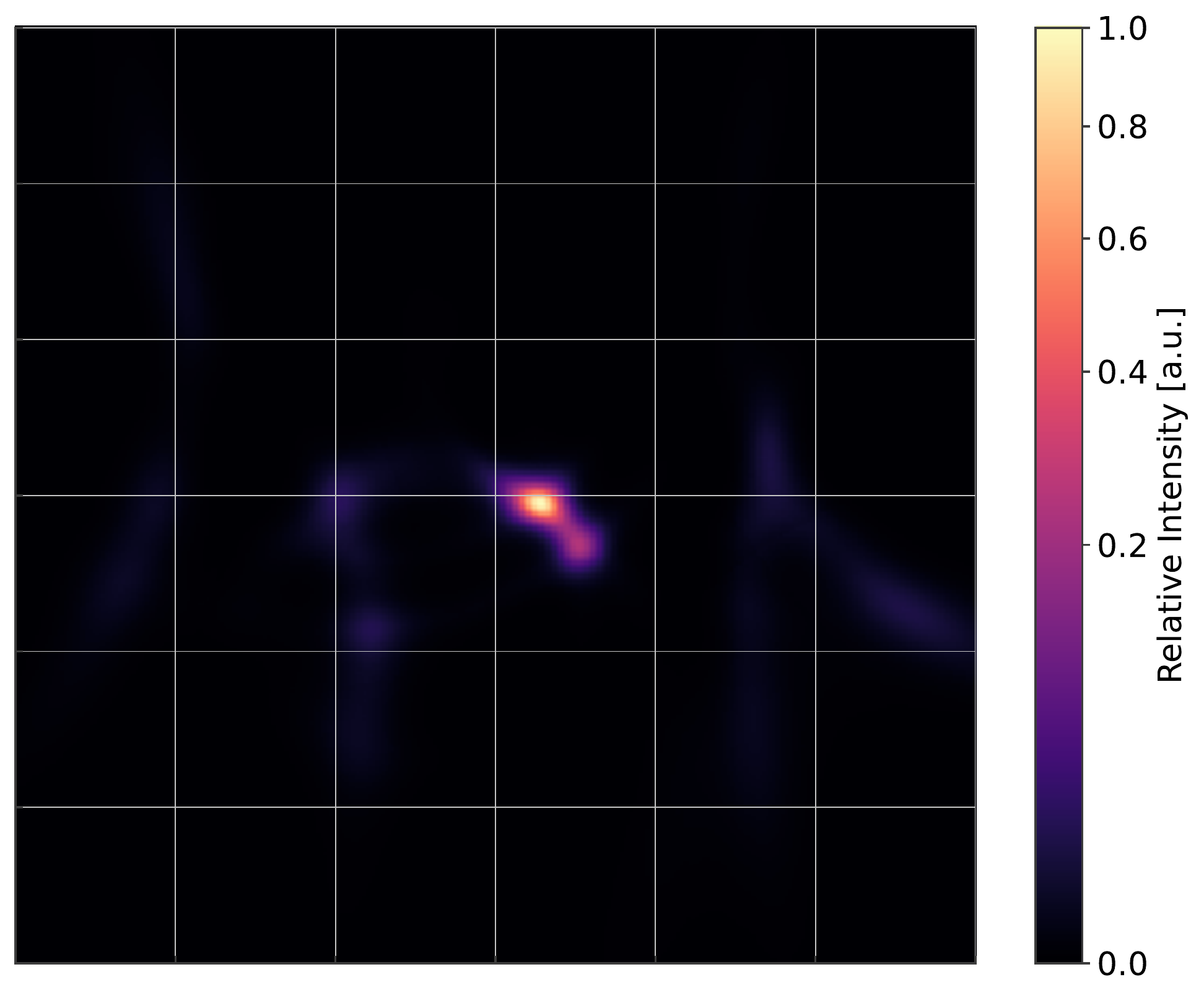}
    \caption{Visualisation of MeV telescope responses. \textbf{\textit{Left}}: Back-projection of the SPI coded-mask response see from its central detector (unit 00) at an energy of 66\,keV. The same mask pattern is visible for an outer detector (\textbf{\textit{second image}}; unit 12; which is equivalent to shifting the source position) at an energy of 5\,MeV, however the edges are not as sharp because the transparency of the material increases with photon energy. The inner edge of SPI's anti-coincidence shield is visible (hexagonal shape outside the central ring). \textbf{\textit{Right}}: Back-projection of Compton circles from the COSI balloon response. Shown are 20 Compton circles at an energy of 511\,keV, overlapping at the source position. An iterative deconvolution algorithm (\textbf{\textit{fourth image}}) is applied to reduce improbable regions so that a point source emerges.}
    \label{fig:back_projection_response}
\end{figure}

\noindent The abstract data spaces of $\gamma$-ray instruments are spanned by reconstructed or inferred variables, such as the three scattering angles in Compton telescopes or the number of pixels (detectors) in coded mask telescopes.
Typically, these data spaces are not necessarily the real instrument spaces because they can have considerable uncertainties which are often omitted in the subsequent data analysis steps, or because data filtering due to quality criteria after reconstruction skewed the true generating process: counting photons.
The real data space of each instrument goes down to the level of its electronics and the specific geometry -- a coding mask, shadow patterns, Compton cones, iterative deconvolutions, polarigrams, scattering angle distributions, etc., and is always an abstraction of one possible way to visualise the raw data.
In the reconstructed variables or data spaces, data analysis should be treated with care because there are often (hidden) assumptions which destroy the character of the measurement, change its likelihood, or are ill-defined because the instruments suffer from dispersion.
%
%

Instead, the method of forward-folding should be used to convolve models with physical units into the raw data space of channel number per detector unit.
Forward modelling is the only statistically proper way to analyse $\gamma$-ray data $\bf d$, which, without loss of generality, can be described as matrix equation
\begin{equation}
    \bf d = \bf R \cdot \bf m\mathrm{,}
    \label{eq:data_generation}
\end{equation}
with $\bf R$ being the response matrix, and $\bf m$ an (unknown) model that is to be inferred.
Except for approximate cases (for example mask coding with fully transparent and opaque elements), $\bf R$ is not invertible which means that a solution shaped like $\bf m = \bf R^{-1} \bf d$ does not exist.
The model $\bf m$ is not measured, only the data $\bf d$ are which means in turn that $\bf m$ must be assumed, i.e. modelled to explain the data.
This is done by parametrising the model with a set of variables, so-called fit-parameters $\bf \varphi$ so that the model becomes a function of unknown parameters: $\bf m(\bf \varphi)$.

\begin{figure}[b]
    \centering
    \includegraphics[width=0.32\textwidth]{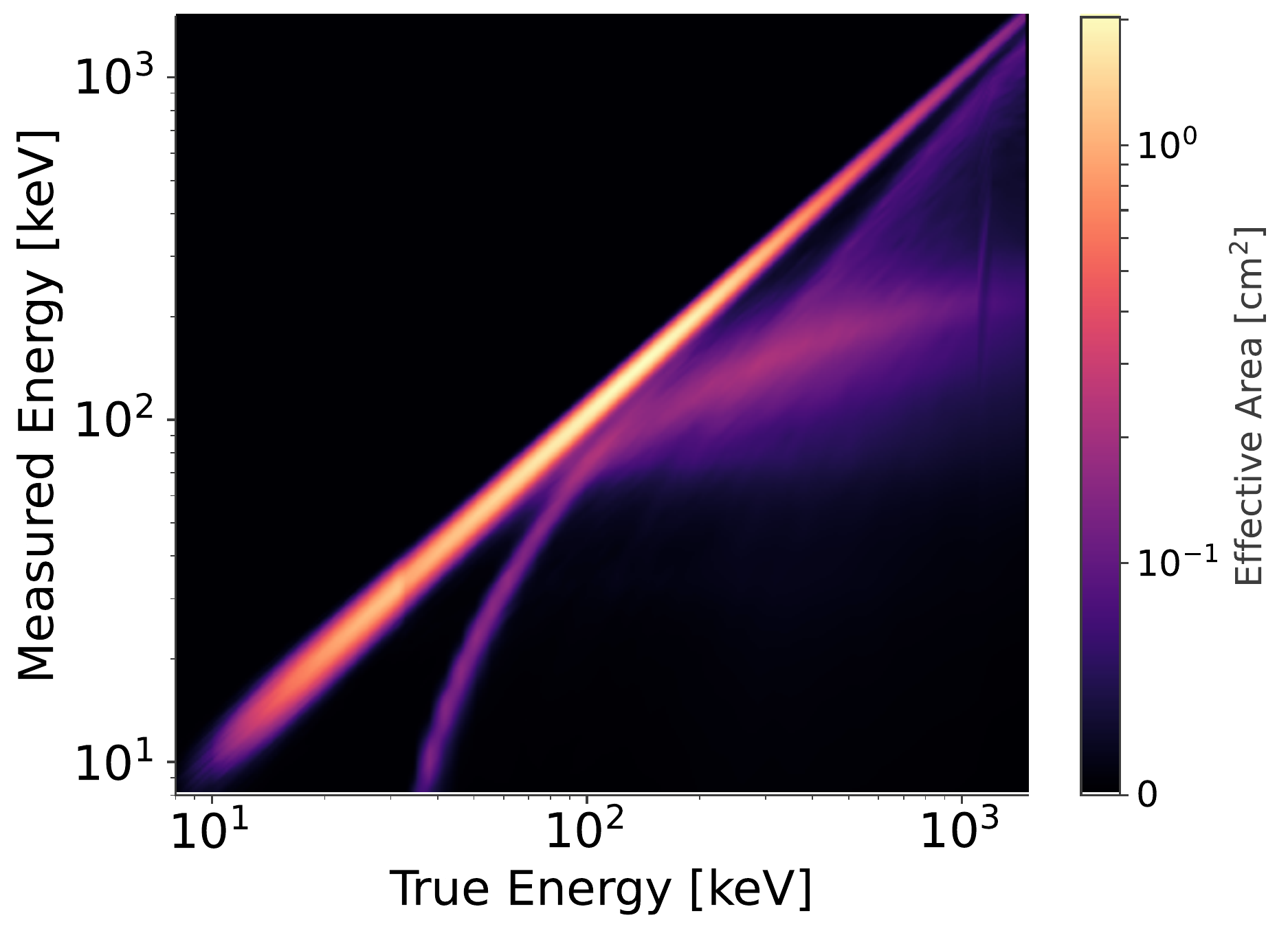}
    \includegraphics[width=0.32\textwidth]{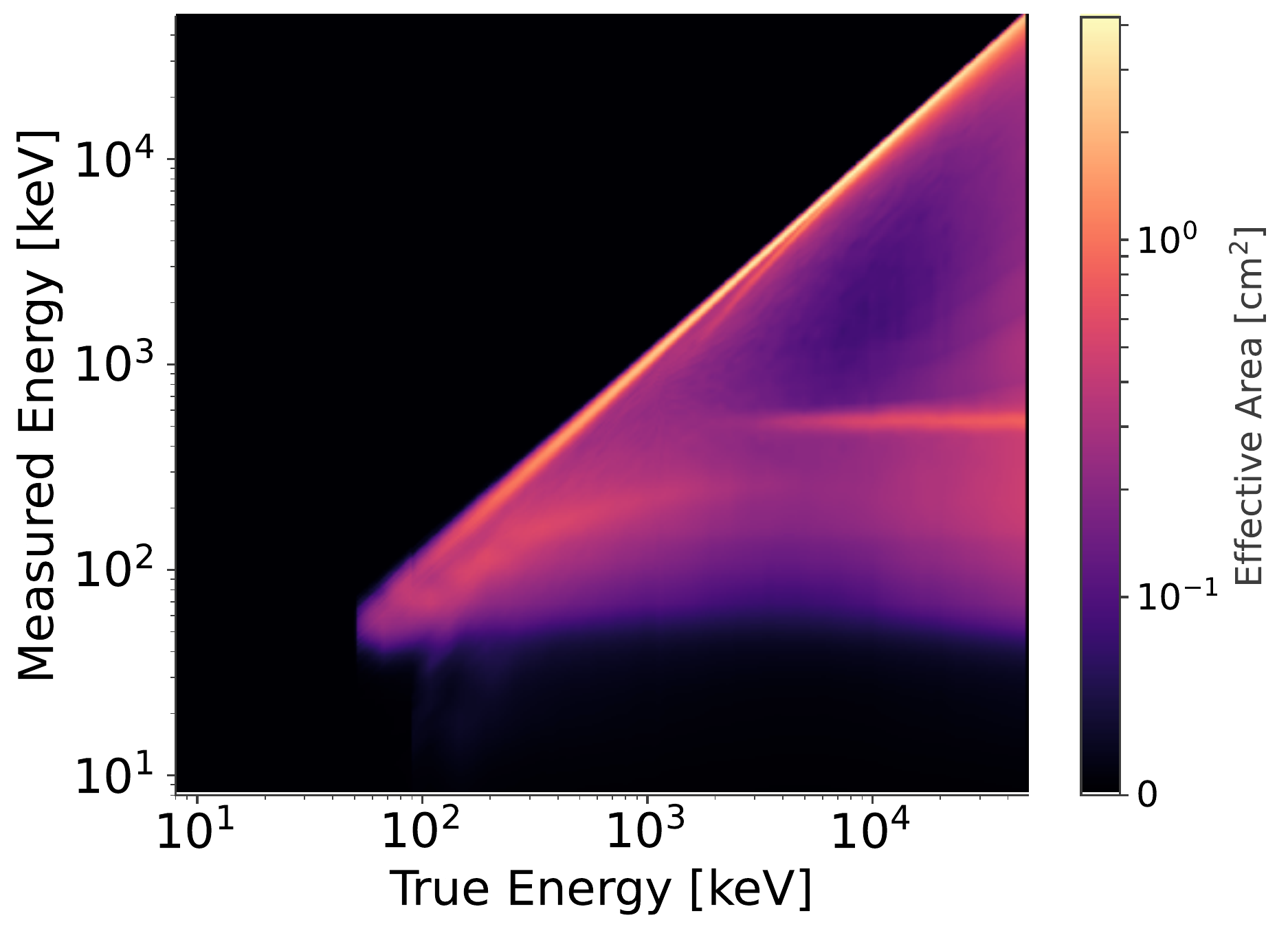}
    \includegraphics[width=0.32\textwidth]{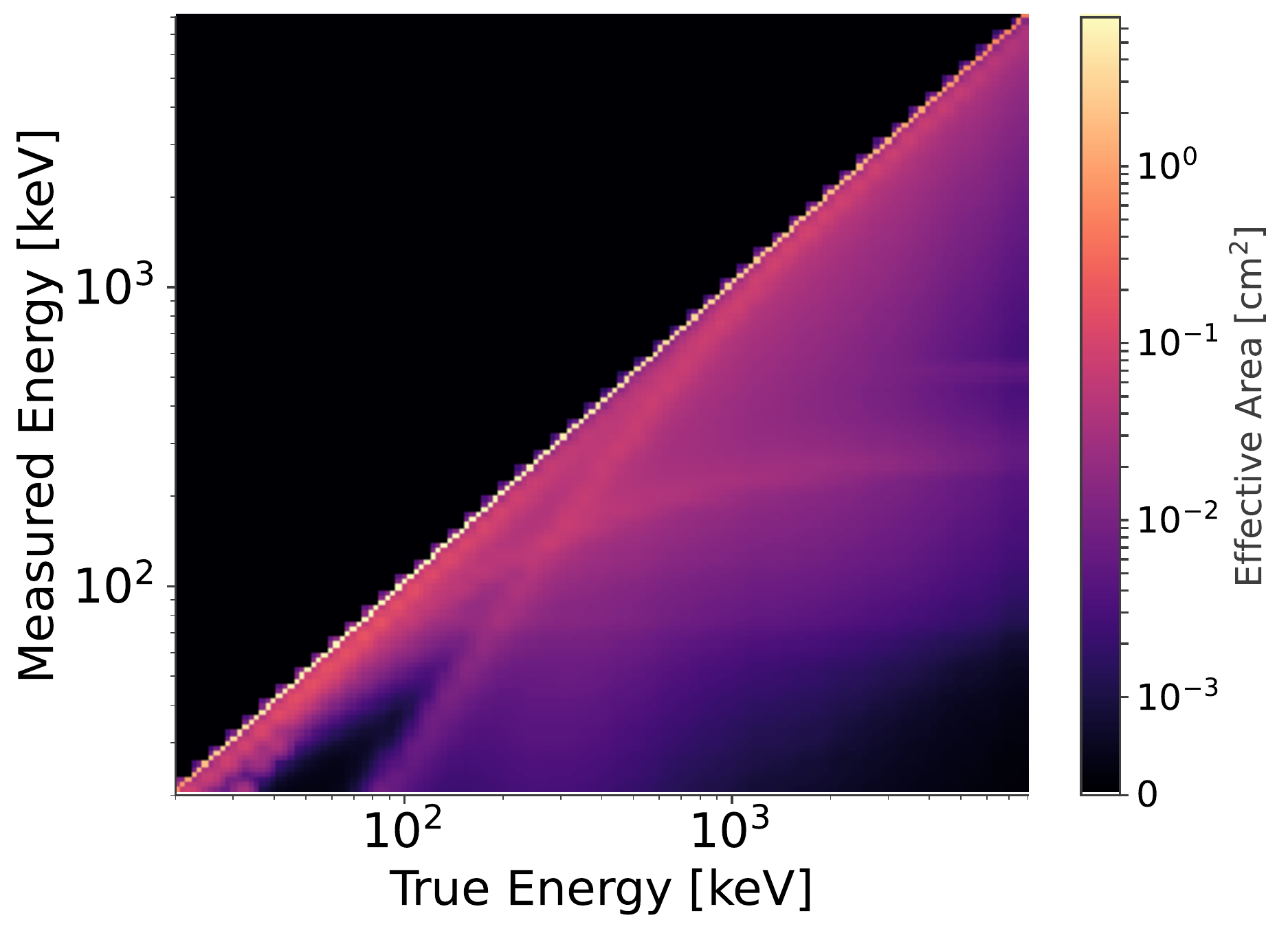}
    \caption{Energy dispersion of different $\gamma$-ray detectors in units of $\mathrm{cm^2}$. Given a photon with true (initial) energy $E_i$, there is a probability that the photon is measured at a lower (final) energy $E_f$. The diagonal represents the photopeak efficiency. \textit{\textbf{Left}}: NaI detector onboard GBM. \textbf{\textit{Middle}}: BGO detector from GBM. \textbf{\textit{Right}}: SPI germanium detector. The scale of the colour bar is enhanced by two orders of magnitude for SPI to indicate the features away from the diagonal.}
    \label{fig:energy_dispersion}
\end{figure}

The model includes everything that is required to describe the astrophysical source of interest.
This means it includes a spectral shape, temporal variability, spatial extent, and polarisation parameters, among others.
These properties can be interdependent, which can be incorporated in the response function $\bf R$ (Fig.\,\ref{fig:back_projection_response}).
The response is described in units of $\mathrm{cm^2}$ and is equivalent to the effective area, i.e. it changes as a function of zenith, azimuth, initial energy, polarisation angle, and instrumental environment parameters (temperature, voltage, etc.).
Typical features visible in spectral dispersion matrices are the Compton edges (at measured energies $E_f = E_i(1 - (1+2E_i/(m_ec^2))^{-1})$, approaching $E_i - 0.25$\,MeV for large $E_i$, Compton continuum (single and multiple scatters), first and second escape peaks (Fig.\,\ref{fig:gamma-ray_measurements}), and the emergence of a 511\,keV line for initial energies above 1.022\,MeV.

\begin{figure}[t]
    \centering
    \includegraphics[width=\textwidth]{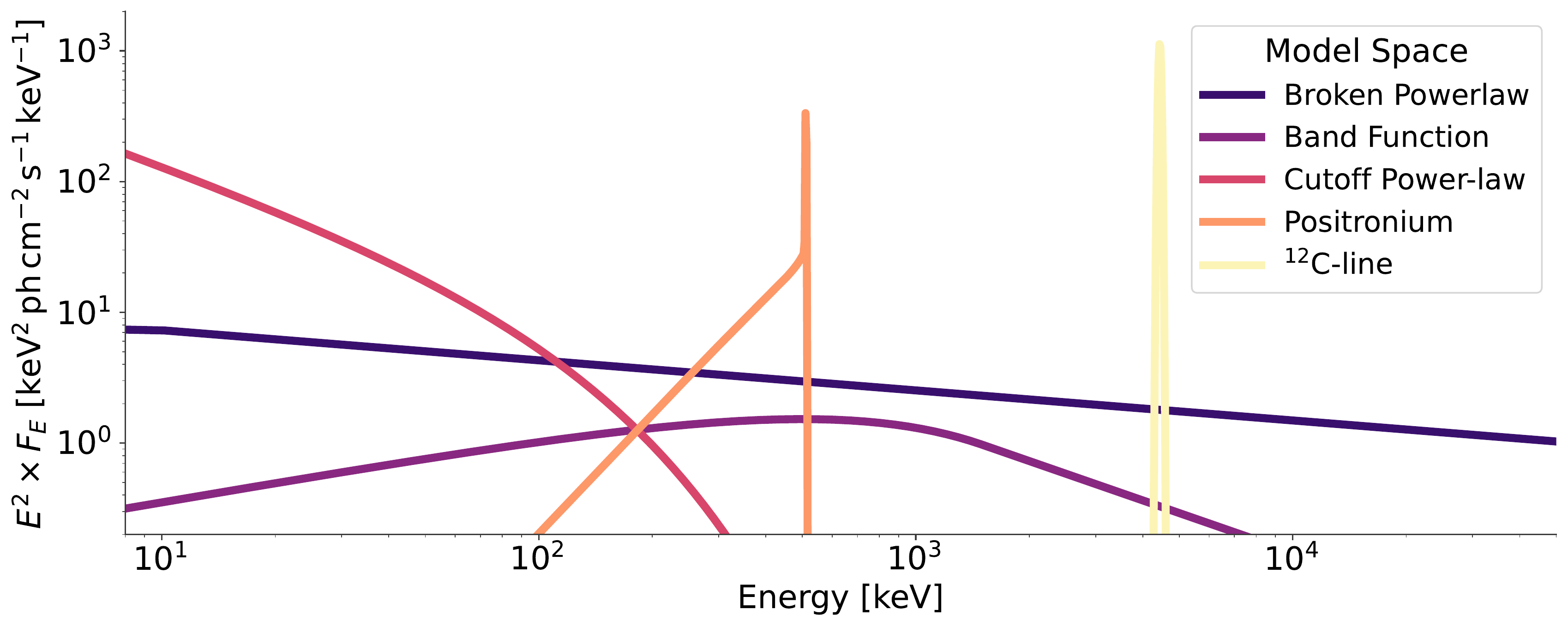}\\
    \includegraphics[width=0.32\textwidth]{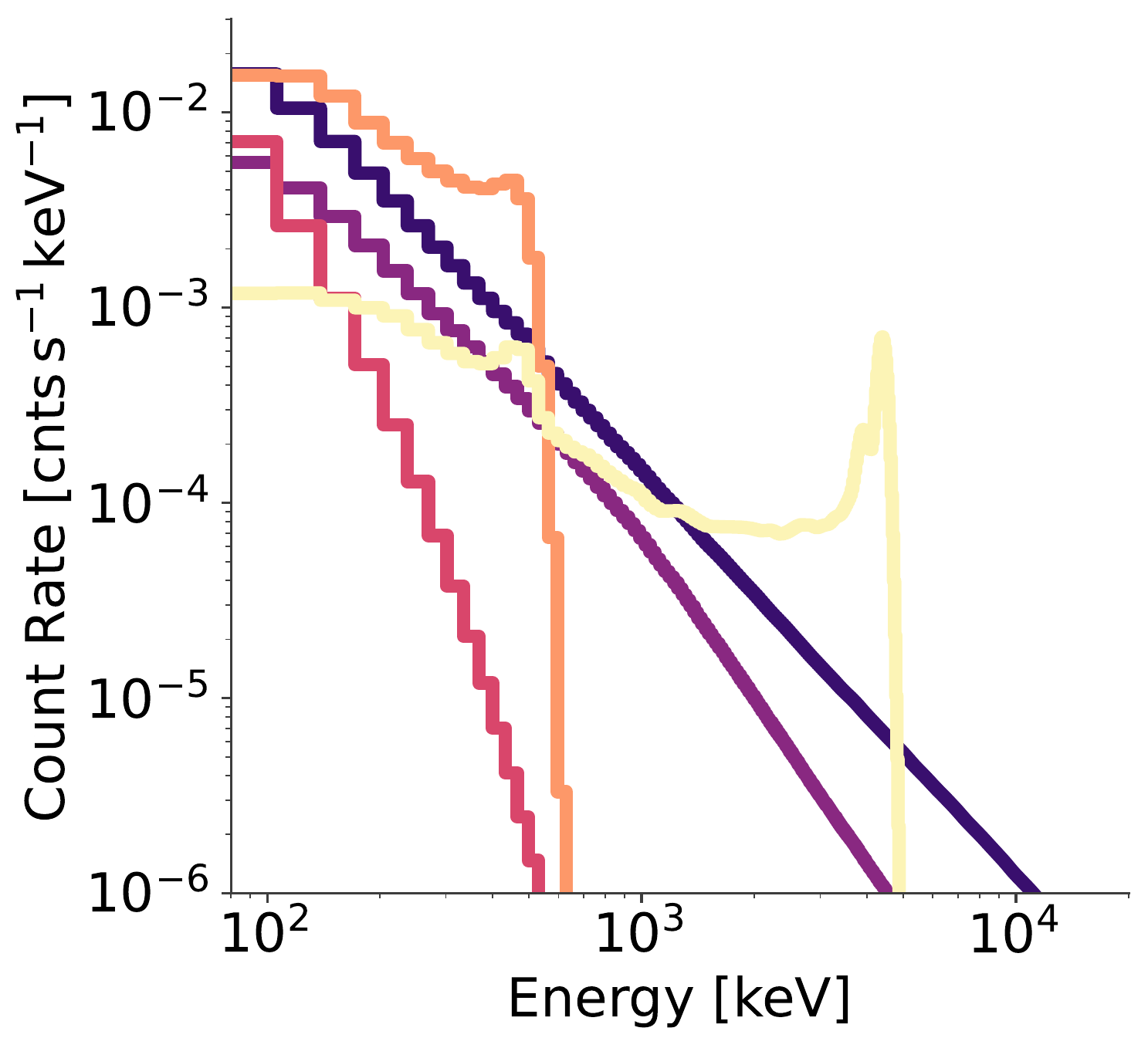}
    \includegraphics[width=0.32\textwidth]{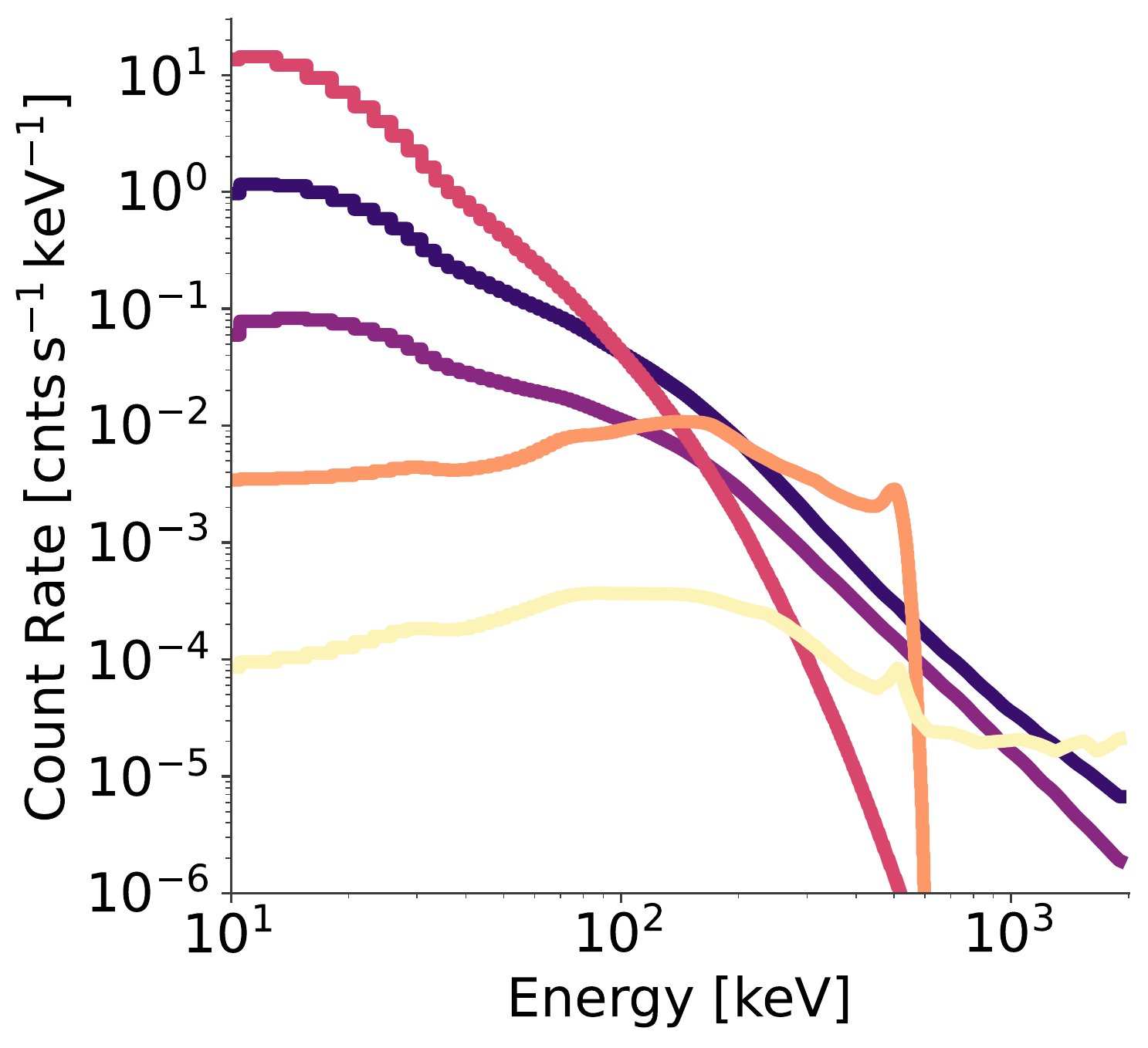}
    \includegraphics[width=0.32\textwidth]{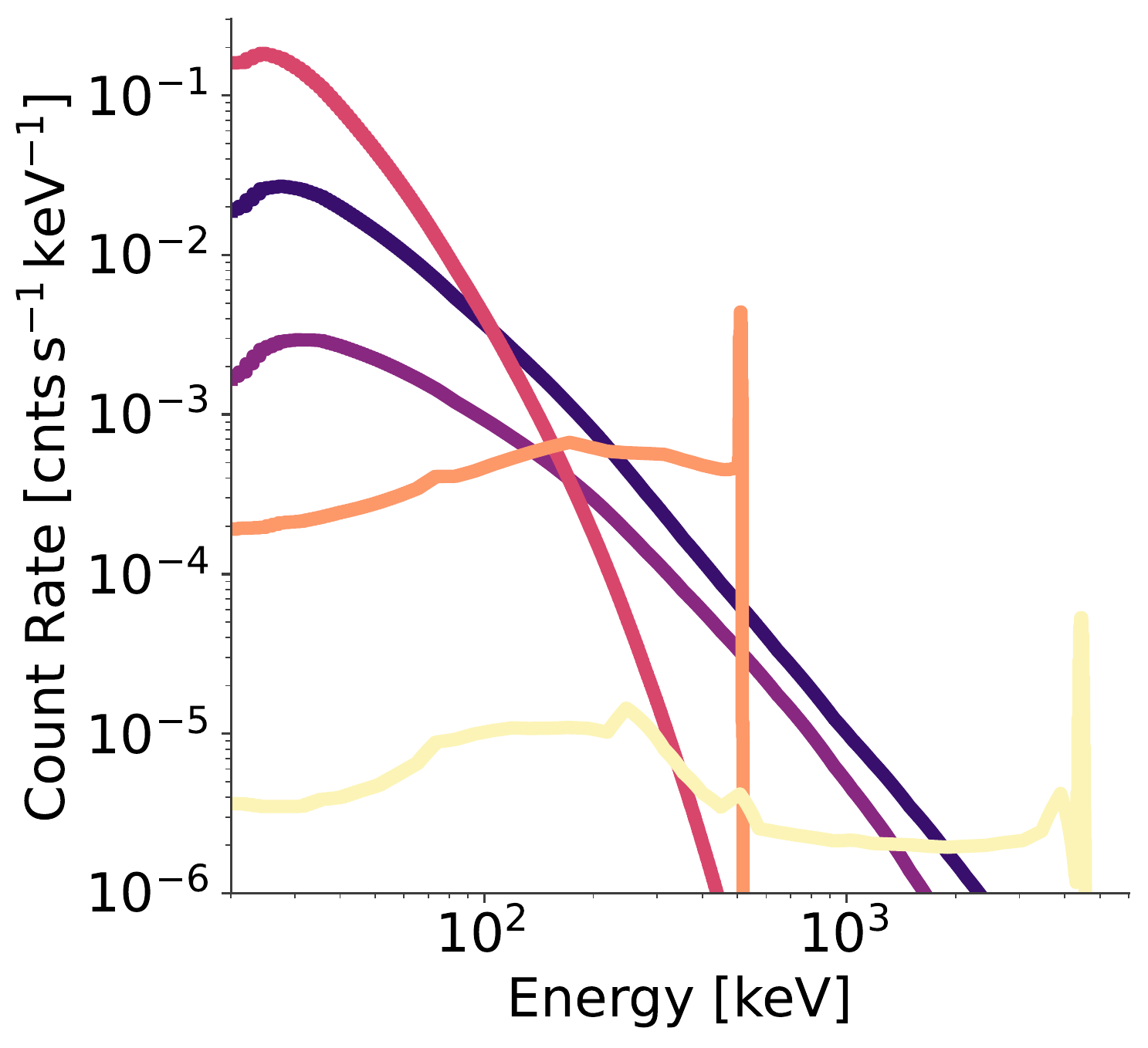}
    \caption{Forward folding of spectral models into the count data space of different detectors. \textbf{\textit{Top}}: Selection of five spectral models. A broken power-law representing the Crab spectrum \citep{Jourdain2009_Crab}, a Band function \citep{Band1993_BandFunction} with 10\,\% of the Crab flux, $\alpha=-1.5$, $\beta=-3.0$, and $E_{\rm peak} = 1.5$\,MeV, a cutoff power-law with 10 times the Crab flux, $\alpha = -3.0$, $E_C = 100$\,keV, a Positronium (ortho+para, \citep{Ore1949_511}) spectrum with a line flux of $6 \times 10^{-3}\,\mathrm{ph\,cm^{-2}\,s^{-1}}$, and a 100\,keV broadened \nuc{C}{12} line with the same line flux. \textbf{\textit{Bottom}}: The models convolved with the energy responses from Fig.\,\ref{fig:energy_dispersion} for BGO (\textbf{\textit{left}}), NaI (\textbf{\textit{middle}}), and Ge (\textbf{\textit{right}}). The resolution of the detectors is increasing from left to right. It is apparent that different spectral models (e.g., the broken power-law and the Band function) can appear very similar in the data space. Likewise, components above the maximum considered energy of the instrument contribute to the counts at lower energies because of dispersion.}
    \label{fig:model_to_data_space}
\end{figure}
Conceptually, the data are generated by assuming the position of a source (in astronomical coordinates) which converts to instrument coordinates (zenith/azimuth) for which an instrumental response is created.
A differential spectrum (in physical units, for example, $\mathrm{ph\,cm^{-2}\,s^{-1}\,keV^{-1}}$) is integrated over the energy range to obtain the expected flux per response element ($\rightarrow \mathrm{ph\,cm^{-2}\,s^{-1}}$), to which the response function is applied, resulting in an expected rate per data space bin ($\rightarrow \mathrm{cnts\,s^{-1}}$).
Note the change of notation here from physical photons to received number of counts after the application of the response.
Given the exposure time, the model counts per data space bin $i$ ($\rightarrow \mathrm{cnts}$) can be compared to the data via the Poisson likelihood
\begin{equation}
    \mathcal{L}(\mathbf{d} | \mathbf{m}(\mathbf{\varphi})) = \prod_{i} \frac{m_i(\mathbf{\varphi})^{d_i}\exp(-m_i(\mathbf{\varphi}))}{d_i!}\mathrm{.}
    \label{eq:Poisson_likelihood}
\end{equation}

\noindent Finally, the once smooth (infinitesimal) model is transferred into the (possibly) binned data space.
Examples of how such a convolution appears are shown in Fig.\,\ref{fig:model_to_data_space}.
Here, the same five models are folded through the responses of one of NaI and BGO detectors on GMB, and through the response of the central detector of SPI.
Depending on the orientation of the detector with respect to the source in the sky (assumed to be identical here), the response changes (Fig.\,\ref{fig:energy_dispersion}) as the incoming photons are dispersed in different ways in the same detector.
It is also seen that different spectral resolutions impact the identification capabilities of different features -- for example lines can be seen more clearly with higher resolution.
However also high-resolution germanium detectors suffer from dispersion, distributing more than 50\,\% of photons with an initial energy of 4.4\,MeV to lower energies.

\subsection{Simulations}\label{sec:simulations}
The above-described response functions, and therefore the entire data analysis and scientific output, rely heavily on simulations.
In this Handbook, an entire chapter is dedicated to the procedures, requirements, and details of particle and photon simulations.
This also includes details about electronics simulations, detector effects, and methods to handle these in real measurements.
Here, a short overview of the basic features and general idea is given to understand the need for simulations as well as a few examples and caveats.

Before instrument prototypes are built, they are often simulated in the Geometry and Tracking (GEANT, \citep[e.g.,][]{Agostinelli2003_GEANT4}) environment.
In GEANT, active and passive volume elements are arranged as similarly as possible to the real geometry of the instrument, and irradiated with particles and photons through a Monte Carlo technique.
This allows the instrument designer to 1) estimate the performance of the instrument, 2) adjust the geometry and mountings before building the prototype, 3) choose appropriate materials, and 4) determine the response functions of the instrument.
Because GEANT includes the most-complete database of cross sections and interactions of many particles, both the instrumental background and the sources of interest can be simulated.
Using appropriate statistics, the sensitivity of the instrument can be estimated for different cases, for example as a function of time, energy, spectral shape, and aspect angle.
Including the anti-coincidence shield in addition to the active detector(s) in the simulation will provide an estimate of the background reduction.
Finally, for the full response as a function of initial energy, zenith, azimuth, and other environmental parameters,  the satellite (or balloon) structure must also be included.
The more detailed the geometrical mass model of the instrument, the better the scientific output in the end.

The more sophisticated the mass model is, however, the longer the simulations will run.
In addition, the computational resources to perform the simulations might also be limited.
For example, the SPI response simulations \citep{Sturner2003_SPI} were performed for only 51 individual energies for the more than two decade spanning energy range.
This was required because a full, highly resolved energy response, would have taken years to simulate.
With the ground calibration \citep{Attie2003_SPI} (Sec.\,\ref{sec:calibration}), the simulations were validated so that a full response could be constructed by interpolation.
Although SPI has an energy resolution of $<0.2\,\%$, finely resolving $\gamma$-ray lines, the details of the spectral response are only inserted after the interpolation step.
Another newer example is the Compton telescope response of COSI \citep{Zoglauer2006_MEGAlib}:
For a single energy bin, here the 511\,keV line, it took about one million CPU hours to obtain enough statistics for the Compton Data Space to be populated.
Even though the response matrix in Compton telescopes is sparse, many photons are required to avoid unphysical jumps in the response.
Interpolations in the response functions from simulations are generally unavoidable.
Typically the assumption is made that the response functions present no jumps in different instrument dimensions.
While this assumption is true to first order, instruments are typically not perfect: they have edges, are asymmetric, and placed in the unpredictable space environment.

Finally, as mentioned above, simulations are iterated with calibration measurements (Sec.\,\ref{sec:calibration}).
This means that once the simulations are performed and a prototype or flight model of the instrument is built, the calibration shows how far off the expectations are from reality.
This offset in different dimensions can be used to adjust the simulations for a better response function.

Another problem for $\gamma$-ray instruments is that the space environment, which produces most of the primary and secondary instrumental background radiation, is challenging to predict using simulations.
While a multitude of instruments in space have measured the CR background, for example, it can hardly be predicted on the time scale of minutes to hours, important for long-term observations of diffuse emission.
In the case of background simulations for SPI, the shape of the spectrum was rather well determined with many instrumental lines, but the absolute amplitude in INTEGRAL's high eccentricity and high inclination orbit was missed by a factor of three \citep{Weidenspointner2003_SPI}.
However, these simulations were performed with an older version of GEANT (v3, now v4 with several updated cross sections and reactions) so that more than two decades have passed, and the prospects to provide a stand-alone background model for low-energy $\gamma$-rays should be investigated again.

\subsection{Calibrations}\label{sec:calibration}
Calibration measurements on Earth are performed to adjust the expectations from simulations to reality.
While the goal is identical for all instruments described in this Chapter, a separation between MeV and GeV telescopes, as well as $\gamma$-ray polarimeters is useful here because the calibration techniques differ in detail.
In general, as many zenith and azimuth angles and photon energies as possible are tested to provide a broad basis to scale and correct the simulated responses.
In practice, only a small set of aspect angles and energies can be tested in a laboratory setting because of available beam time and scheduling.
For values for which no calibration measurement could have been performed, multi-dimensional interpolations are used to obtain intermediate values.

Calibrations in space can also be performed with radioactive sources onboard the spacecrafts.
If those are not available, so-called `standard candles' could potentially be used calibrate the absolute flux of the instrument.
Standard candles are astrophysical sources with known and constant spectra.
For example for cosmology, type Ia supernovae can be made standardised candles to infer the expansion rate of the Universe.
In $\gamma$-ray astronomy, there are no known standard candles:
The typically used Crab pulsar (and nebula) has a very well measured spectrum from different instruments throughout the last 50 years.
However, it was shown that the source is actually variable on the order of a few percent within a timescale of ten years \citep{Wilson-Hodge2011_Crab}.
Instruments are typically calibrated to each other to obtain the same Crab spectrum which provides some level of consistency between the measurements.
However this also introduces a huge bias in the actual performance and response of the instruments, especially if the calibration source being used is variable.
Furthermore, the effective areas of most instruments vary differently as a function of energy which makes inter-instrument-calibrations virtually impossible without a known source spectrum.

To determine the angular resolution and in particular the point spread function (PSF), isolated pulsars are ideal targets to observe.
These are point-like sources for $\gamma$-ray instruments so that no spatial extent would be expected.
Therefore a measurement of the position as a function of energy determines the PSF most precisely.
Environmental conditions that impact the performance of the instruments, in particular the spectral and angular resolution, are also calibrated on Earth before launch.
These include, among others, temperature, voltage, details of the electronics (e.g., baseline noise, dark currents), or vibrations due to the oftentimes required cryo-cooling system.

The calibration of a high-energy telescope based on the detection of single photons is given in terms of the so-called `instrument functions' which describe the measurement process:
\begin{equation}
    N(E_m,{\vec x}_m)\,dE_m\,d{\Omega}_m\;=\;\left\lgroup{\int_{E_t}\!
\int_{{\vec x}_t} I(E_t,{\vec x}_t)\cdot G \cdot\, dE_t\,d{\Omega}_t
\,dt}\right\rgroup\;dE_m\,d{\Omega}_m
\end{equation}
where $N(E_m,{\vec x}_m)\,dE_m\,d{\Omega}_m$ is the number of detected events in `measurement space', $I(E_t,{\vec x}_t)\,dE_t\,d{\Omega}_t\ (photons/cm^2\ s)$ is the incident
flux of photons in `true' (model) space, and $G(E_t,E_m,{\vec x}_t,{\vec x}_m,{\vec x}_a,t)$ is a generalised instrument response function (Sec.\,\ref{sec:measurements}).
The physics of the photon detection process suggests that $G$ can be separated into a sequence of functions whose product is then an approximation of the true instrument function.
One assumes that (i) $G$ is constant over observation time $T$, and that (ii) the incident flux is uniform over the sensitive area of the experiment.
Then,
\begin{equation}
    G(E_t,E_m,{\vec x}_t,{\vec x}_m,{\vec x}_a,t) =
T\cdot A\cdot \varepsilon\cdot R
\end{equation}
where, $T$ is dead time corrected exposure time, $A({\vec x}_t,{\vec x}_a)$ is the geometrical cross-section of the instrument seen from the true incidence direction ${\vec x}_t$ with the telescope pointing at ${\vec x}_a$ ($a$ for attitude, e.g. measured in zenith/azimuth), $\varepsilon(E_t,{\vec x}_t,{\vec x}_a)$ is the efficiency averaged over $A$, and $R(E_t,E_m,{\vec x}_t,{\vec x}_m,{\vec x}_a)$ is the general response (dispersion) in energy, direction, and for attitude ${\vec x}_a$.
The essential parameters are the averaged efficiency $\varepsilon$ and the  response functions $R$.
Equally important for the sensitivity of an astronomical telescope is the suppression of instrumental and environmental background from non-astronomical sources (Sec.\,\ref{sec:instrumental_background}).
This background susceptibility is also a topic of calibration measurements.

\subsubsection{MeV: Radioactive Sources}\label{sec:MeV_calibation}
\begin{table}[h]
    \centering
    \begin{tabular}{c|c|l}
        \hline
        Isotope / Reaction & Half-life Time & Line Energy [keV] \\
        \hline
        \nuc{Am}{241} & 432.6\,yr & 13.9 (0.37), 26.34 (0.02), 59.54 (0.36) \\
        \nuc{Ba}{133} & 10.55\,yr & \parbox{7.25cm}{30.63 (0.34), 30.97 (0.62), 34.92 (0.06), 34.99 (0.11), 35.82 (0.04), 53.16 (0.02), 79.61 (0.03), 81.00 (0.33), 276.40 (0.07), 302.85 (0.18), 356.01 (0.62), 383.85 (0.09)} \\
        \nuc{Co}{57} & 271.74\,d & 14.41 (0.09), 122.06 (0.86), 136.47 (0.11) \\
        \nuc{Ce}{139} & 137.64\,d & \parbox{7.25cm}{33.03 (0.23), 33.44 (0.41), 37.72 (0.04), 37.80 (0.08), 38.73 (0.02), 165.86 (0.80)} \\
        \nuc{Cs}{137} & 30.08\,yr & 31.82 (0.02), 32.19 (0.04), 661.66 (0.85) \\
        \nuc{Mn}{54} & 312.20\,d & 834.85 (1.00) \\
        \nuc{Eu}{152} & 13.517\,yr & \parbox{7.25cm}{344.28 (0.27), 411.12 (0.02), 778.90 (0.13), 1089.74 (0.02), 1299.14 (0.02)} \\
        \nuc{Zn}{65} & 243.93\,d & 1115.54 (0.50), [511.0 (0.03)] \\
        \nuc{Co}{60} & 5.2714\,yr & 1173.23 (1.00), 1332.49 (1.00) \\
        \nuc{Na}{22} & 2.6018\,yr & 1274.54 (1.00), [511.0 (1.80)] \\
        \hline
        \nuc{C}{13}(p,$\gamma$)\nuc{N}{14} & - & \parbox{7.25cm}{1635, 2313, 3948, 5105, 5690, 6445, 6858, 7027, 8062, 9169} \\
        \nuc{Al}{27}(p,$\gamma$)\nuc{N}{28} & - & \parbox{7.25cm}{1522, 1779, 2839, 3063, 4498, 4608, 4743, 6020, 6265, 7924, 7933, 10763} \\
        \hline
        \hline
    \end{tabular}
    \caption{Radioactive isotopes and resonance photons used for the calibration of MeV instruments between 10\,keV and 10\,MeV (including K- and L-shell X-rays) with at least a branching ratio of 0.01 (in parentheses) \citep{NUDAT,Kiener2004_13Cpgamma14N,Anttila1977_27Alpgamma28Si}.}
    \label{tab:radioactive_sources}
\end{table}
\noindent At MeV energies, the instrument calibration is performed either with radioactive sources of precisely known photon energies or particle beams with known resonances in specific reactions.
Table\,\ref{tab:radioactive_sources} gives an overview of typically used sources and reactions as a function of photon energy.

For example, the ground calibration of SPI was performed at Bruyeres-Le-Chatel (BLC; \citep{Attie2003_SPI}).
Low-intensity radioactive sources were mounted at distances of only 8\,m to calibrate the energy resolution, the camera efficiency and to adjust and find inhomogeneities of the instrument.
To test the imaging capabilities of SPI, radioactive sources were placed at greater distances of 125\,m.
In addition to radioactive sources, a Van de Graaf accelerator was also used to reach to higher energies, $E_{\gamma} > 2.7$\,MeV \citep{Mandrou1997_SPI,Schanne2002_SPI}.

SPI imaging was calibrated with high-flux sources of \nuc{Am}{241}, \nuc{Cs}{137}, \nuc{Co}{60}, and \nuc{Na}{24}, located outside the laboratory through a transparent window.
The beams were strongly collimated to avoid radiation in other directions and scatters inside the detector hall.
For energy calibration of SPI, the mask was removed so that all detectors could be illuminated simultaneously.
Although the field of view of SPI is about $16^{\circ} \times 16^{\circ}$, the calibration took place only on axis and no other zenith and azimuth angles were tested.
Instead, the telescope efficiency is derived from the absorption properties of the individually measured transmissivities of opaque and transparent mask elements \citep{Sanchez2003_SPI}.
A mathematical model was then fit to obtain the correct mask properties given the full set of calibration measurements.
For future MeV instruments, such as COSI, a full field of view calibration is anticipated which will reduce the systematic uncertainties from calibrations with a single beam line plus simulations to obtain the instrument response functions.

\subsubsection{GeV: Particle Accelerators}\label{sec:GeV_calibration}
There have been different ways to derive the instrument functions for past and present GeV telescopes:
in pioneering counter-type instruments (e.g., OSO-3, Explorer XI) design and calibrations were based on analytical estimates (geometry, cross sections, and individual detector responses) which were verified on beam tests using $\pi^{0}$ decay photons or a synchrotron beam and the response to CR muons at ground level.
The first imaging telescopes (e.g., SAS-2, COS-B) were calibrated with $\gamma$-ray synchrotron beams (up to a few 100\,MeV), electron beams (to characterise the tracking), and CR muons (to test the anticoincidence counters).

A full scale beam calibration was performed on the next generation EGRET instrument \citep{1987ITNS...34...36T,Kanbach1988_EGRET}.
The $\gamma$-ray beam at the Stanford Linear Accelerator (SLAC) was generated by inverse Compton scattering of laser photons (2.34\,eV) on the high-energy electron pulses from the  accelerator.
Tuning the linac electron energy from about 0.65 to 20\,GeV resulted in back-scattered photons at ten energies between 15\,MeV and 10\,GeV, with an energy dispersion of $\sim 11\%$ FWHM.
The beam was constrained to a collimated pencil beam, with an intensity of $\sim 0.4$ photons per 40\,ns pulse width and about 15 pulses/sec.
EGRET, with a total weight of about 1.8 tons, was mounted to an electro-hydraulic computer-controlled fixture which could position the telescope with 0.2\,mm accuracy laterally and $0.1^{\circ}$ in attitude angles.
A raster-scan over and beyond the sensitive volume of the telescope with scan points spaced by 5\,cm was performed for attitudes out to $40^{\circ}$ off-axis.
These calibration measurements undertaken over a three month period in 1986 provided not only the data for a model of the efficiency of detection, but also the efficiency of `recognition' of good $\gamma$-ray events in data analysis of real events.
This latter point is often difficult to quantify with simulated data (Sec.\,\ref{sec:simulations}).

Since the space shuttle accident in 1986 delayed the launch of {\it{CGRO}} until April 5, 1991, further calibration measurements could be performed with EGRET.
At Brookhaven National Laboratory a proton beam up to 10\,GeV was used to generate background $\gamma$-rays in the material outside the anticoincidence system as a test to see the effect of CR particles on the instrumental background.
This was found to be significantly below the expected cosmic $\gamma$-ray background.
Final $\gamma$-ray measurements at the MIT Bates accelerator up to 830\,MeV were used to verify some technical developments on the instrument after the main calibrations.

With the experience from ground level calibrations and results from EGRET on $CGRO$ using celestial sources, e.g. pulsed photons from strong pulsars for angular resolution, the design, simulation, and calibration of the currently active GeV telescope {\it{Fermi}}-LAT could be undertaken without a full-instrument beam test.
Subsystems, called the LAT Calibration Unit, were exposed to a large variety of beams at CERN and the GSI accelerator facilities to probe $\gamma$-ray detection and background sensitivity.
Beams of photons ($<2.5$\,GeV), electrons (1--300\,GeV), hadrons (pions and protons, at up 100\,GeV) and ions (C, Xe, 1.5\,GeV/n) were used.
But the much refined simulation tools based on GEANT4 and the fine tuning of the instrument and analysis tools provided the necessary instrument functions with increasing accuracy as the Fermi mission continues after its launch in 2008.

\subsubsection{Gamma-Ray Polarimetry}\label{sec:polarimetry_calibration}
The calibration of $\gamma$-ray polarimeters is particularly complicated because naturally, no polarised $\gamma$-ray sources exist (on Earth) and therefore must first be produced.
In particular, a radioactive source is placed inside a lead structure which contains a small plastic component.
The plastic is used as a target for the decay $\gamma$-rays to scatter at an angle of $90^{\circ}$ for the creation of a nearly 100\,\% polarised beam.
The scattered $\gamma$-rays then pass through a collimator before exiting so that the polarisation response can be calibrated.
Alternatively, the POLAR instrument, for example, used the correlated polarisation of two 511\,keV photons from a partially shielded \nuc{Na}{22} source for calibrations \citep{Orsi2011_POLAR}.
The \nuc{Na}{22} $\beta^+$-decay emits positrons which quickly form positronium in the surrounding shield and annihilate by the emission of two 511\,keV photons in opposite directions.
The 511\,keV photons are then linearly polarised with a polarisation degree of 55--60\,\% in this experimental setup.

Another possibility is to use particle accerelator beam lines.
As a by-product of cyclotrons and synchrotrons, for example, polarised photon emission from the acceleration and bending of electrons in the magnetic fields of circular accelerators can be used for calibration.
Here, the radiation is pulsed with an energy in the range of keV to MeV, which depends mostly on the bending radius and magnetic field strength of the accelerator, as well as the mass and charge of the particle that produces the synchrotron radiation.
An example would be the HIGS beam (High Energy Gamma-Ray Source via synchrotron self-Compton) which was 100\,\% polarised up to nearly 50\,MeV and was used in the calibrations of the MEGA telescope to obtain a polarisation response up to about 5\,MeV \citep{Litvinenko1995_HIGS_Duke_beam,Zoglauer2006_PhD}.

\section{Outlook}\label{sec:outlook}
The future of space-based $\gamma$-ray telescopes is discussed in detail in a Chapter in this book.
A brief outline is provided here.
Many of the telescopes under development for $\gamma$-ray astronomy are designed to improve the sensitivity in the MeV gap (Sec.\,\ref{sec:MeV_gap}) and therefore include instruments that are sensitive to $\gamma$-rays interacting through Compton scattering.

The Compton Spectrometer and Imager (COSI; \citep{Tomsick2019_COSI}) is a soft $\gamma$-ray survey telescope that will operate in the energy range from 0.2--5\,MeV after its planned launch in 2025.
It comprises sixteen high-resolution germanium detectors enabling it to perform imaging of the sky, provide high background rejection and measure the polarisation of the incident $\gamma$-ray.
The technologies for COSI have been developed over decades and notably, were tested on a 46-day scientific balloon flight on NASA's new Super Pressure Balloon, launched from Wanaka, New Zealand in May 2016 \citep{Kierans2016_COSI}.
Within its nominal two-year mission, COSI will improve upon the sensitivities of SPI and COMPTEL by at least a factor of 10.

Another Compton telescope that, for the first time, used also the recoil electron information through measurements of the electron tracks in a gas detector on a balloon flight, is SMILE \citep{Takada2022_SMILE_Crab}.
In joint determination of the photon scattering path and the electron recoil energy and direction, the Compton rings reduce to arcs, which allows for a narrower point spread function as well as a clearer background photon rejection.

The All-sky Medium Energy Gamma-ray Observatory (AMEGO; \citep{2020SPIE11444E..31K}) is a mission concept designed to explore the MeV sky with instruments capable of providing sensitive coverage of both the Compton (0.2--10\,MeV) and pair conversion (0.01--10\,GeV) regimes.
In the Compton regime, solid state technology (double-sided silicon detectors) will be used thus providing substantial performance improvements relative to COMPTEL.
In the pair-production regime, AMEGO has been optimised to have its peak performance at lower energy than {\it{Fermi}}-LAT.
This is achieved by minimising the passive material (i.e. the conversion foils) in the tracker and by enhancing the readout at low energies for the calorimeter.
Indeed AMEGO will have two calorimeters, a low-energy precision calorimeter to measure the energy of Compton-scattered events, and a second calorimeter to contain the high-energy events. 

The ASTROGAM satellite concept comprises a $\gamma$-ray telescope designed to operate between 100\,keV and 2\,GeV, being capable of the detection of photons in both the Compton and pair regimes \citep{2021arXiv210202460D}.
Like AMEGO, the proposal is to use double-sided silicon detectors to track both the Compton and pair-production events.
The Calorimeter design comprises a pixelated detector made of a high-Z scintillation material such as thallium-activated cesium iodide or cerium bromide.
ASTROGAM and AMEGO would have comparable sensitivities on the order of 10\,mCrab for an observation time of 1\,Ms, bridging the MeV gap in the range from 0.1\,MeV to 10\,GeV.

ASTENA, the Advanced Surveyor of Transient Events and Nuclear Astrophysics is a mission concept being proposed for a deep study of the transient sky and to perform nuclear astrophysics carrying two instruments onboard \citep{Frontera2019_ASTENA}.
The Narrow Field Telescope (NFT, 50--700\,keV) comprises a 20\,m focal length Laue lens made with bent germanium and silicon crystals with a solid state device made with four layers of CZT at the focus.
The Wide Field Monitor with Imaging and Spectroscopic capabilities (WFM-IS, 0.002--20\,MeV) is composed of 12 position sensitive detectors.
These are distributed around the NFT and are oriented $15^{\circ}$ two by two outwards with respect to the Laue lens axis in order to extend the field of view of the overall instrument.
The continuum sensitivity of NFT would be 2--3 orders of magnitude better than that of currently working low-energy $\gamma$-ray telescopes.
Also, its anticipated angular resolution of 5\,arcsec would be unprecedented in the MeV regime.
ASTENA, ASTROGAM, and AMEGO would all have $\gamma$-ray line sensitivities that would be a factor of 10--100 better than SPI, with better improvements for higher energies.

Other concepts under development for future $\gamma$-ray missions include CubeSats, a small satellite technology that can make space observations more readily accessible thanks to their low cost and fast-delivery design \cite{2021SPIE11852E..6MD}.
COMCUBE, for example,  is a project under development with a goal of using a Compton polarimeter to measure the polarisation of bright GRBs \citep{2021sf2a.conf..105L}.
Another example would be BurstCube \citep{Perkins2020_BurstCube} that is aiming for quick follow-ups to gravitational wave events.

The idea of using a time projection chamber to measure the polarisation of $\gamma$-rays in the MeV range was explored by \citep{2019NIMPA.936..405B} who constructed the HARPO prototype to demonstrate the measurement of polarisation of a linearly polarised beam. A discussion on the use of time projection chambers for $\gamma$-ray astronomy can be found in a Chapter of this book.

At higher energies, the future Cherenkov Telescope Array (CTA; \citep{2019scta.book.....C}), with its extensive energy range will be sensitive to $\gamma$-rays from 0.02--300\,TeV thus increasing the reach of imaging arrays of ground-based telescopes into the energy range traditionally only accessible to space-based instruments.

\bibliographystyle{unsrt}
\bibliography{thomas.bib}

\end{document}